\documentclass[11pt]{article}
\usepackage{axodraw2,pix}
\usepackage{epsfig}
\usepackage{amsfonts}
\usepackage{amsmath}
\usepackage{amssymb}
\usepackage{bbm,bm}
\usepackage{cite}
\allowdisplaybreaks[1]

\usepackage{tikz}
\usetikzlibrary{shapes,arrows,shadows,automata,positioning}
\newdimen\nodeDist
\usepackage{comment}
\usepackage{float} 
\usepackage[normalem]{ulem} 

\usepackage[
  colorlinks=true,
  linkcolor={red!50!black},
  citecolor={blue!50!black},
  filecolor=black,
  urlcolor=black,
  breaklinks=true
  ]{hyperref}
\usepackage{orcidlink}

\renewcommand{\vec}[1]{{\bf #1}}

\newcommand{\gammaE}{\gamma_\rmii{E}}

\newcommand{\Lamd}{\bar\Lambda_{\rmi{3d}}}
\newcommand{\LamD}{\bar\Lambda}
\newcommand\MSbar{$\overline{\rm MS}$}

\renewcommand{\Lb}{L_b}

\newcommand{\Ltd}{L_\rmi{3d}}

\newcommand{\Tc}{T_{\rm c}}
\newcommand{\Tn}{T_{\rm n}}
\newcommand{\dA}{d_\rmii{A}}

\newcommand{\CA}{C_\rmii{\!A}}
\newcommand{\CF}{C_\rmii{F}}

\newcommand{\alphaF}{\widehat\alpha}
\newcommand{\alphaFR}{\alpha}
\newcommand{\mD}{m_\rmii{D}}
\newcommand{\mG}{m_\rmii{G}}
\newcommand{\mB}{m_\rmii{$B$}}
\newcommand{\mh}{m_h}

\newcommand{\xc}{x_{\rm c}}
\newcommand{\yc}{y_{\rm c}}
\newcommand{\yD}{y_\rmii{D}}

\newcommand{\geff}{g_\rmi{eff}}

\def\lsi{\raise0.3ex\hbox{$<$\kern-0.75em\raise-1.1ex\hbox{$\sim$}}}
\def\gsi{\raise0.3ex\hbox{$>$\kern-0.75em\raise-1.1ex\hbox{$\sim$}}}

\renewcommand{\nn}{\nonumber \\}
\renewcommand{\rmi}[1]{{\mbox{\scriptsize #1}}}
\newcommand{\rmii}[1]{{\mbox{\tiny\rm{#1}}}}

\newcommand{\Tint}[1]{{\hbox{$\sum$}\!\!\!\!\!\!\!\int\,}_{\!\!\!\!\raise-0.9ex\hbox{$\scriptstyle{#1}$}}}
\newcommand{\Tinti}[1]{{{\Sigma}\!\!\!\!\raise0.3ex\hbox{$\int$}_\rmii{${#1}$}}}
\newcommand{\Tintip}[1]{{{\Sigma'}\!\!\!\!\!\raise0.3ex\hbox{$\int$}_\rmii{${#1}$}}}

\newcommand{\deltabar}{\raise-0.02em\hbox{$\bar{}$}\hspace*{-0.8mm}{\delta}}

\newcommand{\vev}{vev}

\newcommand{\xiNew}{\xi}
  
\def\scfc{0.7}  
\def\phgt{21}   
\def\pwc{21}    
\def\pwcb{31.5} 
\def\pwcc{42} 
\newcommand{\PIC}[4]{\;\parbox[c]{#2 pt}{\begin{picture}(#2,#3)(0,0)
\SetWidth{1.0}\SetScale{#4} #1 \end{picture}}\;}
\renewcommand{\pic}[1]{\PIC{#1}{\pwc}{\phgt}{\scfc}}
\renewcommand{\picb}[1]{\PIC{#1}{\pwcb}{\phgt}{\scfc}}
\renewcommand{\picc}[1]{\PIC{#1}{\pwcc}{\phgt}{\scfc}}

\makeatletter \@addtoreset{equation}{section} \makeatother
\renewcommand{\theequation}{\arabic{section}.\arabic{equation}}
\makeatletter
\renewcommand\section{\@startsection{section}{1}{\z@}%
  {-5.5ex \@plus -1ex \@minus -.2ex}
  {2.3ex \@plus.2ex}%
  {\normalfont\large\bfseries}}
\renewcommand\subsection{\@startsection{subsection}{2}{\z@}%
  {-3.25ex\@plus -1ex \@minus -.2ex}%
  {1.5ex \@plus .2ex}%
  {\normalfont\normalsize\bfseries}}
\renewcommand\thesection{\@arabic\c@section}
\renewcommand\thesubsection{\thesection.\@arabic\c@subsection}
\renewcommand{\@seccntformat}[1]{%
  \csname the#1\endcsname.\hspace{1.0em}}
\makeatother

\begin{document}

\flushbottom

\begin{titlepage}

\begin{flushright}
July 2026
\end{flushright}
\begin{centering}

\vfill

{\Large{\bf
    Hard thermal contributions to\\
    phase transition observables at NNLO
  }
}

\vspace{0.8cm}

\renewcommand{\thefootnote}{\fnsymbol{footnote}}
Fabio Bernardo%
\orcidlink{0009-0008-0719-3219}%
,$^{\rm a,}$%
\footnote{fabio.bernardo@unige.ch}
Mikael Chala%
\orcidlink{0000-0002-8194-1050}%
,$^{\rm b,}$%
\footnote{mikael.chala@ugr.es}
Luis Gil%
\orcidlink{0009-0007-4502-7521}%
,$^{\rm b,}$%
\footnote{lgil@ugr.es}
and
Philipp Schicho%
\orcidlink{0000-0001-5869-7611} %
$^{\rm a,}$%
\footnote{philipp.schicho@unige.ch}

\vspace{0.8cm}

$^\rmi{a}$%
{\em
D\'epartement de Physique Th\'eorique, Universit\'e de Gen\`eve,\\
24 quai Ernest Ansermet, CH-1211 Gen\`eve 4,
Switzerland\\}

\vspace{0.3cm}

$^\rmi{b}$%
{\em
Departamento de F\'isica Te\'orica y del Cosmos, Universidad de Granada,\\
Campus de Fuentenueva, E–18071 Granada, Spain\\}

\vspace*{0.8cm}

\mbox{\bf Abstract}

\end{centering}

\vspace*{0.3cm}

\noindent
To construct the high-temperature effective field theory of
gauge-Higgs models up to $\mathcal{O}(g^6)$ in the gauge coupling,
we integrate out hard modes to three-loop level and use 
the next-to-next-to-leading order effective potential.
For the Abelian Higgs model,
we quantify the impact of both
higher-dimensional operators and
higher-loop corrections
on thermodynamic parameters relevant
for gravitational-wave observables,
finding that one-loop dimension-six effects typically dominate over
two- and three-loop corrections to super-renormalizable parameters
for the strongest transitions.
We derive
the three-loop scalar and Debye masses for
the ${\rm U(1)}$ and ${\rm SU}(N)$ gauge-Higgs models,
as well as
the two-loop quartic couplings for the Abelian case,
show gauge independence of physical parameters, and
demonstrate that no new master integrals are required for the matching,
while consistency of 4d and 3d renormalizability
points to previously missing contributions in these master integrals.
As a byproduct, we report a previously missing contribution to
the three-loop QCD Debye mass.

\vfill
\end{titlepage}

{\hypersetup{hidelinks}
\tableofcontents
}
\clearpage

\renewcommand{\thefootnote}{\arabic{footnote}}
\setcounter{footnote}{0}

%
\section{Introduction}
\label{sec:intro}

The potential future observation of a stochastic gravitational wave (GW)
background of cosmological origin has opened the door
to entirely new avenues in high energy physics research.
In particular, one type of source that has drawn
a lot of interest in recent years are
strong first-order phase transitions (PTs)
associated to symmetry breaking
in the early universe~\cite{Harry:2006fi,Kawamura:2006up,Ruan:2018tsw,LIGOScientific:2014pky,Caprini:2019egz,NANOGrav:2020bcs}.
The reason is that the Standard Model (SM) does not predict any PT
neither in the electroweak~\cite{Kajantie:1996mn, Gurtler:1997hr, Csikor:1998eu}
nor in the QCD sectors~\cite{%
  Braaten:1995cm,Braaten:1994na,Braaten:1995jr,Kajantie:1997tt,Laine:2019uua,
  Laine:2018lgj,Ghiglieri:2021bom,Navarrete:2024ruu,Gorda:2025cwu},
so observing a GW background compatible with such a transition
would be an undeniable proof of new physics.

Studying these thermally-induced transitions typically relies on effective field theory (EFT) methods,
such as high-temperature dimensional reduction,
which systematically integrates out the effect from heavy thermal excitations~\cite{Matsubara:1955ws} in favor of
a three-dimensional EFT (3d~EFT) for the infrared bosonic modes.
The systematics of this EFT framework were clarified long ago~\cite{Ginsparg:1980ef,Appelquist:1981vg}, and
it has since become a cornerstone in the study of physics of
thermal equilibrium. In this construction, higher-order effects
from heavy modes are encoded in a tower of local operators that arise in
a perturbative expansion in $m_\rmi{eff}/(\pi T)$, where
$m_\rmi{eff}$ is the mass of the field driving
the transition and $\pi T$ is the heavy thermal scale that is integrated out.

Most existing studies that explore PTs within
dimensional reduction~\cite{%
  Brauner:2016fla,Andersen:2017ika,Niemi:2018asa,Gorda:2018hvi,Kainulainen:2019kyp,
  Croon:2020cgk,Gould:2019qek,Niemi:2020hto,Gould:2021ccf,Gould:2021dzl,
  Schicho:2021gca,Niemi:2021qvp,Camargo-Molina:2021zgz,Niemi:2022bjg,
  Ekstedt:2022ceo,Gould:2022ran,Ekstedt:2022zro,Biondini:2022ggt,
  Schicho:2022wty,Lofgren:2021ogg,Gould:2023jbz,Kierkla:2023von,
  Aarts:2023vsf,Niemi:2024axp,Chala:2024xll,Qin:2024idc,Gould:2024jjt,
  Chakrabortty:2024wto,Niemi:2024vzw,Kierkla:2025qyz,Bhatnagar:2025jhh,
  Bernardo:2025vkz,Chala:2025aiz,Zhu:2025pht,Chala:2025oul,Li:2025kyo,
  Annala:2025aci,Navarrete:2025yxy,Chala:2025xlk,Chala:2025cya,Biekotter:2025npc,
  Chakrabortty:2026swu
  }
focus on leading-order contributions in the 3d~EFT while omitting operators beyond the super-\linebreak renormalizable level. While this approximation suffices for weaker PTs driven by thermal loop effects, recent studies~\cite{Chala:2024xll,Niemi:2024vzw,Chala:2025oul,Chala:2025aiz,Bernardo:2025vkz,Chala:2025cya,Biekotter:2025npc}
provide increasing evidence
that an accurate description of moderately strong PTs requires including marginal operators,
such as dimension-six terms.%
\footnote{%
  Henceforth, unless otherwise stated,
  we use four-dimensional (4d) units when discussing energy dimensions,
  even though these operators live in 3d Euclidean space.
}
Their inclusion turns out to be 
crucial to reduce
the large uncertainties that populate nucleation computations
for strong transitions~\cite{Croon:2020cgk}, namely gauge and scale dependence,
which become relevant if one truncates the EFT expansion
at low orders~\cite{Bernardo:2025vkz,Chala:2025aiz}.
In the limit of very strong PTs,
the high-temperature expansion breaks altogether and
alternative approaches have recently been suggested~\cite{Navarrete:2025yxy}.
For weaker transitions,
the perturbative series must be reorganized until non-perturbative effects become relevant.
In this regime, thermal loop corrections are dominant.
In the intermediate regime,
higher-order operators and loop corrections within
the 3d~EFT framework can bridge
the gap between these two limiting cases,
while simultaneously scrutinizing the validity of
broad-temperature approaches.

In this work,
we build upon this direction by providing
a thorough analysis of the high-temperature limit of
the Abelian Higgs model at $\mathcal{O}(g^6)$ in the gauge coupling constant.
Employing the usual power counting~\cite{Arnold:1992rz,Ekstedt:2022zro,Ekstedt:2024etx},
this implies the computation of
three-loop matching contributions to the masses,
two-loop contributions to couplings, and
one-loop contributions to dimension-six operators in the 3d~EFT.
The Abelian Higgs model is the simplest model that captures the essential features of more complex gauge–Higgs theories, and
serves as a playground to explore the difficulties and consequences of such high-order computations.
In particular, extending~\cite{Chala:2025oul,Bernardo:2025vkz},
our main goals are threefold:
\begin{itemize}
    \item[(a)]
    to prove the cancellation of gauge dependence order by order in loops,
    \item[(b)]
    to show the convergence of 3d parameters, by inspecting their increasingly small renormalization scale dependence at higher orders, and
    \item[(c)]
    to quantify the relevance of higher-dimensional operators in strong PTs
    and compare it to the contribution of the three-loop matching.
\end{itemize}
By using
an in-house
{\tt FORM}~\cite{Davies:2026cci}
implementation of the standard Laporta algorithm~\cite{Laporta:2000dsw} for
the reduction of sum-integrals~\cite{Nishimura:2012ee},
we recover a series of integration-by-part relations~\cite{Ghisoiu:2015uza} for
three-loop bosonic sum-integrals and
show that no new master integrals appear in
the computation of the thermal scalar masses in
generic ${\rm SU}(N)$ gauge theories with a fundamental scalar.
The integral reduction also agrees 
with results obtained with
{\tt SIRENA}~\cite{Gil:2026cqz}.
Additionally,
we also identify an error in the original evaluation of one of
the three-loop sum-integrals entering our results, and
consequently report on a previously missing contribution to
the three-loop QCD Debye mass.

The article is organized as follows.
Section~\ref{sec:setup} introduces the Abelian Higgs model.
The details of its dimensional reduction are presented in sec.~\ref{sec:dr}.
Section~\ref{sec:higherorder} describes
the computation of the critical parameters and assesses
the contribution from different orders in perturbation theory.
Section~\ref{sec:outlook} summarizes our results and outlines future lines of work.
Technical details of our computations are relegated to
appendices~\ref{app:4d rges}, \ref{app:masters} and~\ref{app:DR:relations}.

%
\section{Setup}
\label{sec:setup}

We consider the high-temperature limit of a ${\rm U}(1)$ gauge theory
coupled to a complex fundamental scalar $\phi$ with unit charge.
In 
four-dimensional (4d)
Euclidean space,
the theory is described by the Lagrangian
\begin{equation}
\label{eq:lag4d}
\mathcal{L}_\rmi{4d} =
    \frac{1}{4} F_{\mu\nu} F_{\mu\nu}
  + (D_\mu \phi)^\dagger (D_\mu \phi)
  + \mu^2 \phi^\dagger \phi
  + \lambda (\phi^\dagger \phi)^2
  + \mathcal{L}_\rmii{GF}
  \,,
\end{equation}
where
$F_{\mu\nu} = \partial_\mu B_\nu - \partial_\nu B_\mu$
is the field strength tensor and
the covariant derivative is defined as
$D_\mu = \partial_\mu - i g B_\mu$,
with
$g$ being the gauge coupling.
The $R_\xi$ gauge-fixing term reads
\begin{equation}
\label{eq:gf:U1}
    \mathcal{L}_\rmii{GF} = \frac{1}{2 \xi} \left( \partial_\mu B_\mu \right)^2
    \,.
\end{equation}

Far from a toy model,
this theory plays a significant role in the study of dark-sector PTs
that are decoupled from the SM~\cite{Jaeckel:2016jlh,Addazi:2017gpt,Croon:2018erz,Breitbach:2018ddu,Christiansen:2025xhv}.
Moreover, it is closely related to condensed matter physics, where
its 3d counterpart
provides
a mean-field description of superconductivity~\cite{Halperin:1973jh,Dasgupta:1981zz,Kajantie:1998zn}.
More recently,
the classically conformal Abelian Higgs model has been used
to study primordial black hole formation
from supercooled PTs~\cite{Lewicki:2024sfw},
though their predicted abundances remain limited
by the constraints of the formation mechanism~\cite{Franciolini:2025ztf,Kierkla:2025vwp}.

%
\section{High-temperature effective field theory}
\label{sec:dr}

One major challenge in finite-temperature computations is
the emergence of a hierarchy of dynamically generated scales due
to in-medium effects.
These scales are defined relative to the hard scale $\pi T$,
which sets the typical energy of on-shell particles at high temperatures, and
in a gauge theory they read~\cite{Gould:2023ovu}
\begin{equation}
    \underbrace{\vphantom{g}\pi T}_{\text{hard scale}}\gg
    \underbrace{g T}_{\text{soft scale}}\gg
    \underbrace{g^{3/2} T}_{\text{softer scale}}\gg
    \underbrace{g^2 T}_{\text{ultrasoft scale}}
    \,,
\end{equation}
where $g$ is the largest gauge coupling, which we use as the power counting parameter.%
\footnote{%
  Following~\cite{Bernardo:2025vkz},
  we adopt the power counting
  $\mu^2/T^2 \sim \lambda \sim g^2$ for 4d parameters.
}

The first scale is that of the heavy Matsubara modes,
whose masses are of $\mathcal{O}(\pi T)$~\cite{Matsubara:1955ws}.
The second is associated with the screening of bosonic zero modes, which
induces a Debye mass of $\mathcal{O}(g T)$ for the temporal component of gauge bosons.
The third corresponds to the thermal effective mass of the scalar driving the transition,
which can become parametrically smaller than the soft scale near the critical temperature.
Finally, the ultrasoft scale corresponds to non-perturbative contributions
from spatial gauge bosons in the deep infrared~\cite{Linde:1980ts}.

This plethora of hierarchies is problematic,
since it leads to the appearance of large logarithms in perturbative computations.
One way to systematically handle them is to use
finite-temperature
EFT techniques~\cite{Kajantie:1995dw,Braaten:1995cm,Gould:2023ovu}.
This framework consists in building a tower of effective theories by successively integrating out
heavy degrees of freedom with respect to each scale,
through a process called~\textit{matching}.
With this, at lower scales
the resulting effective Wilson coefficients 
are renormalization-group (RG) improved,
ensuring the resummation of large logarithms. 

In going from the hard to the soft scale,
one performs {\em dimensional reduction} (DR),
constructing a static 3d~EFT for bosonic zero modes.
The EFT parameters encode all temperature dependence.
At scales below the hard scale,
the matching between different 3d theories follows
the standard vacuum procedure and is temperature-independent.
Schematically, there exist 
two alternative constructions~\cite{Kajantie:1995dw},
\begin{align}
\label{eq:DR:A}
  \mathcal L_\rmi{4d} &\stackrel{\mathrm{DR}}{\longmapsto}
  \mathcal L_{\text{soft}} \mapsto
  \mathcal L_{\text{softer}} \mapsto
  V_{\text{eff}}
  \tag{DR-A}
\,, \\
\label{eq:DR:B}
  \mathcal L_\rmi{4d} &\stackrel{\mathrm{DR}}{\longmapsto}
  \mathcal L_{\text{soft}} \mapsto
  V_{\text{eff}}
  \,,
  \tag{DR-B}
\end{align}
where the final mapping indicates that the effective potential $V_\mathrm{eff}$ 
is computed at the lowest energy scale in each respective approach.

The dimensional reduction of the Abelian Higgs model,
including higher-dimensional operators,
was carried out in~\cite{Bernardo:2025vkz}.
The latter demonstrated that
the approach in~\eqref{eq:DR:A} becomes unreliable in
the regime of strong PTs, where nucleation takes place already at
the soft scale.
Indeed, the softer EFT Lagrangian can be written as
an expansion in operators of increasing dimension,
\begin{equation}
    \mathcal L_{\text{softer}} \supset \text{const.} \times \frac{\Lambda_{\text{soft}}^3}{4\pi}
    \Bigl( \frac{D_i}{\Lambda_{\text{soft}}} \Bigr)^m
    \Bigl( \frac{g_3 \phi}{\Lambda_{\text{soft}}} \Bigr)^n
    \ ,
\end{equation}
where $\Lambda_{\text{soft}} \sim gT$ is the soft scale,
$g_3$ is the effective gauge coupling,
$D_i$ denotes the gauge covariant derivative, and
$\phi$ is the scalar driving the transition.
For strong PTs,
the scalar vacuum expectation value (\vev{}) in
the broken phase typically becomes much larger than $\Lambda_{\text{soft}}/g_3$,
leading to a breakdown of the operator expansion,
which invalidates the softer scale description.

In~\cite{Bernardo:2025vkz}
the matching was performed at
two-loop level (or $\mathcal{O}(g^4)$) for dimension-two operators,
namely for the scalar and Debye masses; and
at one-loop level (or $\mathcal{O}(g^4)$) for dimension-four operators as
well as for higher-dimensional operators, which first appear at $\mathcal{O}(g^6)$.
While gauge dependence was shown to cancel consistently
for dimension-six operators upon redefinition onto a physical basis,
a residual gauge-dependent contribution was found at $\mathcal{O}(g^6)$ in
the scalar mass;
see eq.~(B.68) in~\cite{Bernardo:2025vkz}.

As we shall see, the gauge-dependent contribution arising from the three-loop matching of the mass parameters, which enters at $\mathcal{O}(g^6)$, precisely cancels this residual gauge dependence;
see also~\cite{Balui:2025yvd} for a discussion on gauge independence of effective potentials.
As a result, the 3d~EFT is fully gauge independent up to $\mathcal{O}(g^6)$.
Furthermore, our analysis provides new results for the two-loop matching of
the quartic operators up to $\mathcal{O}(g^6)$.

\subsection{Effective theory at the soft scale}
\label{sec:drgw}

In the soft-scale 3d~EFT Lagrangian~\eqref{eq:DR:B},
\begin{align}
\mathcal{L}_{\rmi{soft}}&=
  \label{eq:soft lag}
    \frac{1}{4}F_{ij}F_{ij}
  + (D_i\phi)^{\dagger}(D_i\phi)
  + \mu_3^{2}(\phi^{\dagger} \phi)
  + \frac{1}{2}(\partial_i B_0)^2
  + \frac{1}{2} m_\rmii{D}^{2} (B_0)^2
  \nn &
  + \lambda_3(\phi^{\dagger}\phi)^2
  + h_3(\phi^{\dagger}\phi)(B_0)^2
  + \kappa_3(B_0)^4
  + \frac{1}{2 \xi}(\partial_i B_i)^2
  + \mathcal{L}_\text{soft}^{(6)}
  \,,
\end{align}
with a little abuse of notation,
we have introduced the zero modes of the scalar, $\phi$,
the spatial gauge field $B_i$, and the temporal gauge field $B_0$,
which have (3d) mass dimension
$[\phi] = [B_i] = [B_0] = 1/2$ when normalized canonically.
The 3d covariant derivative is given by $D_i = \partial_i - i g_3 B_i$,
with effective gauge coupling $g_3$.
Finally, $\mathcal{L}_\text{soft}^{(6)}$ contains
a basis of dimension-six operators.
See~\cite{Chakrabortty:2026swu}
for a complete, non-redundant basis of dimension-six operators in 3d SMEFT.

In this section, we focus on the effective scalar mass $\mu_3^2$ and Debye mass
$\mD^2$, since these are the only parameters that must be computed at three-loop
order to achieve $\mathcal{O}(g^6)$ accuracy.
The matching relations for the static
screening masses $\Pi(k_0=0,\vec{k}\to 0)$ of the scalar field and the temporal gauge
field are obtained by matching the poles of the corresponding
static propagators~\cite{Ghisoiu:2015uza},
{\em viz.}
\begin{align}
\label{eq:scalar:mass:def}
  \mu_3^2 &=
    \mu^2
    + \Pi_{\phi^\dagger \phi}^{1\ell}
    + \Pi_{\phi^\dagger \phi}^{2\ell}
    - \Bigl(
          \mu^2
        + \Pi_{\phi^\dagger \phi}^{1\ell}
        + \Pi_{\phi^\dagger \phi}^{2\ell}
      \Bigr)\Pi_{\phi^\dagger \phi}^{1\ell,\rmii{(1)}}
  \nn &
    + \Bigl(
          \mu^2
        + \Pi_{\phi^\dagger \phi}^{1\ell}
      \Bigr)\Bigl[
        \bigr(\Pi_{\phi^\dagger \phi}^{1\ell,\rmii{(1)}}\bigr)^2
      + \Bigl(
            \mu^2
          + \Pi_{\phi^\dagger \phi}^{1\ell}
        \Bigr)\Pi_{\phi^\dagger \phi}^{1\ell,\rmii{(2)}}
      - \Pi_{\phi^\dagger \phi}^{2\ell,\rmii{(1)}}
    \Bigr]
    + \Pi_{\phi^\dagger \phi}^{3\ell}
    - \delta \mu_3^2
  \,,\\[2mm]
\label{eq:Debye:mass:def}
  \mD^2 &=
      \Pi_{B_0 B_0}^{1\ell}
    + \Pi_{B_0 B_0}^{2\ell}
    - \Bigl(
        \Pi_{B_0 B_0}^{1\ell}
      + \Pi_{B_0 B_0}^{2\ell}
      \Bigr)\Pi_{B_0 B_0}^{1\ell,\rmii{(1)}}
  \nn &
  + \Pi_{B_0 B_0}^{1\ell}\Bigl[
        \bigl(\Pi_{B_0 B_0}^{1\ell,\rmii{(1)}}\bigr)^2
      + \Pi_{B_0 B_0}^{1\ell}\Pi_{B_0 B_0}^{1\ell,\rmii{(2)}}
      - \Pi_{B_0 B_0}^{2\ell,\rmii{(1)}}
    \Bigr]
    + \Pi_{B_0 B_0}^{3\ell}
    - \delta \mD^2
  \,,
\end{align}
where $n\ell = 1\ell,\dots,3\ell$ indicates the different loop levels,
$\Pi^{n\ell}$ is the $n$-loop renormalized correlator, and
the $\Pi^{(n)} = \Pi^{(1)},\dots,\Pi^{(2)}$ superscripts indicate
the order of momentum derivatives.
Finally, $\delta \mu_3^2$ and $\delta \mD^2$ are
the corresponding mass counterterms in the 3d EFT;
see eqs.~\eqref{eq:mu32 3d CT} and~\eqref{eq:mD2 3d CT}.

\begin{figure}[t]
  \centering
  \begin{eqnarray*}
  \SPropCircPropn(\Lsc1,\phi,3\ell)
    &=&
    \ToprSY(\Lsc1,\Aglx,\Acs1,\Aglx,\Asc1,\Axx,\Lcs1,\Lxx,\Lglx)
    \ToprSX(\Lsc1,\Aglx,\Asc1,\Aglx,\Asc1,\Aglx,\Asc1,\Lcs1,\Lsc1)
    \ToprSV(\Lsc1,\Axx,\Aglx,\Axx,\Aglx,\Aglx,\Acs1,\Acs1)
    \ToprSMM(\Lsc1,\Aglx,\Acs1,\Aglx,\Asc1,\Aglx,\Asc1,\Lsc1,\Lcs1)
    \ToprSTS(\Lsc1,\Aglx,\Aglx,\Aglx,\Axx,\Asc1)
    \ToprSMB(\Lsc1,\Aglx,\Acs1,\Aglx,\Asc1,\Lxx,\Aglx,\Acs1,\Lxx)
    \ToprSTBBlr(\Lsc1,\Aglx,\Axx,\Acs1,\Aglx,\Asc1,\Axx,\Aglx)
    \ToprSSB(\Lsc1,\Aglx,\Aglx,\Lxx,\Aglx,\Asc1,\Lxx)
    \ToprSBB(\Lsc1,\Acs1,\Axx,\Aglx,\Axx,\Axx,\Aglx,\Acs1,\Aglx)
    \nn[2mm] &+&
  \text{(246 diagrams)}
  \nn[3mm]
  \SPropCircPropn(\Lglx,B_\mu,3\ell)
    &=&
    \ToprSY(\Lglx,\Asc1,\Axx,\Axx,\Aglx,\Axx,\Lglx,\Lcs1,\Lsc1)
    \ToprSX(\Lglx,\Asc1,\Aglx,\Axx,\Acs1,\Aglx,\Axx,\Lsc1,\Lsc1)
    \ToprSV(\Lglx,\Axx,\Aglx,\Axx,\Asc1,\Axx,\Axx,\Asc1)
    \ToprSMM(\Lglx,\Axx,\Aglx,\Axx,\Axx,\Aglx,\Axx,\Lsc1,\Lsc1)
    \ToprSTS(\Lglx,\Axx,\Aglx,\Axx,\Aglx,\Acs1)
    \ToprSMbl(\Lglx,\Asc1,\Axx,\Aglx,\Axx,\Asc1,\Axx,\Lglx)
    \ToprSTBBlr(\Lglx,\Axx,\Aglx,\Acs1,\Asc1,\Aglx,\Axx,\Axx)
    \ToprSSB(\Lglx,\Aglx,\Acs1,\Lxx,\Aglx,\Axx,\Lxx)
    \ToprSBBal(\Lglx,\Axx,\Axx,\Aglx,\Axx,\Aglx,\Axx,\Acs1)
    \nn[3mm] &+&
    \text{(240 diagrams)}
\end{eqnarray*}
  \caption{%
    Three-loop contributions to the 
    bare two-point functions in the Abelian Higgs model at the soft scale.
    Dashed directed lines denote
    scalars ($\phi$) and
    wiggly lines denote gauge fields ($B_\mu$).
  }
  \label{fig:3loop:diagrams}
\end{figure}
\begin{figure}[t]
  \centering
  \begin{eqnarray*}
    \VPropCircPropn(\Lglx,\Lglx,\Lglx,\Lglx,\;\,B_\mu,2\ell)
    &=&
    \ToptVDM(fex(\Lglx,\Lglx,\Lglx,\Lglx),\Asc1,\Axx,\Axx,\Axx,\Axx,\Axx,\Lglx)
    \ToptVMlud(fex(\Lglx,\Lglx,\Lglx,\Lglx),\Asc1,\Axx,\Axx,\Lglx)
    \ToptVElr(fex(\Lglx,\Lglx,\Lglx,\Lglx),\Asc1,\Axx,\Axx,\Acs1)
    \ToptVBTlr(fex(\Lglx,\Lglx,\Lglx,\Lglx),\Axx,\Axx,\Asc1,\Aglx)
    + \text{(560 diagrams)}
  \nn[3mm]
  \VPropCircPropn(\Lsc1,\Lsc1,\Lcs1,\Lcs1,\phi,2\ell)
    &=&
    \ToptVDM(fex(\Lsc1,\Lcs1,\Lsc1,\Lcs1),\Aglx,\Axx,\Axx,\Aglx,\Axx,\Axx,\Lglx)
    \ToptVMlud(fex(\Lsc1,\Lcs1,\Lsc1,\Lcs1),\Aglx,\Aglx,\Aglx,\Lsc1)
    \ToptVElr(fex(\Lsc1,\Lsc1,\Lcs1,\Lcs1),\Acs1,\Asc1,\Acs1,\Asc1)
    \ToptVElr(fex(\Lsc1,\Lcs1,\Lsc1,\Lcs1),\Aglx,\Aglx,\Asc1,\Axx)
    \ToptVBTlr(fex(\Lsc1,\Lsc1,\Lcs1,\Lcs1),\Axx,\Acs1,\Asc1,\Aglx)
    + \text{(525 diagrams)}
  \nn[3mm]
  \VPropCircPropn(\Lsc1,\Lglx,\Lglx,\Lcs1,,2\ell)
    &=&
    \ToptVDM(fex(\Lglx,\Lsc1,\Lcs1,\Lglx),\Asc1,\Axx,\Axx,\Aglx,\Axx,\Axx,\Lglx)
    \ToptVMlud(fex(\Lglx,\Lglx,\Lsc1,\Lcs1),\Asc1,\Axx,\Aglx,\Lglx)
    \ToptVElr(fex(\Lglx,\Lglx,\Lsc1,\Lcs1),\Asc1,\Axx,\Aglx,\Aglx)
    \ToptVElr(fex(\Lglx,\Lglx,\Lsc1,\Lcs1),\Asc1,\Axx,\Axx,\Acs1)
    \ToptVBTlr(fex(\Lglx,\Lglx,\Lsc1,\Lcs1),\Axx,\Axx,\Asc1,\Aglx)
    + \text{(401 diagrams)}
\end{eqnarray*}
  \caption{%
    Two-loop contributions to the
    bare four-point functions in the Abelian Higgs model at the soft scale
    including all possible permutations of external legs.
    Dashed directed lines denote scalars ($\phi$) and
    wiggly lines denote gauge fields ($B_\mu$).
  }
  \label{fig:4pt:2loop:diagrams}
\end{figure}
As seen from eqs.~\eqref{eq:scalar:mass:def} and~\eqref{eq:Debye:mass:def},
to reach $\mathcal{O}(g^6)$ accuracy,
self-energies are required up to three-loop order,
including higher-order momentum derivatives of
the one- and two-loop contributions.
The diagrams for
the bare three-loop correlators
$\Pi_{\phi^\dagger \phi}^{3\ell,\rmii{$(B)$}}$ and
$\Pi_{B_0 B_0}^{3\ell,\rmii{$(B)$}}$
are listed in fig.~\ref{fig:3loop:diagrams}
while their corresponding analytic expressions
are collected
in appendix~\ref{app:DR:relations}.
In addition to the three-loop contributions to the scalar and Debye masses,
two-loop contributions to the quartic couplings
$\lambda_3$, $h_3$, and $\kappa_3$ are required
for $\mathcal{O}(g^6)$ accuracy.
These are obtained by matching the corresponding
four-point functions at zero external momenta.
The two-loop diagrams for the bare quartic correlators 
are shown in fig.~\ref{fig:4pt:2loop:diagrams},
and their expressions are also collected in appendix~\ref{app:DR:relations}.

The masses are renormalized upon the introduction of the renormalized 4d parameters in
eqs.~\eqref{eq:scalar:mass:def} and~\eqref{eq:Debye:mass:def};
see appendix~\ref{app:4d rges}.
We do this perturbatively as follows.
Let $L=3$ be the loop order of the computation, and let
$\Pi^{n \ell , \rmii{$(B)$}} = f(c_i^\rmii{$(B)$})$ be 
the bare $n$-loop ($n \leq L$) piece of a given correlator,
which is an analytic function $f$ of the bare parameters $c_i^\rmii{$(B)$}$.
The corresponding renormalized parameters can be written as
$c_i = c_i^\rmii{$(B)$} + \sum_{k=1}^L \delta c_i^{k \ell}$, where
$\delta c_i^{k \ell}$ are the $k$-loop associated counterterms.
Now, we obtain the renormalized masses upon replacing
$c_i^\rmii{$(B)$} \to c_i - \sum_{k=1}^L \delta c_i^{k \ell}$ perturbatively
in loops in each of the $n$-loop correlator pieces, up to $L$-loop order.
This well-known perturbative renormalization procedure naturally mixes
lower-loop correlators with counterterm insertions and
higher-loop correlators without counterterm insertions.
Finally, we subtract the 3d mass counterterms to remove all leftover divergences from
the hard region expansion of the 4d and 3d correlators,
which yields a fully renormalized 3d effective mass.

The expressions for the (renormalized) scalar and Debye masses,
up to three-loop level and $\mathcal{O}(g^6)$ accuracy,
read
\begin{align}
  \label{eq:scalar mass 3 loop}
  \mu_3^2&=
      \left[\mu ^2\right]_{\text{tree level}}
    + T^2\left[\frac{g^2}{4}+\frac{\lambda}{3}\right]_{\text{1 loop}}
    \nn &
    + \frac{T^2}{(4\pi)^2}\bigg[
          g^4 \left(-\frac{13}{12} \Lb
        + 3 L_\rmi{3d}
        +\frac{25}{9}
        \right)
    + g^2 \lambda \biggl(
          2 \Lb
        - 4 L_\rmi{3d}
        - \frac{10}{3}
      \biggr)
    \nn &
    \hphantom{+\frac{T^2}{(4\pi)^2}\bigg[}
    + \lambda^2 \biggl(
        - \frac{10}{3}\Lb
        + 4 L_\rmi{3d}
        + 4
      \biggr)
    + \frac{\mu^2}{T^2} \left(
        3 g^2
      - 4 \lambda
    \right)\Lb
    \bigg]_{\text{2 loop}}
    \nn &
    + \frac{T^2}{(4\pi)^4}\bigg[
        g^4 \lambda \left(32\Ltd-\frac{140}{3}\Lb \Ltd+\frac{47}{3} \Lb^2-\frac{544}{9} \Lb+C_1\right)
      \nn &
      \hphantom{+\frac{T^2}{(4\pi)^4}\bigg[}
      + \lambda^3 \left(-80 \Lb \Ltd+\frac{100}{3}\Lb^2-40 \Lb+C_2\right)
      \nn &
      \hphantom{+\frac{T^2}{(4\pi)^4}\bigg[}
      + g^6 \left(10 \Lb \Ltd-\frac{16}{3}\Ltd-\frac{95}{36}\Lb^2+\frac{505}{27} \Lb+C_3\right)
      \nn &
      \hphantom{+\frac{T^2}{(4\pi)^4}\bigg[}
      + g^2 \lambda^2 \left(88 \Lb \Ltd-30 \Lb^2+\frac{188}{3}\Lb+C_4\right)
      \nn &
      \hphantom{+\frac{T^2}{(4\pi)^4}\bigg[}
      + g^4 \frac{\mu^2}{T^2} \left(10 \Lb^2-\frac{43}{3}\Lb-\frac{7}{36} \left(12 \zeta_3+25\right)\right)
      \nn &
      \hphantom{+\frac{T^2}{(4\pi)^4}\bigg[}
      + g^2 \lambda \frac{\mu^2}{T^2} \left(-24 \Lb^2-32 \Lb+\frac{16}{9} \left(\zeta_3-27\right)\right)
      \nn &
      \hphantom{+\frac{T^2}{(4\pi)^4}\bigg[}
      + \lambda^2 \frac{\mu^2}{T^2} \left(28 \Lb^2+20 \Lb+\frac{1}{3} \left(16 \zeta_3+89\right)\right)
      \nn &
      \hphantom{+\frac{T^2}{(4\pi)^4}\bigg[}
      - g^2 \frac{\mu^4}{T^4} \left( \frac{10\zeta_3}{3} \right)
      + \lambda \frac{\mu^4}{T^4} \left( 8 \zeta_3 \right) \bigg]_{\text{3 loop}}
    + \mathcal{O}(g^8)
  \,,
  \\[2mm]
  \label{eq:debye mass 3 loop}
  \mD^2&=
      T^2\left[\frac{g^2}{3}\right]_{\text{1 loop}}
    + \frac{T^2}{(4\pi)^2}\left[
      - g^4 \frac{\Lb - 7}{9}
      + \frac{4}{3} g^2 \lambda
      + 4 \frac{\mu^2}{T^2} g^2
    \right]_{\text{2 loop}}
    \nn &
    + \frac{T^2}{(4\pi)^4}\bigg[
        g^6 \left(\frac{1}{27}\Lb^2-\frac{158}{27}\Lb+\frac{8}{3}\Ltd+\frac{1}{270} (3220-57 \zeta_3)\right)
      \nn &
      \hphantom{+\frac{T^2}{(4\pi)^4}\bigg[}
      + g^4 \lambda \left(\frac{68}{9}\Lb+16 \Ltd-\frac{8 (\zeta_3+34)}{9}\right)
      - g^2\lambda^2 \left(\frac{40}{3}\Lb+\frac{4}{9} (2 \zeta_3-21)\right)
      \nn &
      \hphantom{+\frac{T^2}{(4\pi)^4}\bigg[}
      + g^4 \frac{\mu^2}{T^2} \left(\frac{32}{3}\Lb-\frac{8}{3} (\zeta_3-11)\right)
      \nn &
      \hphantom{+\frac{T^2}{(4\pi)^4}\bigg[}
      - g^2\lambda \frac{\mu^2}{T^2} \left(16 \Lb+\frac{16 \zeta_3}{3}\right)
      - 8\zeta_3 g^2 \frac{\mu^4}{T^4}
    \bigg]_{\text{3 loop}}
    + \mathcal{O}(g^8)
    \,.
\end{align}
In analogy to~\cite{Ghisoiu:2015uza},
we expressed the results in terms of
\begin{align}
    \Lb &\equiv 2\ln\frac{\LamD e^{\gammaE}}{4\pi T}
    \,,&
    \Ltd &\equiv \ln\frac{\Lamd^2 e^{Z_1}}{4\pi T^2} 
    \,,
\end{align}
where $\LamD$ is the 4d $\overline{\text{MS}}$ matching scale,
$\Lamd$ is the 3d~EFT running scale, and
we abbreviate derivatives of
the Riemann $\zeta$-function
$\zeta_s \equiv \zeta(s) $ by $Z_s^{(n)} = \zeta^{(n)}_{-s} / \zeta_{-s}$.
Finally, $C_1,\dots,C_4$
are constants coming from
the three-loop integrals that have been numerically evaluated (cf.\ appendix~\ref{app:masters}),
\begin{align}
    C_1&\simeq  -217.26192980(7)
    \,,&
    C_2&\simeq  -592.606749846480(16)
    \,,\nn
    C_3&\simeq  -23.808880(4)
    \,,&
    C_4&\simeq  470.903136214018(4)
    \,.
\end{align}
As a crosscheck,
we observe that the matching equations for both physical masses are
finite and gauge independent.
This result is highly non-trivial, and it was essential to spot an error in
the original evaluation~\cite{Ghisoiu:2012yk}%
\footnote{%
    Soon after the submission of the first version of this manuscript,
    this error was acknowledged by the authors of~\cite{Ghisoiu:2012yk} and
    an erratum was published.
}
of the three-loop master sum-integral $\mathcal{I}_{31111-2}$,
for which we report the correct divergent piece
(see appendices~\ref{app:masters} and~\ref{app:mismatch}).
This correction
introduces a finite shift to
the three-loop Debye mass of hot ${\rm SU}(N)$ Yang-Mills theories~\cite{Ghisoiu:2015uza}.
Following the notation in~\cite{Ghisoiu:2015uza},
the difference in the evaluation of $J_{11} = \mathcal{I}_{31111-2}$ in
the non-renormalized Debye mass
eq.~(3.3) of~\cite{Ghisoiu:2015uza} leads to
the following new evaluation of the $\mathcal{O}(g_\rmii{R}^6)$ piece of
the renormalized Debye mass in their eq.~(3.10),
{\em viz.}
\begin{equation}
  \label{eq:new debye}
      \frac{\Delta m_\rmii{ER}^2}{(4\pi T)^2} \equiv
      \frac{m_{\rmii{ER},\rmi{new}}^2 - m_{\rmii{ER}}^2}{(4\pi T)^2}
      = \left[\frac{\CA}{3} \frac{g_\rmii{R}^2}{(4 \pi)^2}\right]^3 \times [-24]
    \,,
\end{equation}
where $g_\rmii{R}$ is the renormalized gauge coupling and
$\CA = N$ is the adjoint Casimir of the gauge group.
This correction yields
a $\Delta m_\rmii{ER}^2/ m_\rmii{ER}^2 \big|_{\mathcal{O}(g_\rmii{R}^6)} \sim 7\%$
correction to the three-loop Debye mass in pure Yang-Mills theory.

\begin{figure}[t!]
    \centering
    \includegraphics[width=0.5\linewidth]{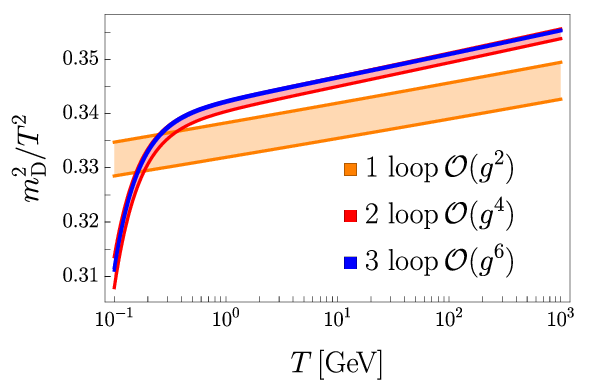}%
    \includegraphics[width=0.5\linewidth]{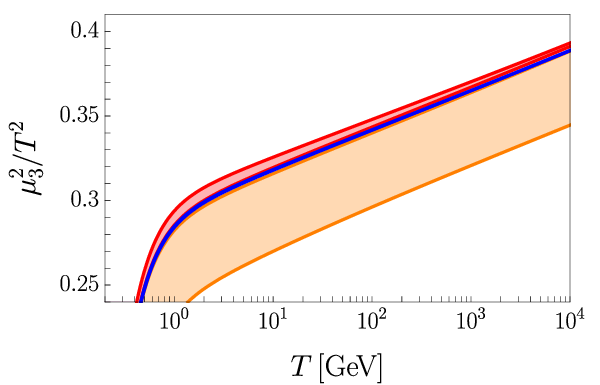}
    \\
    \includegraphics[width=0.5\linewidth]{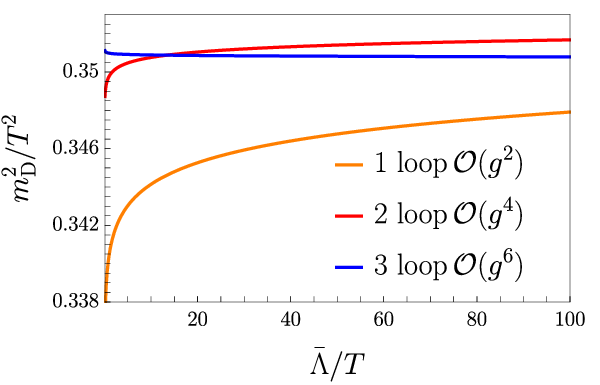}%
    \includegraphics[width=0.5\linewidth]{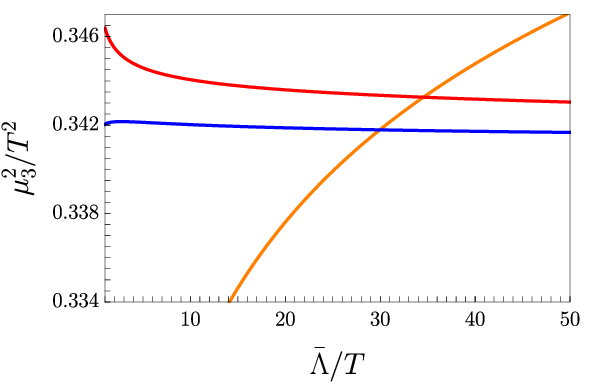}
    \caption{%
    Debye (left panels) and
    scalar (right panels) masses
    as functions of temperature (top row) and matching scale $\LamD$ at $T=100$~GeV (bottom row).
    The 4d couplings are fixed at~\eqref{eq:BM1}.
    The 3d renormalization scale is set to
    the optimized value $\bar\Lambda_\rmi{3d,opt}=2.85\,T$~\cite{Stevenson:1981vj,Laine:2005ai,Ghisoiu:2015uza},
    and we assume $X=0$. Setting $X$ to slightly different values (for instance, $X=-2/5$) leads to only imperceptible deviations in the results. In the upper panels,
    the matching scale is varied in
    the range $\LamD \in [2^{-3},2^{3}]\,\pi T$.
    The $n$-loop curves show the $n$-loop matching results for
    the Debye and scalar masses,
    with
    the 4d
    parameters evolved using two-loop running~\eqref{4d running}.
    The plots illustrate that
    the dependence on the matching scale is further suppressed at
    higher orders when two-loop running and three-loop matching contributions are included.
    }
    \label{fig:debye_scalar_combined}
\end{figure}

As shown in eq.~\eqref{eq:scalar mass 3 loop},
the gauge dependence cancels completely once the three-loop contribution is included.
The corresponding plots for expressions~\eqref{eq:debye mass 3 loop}
and~\eqref{eq:scalar mass 3 loop} as functions of
the temperature are shown in the upper panels of fig.~\ref{fig:debye_scalar_combined}.
Using the following benchmark (BM) values
of the \MSbar{} parameters
at the input scale $\LamD_\rmi{in}$,
\begin{align}
  \label{eq:BM1}
    \LamD_{\rmi{in}}&=1\,\mathrm{GeV}
    \,,&
    g(\LamD_{\rmi{in}})&=1
    \,,&
    \lambda(\LamD_{\rmi{in}})&=0.1
    \,,&
    \mu^2(\LamD_{\rmi{in}})&=-0.01\,\mathrm{GeV}^2
    \,,
  \tag{BM1}
\end{align}
the 4d
couplings are evolved
to the 4d matching scale $\LamD$
using two-loop RG equations given in eq.~\eqref{4d running},
while varying $\LamD$
in the range $\LamD \in [2^{-3},2^{3}]\,\pi T$
to illustrate the reduced dependence on the matching scale
at higher orders in the EFT matching.
The 3d
renormalization scale is set to
the optimized value $\bar\Lambda_\rmi{3d,opt}=2.85\,T$,
determined via the principle of minimal sensitivity~\cite{Stevenson:1981vj},
following the approach used for the QCD Debye mass~\cite{Laine:2005ai,Ghisoiu:2015uza}.

As a crosscheck, we verify that the 3d
mass parameters
in eqs.~\eqref{eq:scalar mass 3 loop} and~\eqref{eq:debye mass 3 loop}
are independent of the matching scale $\bar{\Lambda}$ order by order in $g$.
Using the running of the 4d couplings
$g$, $\lambda$, and $\mu^2$ given in eq.~\eqref{4d running},
we find
\begin{align}
\label{eq:masses_scale_dependence}
    \partial_t^{ } \mu^2_3&=\mathcal{O}(g^8)
    \,,&
    \partial_t^{ } \mD^2&=\mathcal{O}(g^8)
    \,,
\end{align}
where $t \equiv \ln\bar{\Lambda}$.
We show the reduced dependence of the 3d masses on the matching scale $\bar{\Lambda}$,
which is pushed to the next perturbative order,
in the lower panels of fig.~\ref{fig:debye_scalar_combined}.
Similar to~\cite{Ghisoiu:2013zoj,Ghisoiu:2015uza},
the convergence of the matching from the hard
to the soft scale occurs already at two-loop level such
that the improvement from the three-loop contribution
is small.

The matching procedure for the 3d masses
as outlined in
eqs.~\eqref{eq:scalar:mass:def} and~\eqref{eq:Debye:mass:def},
is consistent with performing the EFT matching in
an off-shell operator basis and subsequently carrying out field redefinitions
to eliminate redundancies and extract the effective parameters in
an on-shell operator basis as shown in~\cite{Chala:2024llp}, and applied at finite temperature in~\cite{Bernardo:2025vkz,Chala:2025aiz,Chala:2025oul}.
To determine the rest of the effective parameters besides the thermal masses,
we follow the off-shell matching procedure.
Further details are provided in appendix~\ref{app:DR:relations}.

Having determined the full $\mathcal{O}(g^6)$ soft-scale EFT,
we now aim to determine the soft-scale effective potential
to compute all critical parameters. 

%
\section{Higher-order corrections to equilibrium thermodynamics}
\label{sec:higherorder}

We organize the perturbative expansion of
the effective potential following the power counting
presented in~\cite{Ekstedt:2024etx}.
Higher-dimensional operators are left unassigned to a specific perturbative order,
but are assumed to be subleading relative to
the leading-order (LO) effective potential,
{\em viz.}%
\footnote{%
  Henceforth, we assume that $\lambda \sim g^3$~\cite{Arnold:1992rz},
  which is more appropriate in the region of the 4d parameter space that we explore.
  Upon this change, our matching equations are
  still correct up to $\mathcal{O}(g^6)$,
  but terms involving more powers of $\lambda$ will be in practice further suppressed.
}
\begin{equation}
\label{effective potential 1}
    \frac{V_{\text{eff}}}{T^3} \sim
\boxed{\colorbox{gray!20}{$\displaystyle \frac{g^3}{\pi}$}}
_{\;\text{LO}}
\;+\;
\boxed{\colorbox{gray!20}{$\displaystyle \frac{g^4}{\pi^2}$}}
_{\;\text{NLO}}
\;+\;
\boxed{\colorbox{gray!20}{$\displaystyle \frac{g^{9/2}}{\pi^{5/2}}$}}
_{\;\text{N$^2$LO}}
\;+\;
\boxed{\colorbox{gray!20}{$\displaystyle \frac{g^{6}}{\pi^4}$}}
_{\text{\;dim6}}
\;+\;
\boxed{\colorbox{gray!20}{$\displaystyle \mathcal{O}\Bigl( \frac{g^5}{\pi^3} \Bigr)$}}
_{\text{\;N$^3$LO}}
\,,
\end{equation}
where we neglect N$^3$LO super-renormalizable contributions to $V_\mathrm{eff}$. 
Although the dimension-six term may appear subleading compared to the neglected N$^3$LO contribution,
its $g^6$ scaling reflects the parametric size of
the corresponding Wilson coefficients
rather than the operator itself,
which involves higher powers of fields than super-renormalizable operators.
We therefore remain agnostic about its numerical importance.

The terms of eq.~\eqref{effective potential 1}
originate from
\begin{align*}
    V^{\rmii{LO}}_{\text{eff}}&\sim g^3
    &&
    \text{tree-level scalar and one-loop vector and temporal scalar}
    \,,
    \nn[1mm]
    V^{\rmii{NLO}}_{\text{eff}}&\sim g^4
    &&
    \text{two-loop vector and temporal scalar}
    \,,
    \nn[1mm]
    V^{\rmii{N$^2$LO}}_{\text{eff}}&\sim g^{\frac{9}{2}}
    &&
    \text{one-loop scalar}
    \,,
    \nn[1mm]
    V^{\rmi{dim6}}_{\text{eff}}&\sim g^6
    &&
    \text{tree-level higher-dimensional operators}
    \,.
\end{align*}

Our choice to include higher-dimensional operators as
a correction to the LO effective potential is not arbitrary.
While one cannot, in general, assign an all-encompassing power counting
for higher-dimensional operators in this context,
the validity of the high-temperature expansion does impose
strict constraints~\cite{Camargo-Molina:2024sde}.

For a general $N$-dimensional scalar multiplet $\phi$
in the broken phase,
we shift the field by a real background $v$
defined by $\phi\to\frac{v}{\sqrt{2}}\delta_{i,N} + \phi$.
Parametrically,
the background field
scales as $v \sim \sqrt{T}/g^n$ for some
$n \in \mathbb{R}$, with $[v]=1/2$.
From the requirement that the high-temperature expansion converges,
soft-scale dimension-eight contributions must be subleading compared to dimension-six.
This implies
(omitting factors of $\pi$):
\begin{equation}
    c_8 v^8 \sim \frac{g^8}{T} \times \frac{T^4}{g^{8 n}} \ll c_6 v^6 \sim g^6 \times \frac{T^3}{g^{6 n}} \iff g^{2(1-n)} \ll 1 \iff n < 1\,.
    \label{eq:highT validity 1}
\end{equation}
On the other hand,
from the matching (see eq.~\eqref{eq:lambda3 matching}),
the quartic must satisfy $\lambda_3/T \sim g^m$ with $2 < m < 4$,
so if we assume that dimension-six terms enter $V_\mathrm{eff}^\rmii{LO}$, then
\begin{equation}
    \lambda_3 v^4 \sim c_6 v^6 \iff g^m T \times \frac{T^2}{g^{4 n}} \sim g^6 \times \frac{T^3}{g^{6 n}} \iff g^{6-2n+m} \sim 1 \iff n = 3-\frac{m}{2}\,.
    \label{eq:highT validity 2}
\end{equation}
No value in $2 < m < 4$ is compatible with the condition in
eq.~\eqref{eq:highT validity 1}, and
therefore it is only consistent to include hard-scale generated
dimension-six terms along with soft-scale quantum corrections to
the effective potential.

Assuming therefore the perturbative expansion in
eq.~\eqref{effective potential 1},
the broken-phase field-dependent masses,
\begin{align}
    \mB^2 &\equiv g_3^2v^2
    \,,&
    m_{\rmii{$B_0$}}^2&\equiv \mD^2+h_3^{ }v^2
    \,,\nn[1mm]
    \mh^2&\equiv \partial^2_v\,V^{\rmii{LO}}
    \,,&
    \mG^2&\equiv v^{-1}\partial_v V^{\rmii{LO}}
    \,,
\end{align}
enter directly
into the effective potential in
the broken phase
\begin{align}
    V_{\text{eff}}\big|_{\rmii{LO}}^{\text{bro}}&=
        \frac{1}{2}\mu^2_3v^2
      + \frac{1}{4}\lambda_3v^4
      - \frac{1}{6\pi}\mB^3
      - \frac{1}{12\pi}m_{\rmii{$B_0$}}^3
    \,,\nn[2mm]
    V_{\text{eff}}\big|_{\rmii{NLO}}^{\text{bro}}&=
      \frac{1}{(4\pi)^2}\biggl[
          - g_3^4v^2\biggl(
            1
            +\ln\frac{\Lamd^2}{4\mB^2}
          \biggr)
          - \frac{1}{2}h_3^2v^2\biggl(
              1
            + \ln\frac{\Lamd^2}{4m_{\rmii{$B_0$}}^2}
          \biggr)
          + 3\kappa_3\,m_{B_0}^2
        \biggr]
    \,,\nn[2mm]
    V_{\text{eff}}\big|_{\rmii{N$^2$LO}}^{\text{bro}}&= -\frac{1}{12\pi}\bigg[\mh^3+\mG^3\bigg]
    \,,\nn[2mm]
    V_{\text{eff}}\big|_{\rmi{dim6}}^{\text{bro}}&=\frac{1}{8}c_6v^6
    \,.
\end{align}
The potential up to NNLO can be obtained 
using either
{\tt DRalgo}~\cite{Ekstedt:2024etx} or 
results from~\cite{Hirvonen:2021zej}.
In the symmetric phase, the effective potential takes the form
\begin{align}
    V_{\text{eff}}\big|_{\rmii{LO}}^{\text{sym}}&= -\frac{1}{12\pi}(\mD^2)^{\frac{3}{2}}
    \,,&
    V_{\text{eff}}\big|_{\rmii{NLO}}^{\text{sym}}&= \frac{3}{(4\pi)^2}\kappa_3\, \mD^2
    \,,\nn[1mm]
    V_{\text{eff}}\big|_{\rmii{N$^2$LO}}^{\text{sym}}&= -\frac{1}{6\pi}(\mu_3^2)^{\frac{3}{2}}
    \,,&
    V_{\text{eff}}\big|_{\rmi{dim6}}^{\text{sym}}&=0
    \,.
\end{align}
As a practical step, we recast the potential into a dimensionless form
by defining the dimensionless variables
\begin{align}
    x &\equiv \frac{\lambda_3}{g_3^2}
    \,, &
    y &\equiv\frac{\mu_3^2}{g_3^4}
    \,, &
    \yD &\equiv\frac{\mD^2}{g_3^4}
    \,, &
    \tilde{h}_3 &\equiv\frac{h_3}{g_3^2}
    \,, &
    \tilde{\kappa}_3 &\equiv\frac{\kappa_3}{g_3^2}
    \,, &
    \varphi &\equiv\frac{v}{g_3}
    \,,
\end{align}
and rescaling
both the effective potential
$V_{\text{eff}}\to g_3^{-6}\,V_{\text{eff}}$ and
the scalar field background $v \to g_3 \varphi$.
In practice,
we utilize the difference between the effective potential
in the broken and symmetric phases,
\begin{align}
  \label{eq:Delta:V:eff:N0LO}
    \Delta V_{\text{eff}}\big|_{\rmii{LO}}&=
        \frac{1}{2}y\varphi^2
      + \frac{1}{4}x\varphi^4
      - \frac{1}{6\pi}\frac{3}{2}\varphi^3
      - \frac{1}{12\pi}\Bigl[
        ( \yD+\tilde{h}_3\varphi^2)^{\frac{3}{2}}
        - \varphi^3
        - \yD^{\frac{3}{2}}
      \Bigr]
      \,,\\[1mm]
  \label{eq:Delta:V:eff:N1LO}
    \Delta V_{\text{eff}}\big|_{\rmii{NLO}}&=
      \frac{1}{(4\pi)^2}\biggl[
          - \frac{3}{2}\varphi^2\biggl(
              1
            + \ln\frac{\Lamd^2}{4\mB^2}
            \biggr)
          - \frac{\varphi^2}{2}\biggl(
            \tilde{h}_3^2\biggl(
                1
              + \ln\frac{\Lamd^2}{4m_{\rmii{$B_0$}}^2}
              \biggr)
              - \biggl(
                  1
                + \ln\frac{\Lamd^2}{4\mB^2}
              \biggr)
            \biggr)
        \nn &
        \phantom{{}= \frac{1}{(4\pi)^2}\biggl[ }
      + 3\tilde{\kappa}_3\,\tilde{h}_3\varphi^2
    \biggr]
    \,,\\
  \label{eq:Delta:V:eff:N2LO}
    \Delta V_{\text{eff}}\big|_{\rmii{N$^2$LO}}&= -\frac{1}{12\pi g_3^6}\bigg[
        (\mh^2)^{\frac{3}{2}}
      + (\mG^2)^{\frac{3}{2}}
      - 2(\mu_3^2)^{\frac{3}{2}}
    \bigg]
    \,,\\[1mm]
  \label{eq:Delta:V:eff:dim6}
  \Delta V_{\text{eff}}\big|_{\rmi{dim6}}&=\frac{1}{8}c_6\varphi^6
    \,,
\end{align}
where
for a thermodynamic expression $P$,
we define
$\Delta P \equiv P^{\rmi{bro}} - P^{\rmi{sym}}$
as the difference between the broken and symmetric phases.
Since close to the critical temperature,
the parameter $y$ and
the dimensionless field $\varphi$ scale as
$y\sim x^{-1}$ and
$\varphi\sim x^{-1}$,
one can reorganize the effective potential in powers of $x$
\begin{equation}
\label{effective potential}
  \frac{V_{\text{eff}}}{g_3^6} \sim
  \boxed{\colorbox{gray!20}{$\displaystyle x^{-3}$}}
  _{\,\rmi{LO}}
  \;+\;
  \boxed{\colorbox{gray!20}{$\displaystyle x^{-2}$}}
  _{\,\rmi{NLO}}
  \;+\;
  \boxed{\colorbox{gray!20}{$\displaystyle x^{-\frac{3}{2}}$}}
  _{\,\rmi{N$^2$LO}}
  +\;
  \boxed{\colorbox{gray!20}{$\displaystyle c_6\, x^{-6}$}}
  _{\,\rmi{dim6}}
  +\;
  \boxed{\colorbox{gray!20}{$\displaystyle \mathcal{O} ( x^{-1} )$}}
  _{\,\rmi{N$^3$LO}}
  \,.
\end{equation}
While the $x^{-6}$ term appears dominant,
the effects of dimension-six operators are suppressed by
the associated effective parameters $c_6 \sim g^6/\pi^4$
rendering them parametrically smaller in the gauge coupling compared to other operators.

To assess the impact of the higher-order corrections on the PT parameters,
it is practical to further
recast the soft-scale effective potential in a form similar to
the softer-scale potential~\eqref{eq:DR:A},
as it allows for the existence of analytical expressions for the critical parameters;
see~\cite{Kajantie:1995kf,Ekstedt:2022zro,Ekstedt:2024etx}.
To this end,
we rewrite the bracketed terms in
$\Delta V_{\text{eff}}\big|_{\rmii{LO}}$ in eq.~\eqref{eq:Delta:V:eff:N0LO} and
$\Delta V_{\text{eff}}\big|_{\rmii{NLO}}$ in eq.~\eqref{eq:Delta:V:eff:N1LO} using
\begin{align}
  \Bigl[(\yD+\tilde{h}_3\varphi^2)^{\frac{3}{2}}-\varphi^3-\yD^{\frac{3}{2}}\Bigr]&\simeq
  (\tilde{h}_3^{3/2}-1)\varphi^3+\mathcal{O}(x^{-1})
  \,,\nn
  \varphi^2\,\ln\frac{\Lamd^2}{4m_{\rmii{$B_0$}}^2}&\simeq
      \varphi^2\,\ln\frac{\Lamd^2}{4\mB^2}
    - \varphi^2\, \ln\,\tilde{h}_3
    + \mathcal{O}(x^{0})
    \,.
\end{align}
Hence,
the effective potential differences
of eqs.~\eqref{eq:Delta:V:eff:N0LO}--\eqref{eq:Delta:V:eff:N2LO} are rewritten as
\begin{align}
  \Delta V_{\text{eff}}\big|_{\rmii{LO}}&=
      \frac{1}{2}y\varphi^2
    + \frac{1}{4}x\varphi^4
    - \frac{1}{6\pi}\mathcal{E}_1\varphi^3
  \,,\nn[1mm]
  \Delta V_{\text{eff}}\big|_{\rmii{NLO}}&=
    - \frac{\varphi^2}{(4\pi)^2}\bigg[
        \mathcal{E}_2
    + \mathcal{E}_3\,\ln\frac{\Lamd^2}{4\mB^2}
    \bigg]
  \,,\nn[1mm]
  \Delta V_{\text{eff}}\big|_{\rmii{N$^2$LO}}&= -\frac{1}{12\pi}\bigg[
      \bigl(\tilde{m}_h^2\bigr)^{\frac{3}{2}}
    + \bigl(\tilde{m}_\rmii{G}^2\bigr)^{\frac{3}{2}}
    - 2y^{\frac{3}{2}}
    \bigg]
  \,,
\end{align}
using the enhancement factors
\begin{align}
    \mathcal{E}_1 &\equiv 1+\frac{1}{2}\tilde{h}_3^{\frac{3}{2}}
    \,,&
    \mathcal{E}_2 &\equiv
        1
      + \frac{1}{2}\tilde{h}_3^2\bigl(1 - \ln\tilde{h}_3\bigr)
      + 3\tilde{\kappa}_3\tilde{h}_3
    \,,&
    \mathcal{E}_3 &\equiv 1+\frac{1}{2}\tilde{h}_3^2
    \,,
\end{align}
and the dimensionless field-dependent and enhanced scalar masses 
\begin{align}
    \tilde{m}_h^2 &\equiv \frac{\mh^2}{g_3^4}=
        y
      + 3x\varphi^2
      - \frac{\mathcal{E}_1\varphi}{\pi}
    \,,&
    \tilde{m}_\rmii{G}^2 &\equiv \frac{\mG^2}{g_3^4}=
          y
        + x\varphi^2
        - \frac{\mathcal{E}_1\varphi}{2\pi}
    \,.
\end{align}

In eq.~\eqref{eq:Delta:V:eff:dim6},
we isolate the contribution of the dimension-six operators,
as their effect on equilibrium quantities varies significantly with $x$.
As mentioned above,
we treat this contribution as subleading relative to the LO effective potential
to maintain EFT consistency,
as discussed in eqs.~\eqref{eq:highT validity 1} and~\eqref{eq:highT validity 2}.
At the same time, we
remain agnostic about its precise perturbative order and
retain only the linear correction in $c_6$.
To determine the critical temperature $\Tc$ or,
equivalently, the critical mass $\yc$,
we find the values of $y$ and $x$ for
which the free energies in the two phases coincide,
{\em viz.}
\begin{equation}
    \Delta F (\yc(x),x)=[F_{\text{bro}}-F_{\text{sym}}](\yc(x),x)=0
    \,,
\end{equation}
where the free energy density
is defined as
\begin{align}
    F
      &=
      V_{\text{eff}}(v_{\text{min}})
    \,,
\end{align}
with $v_\rmi{min}$ denoting
the global minimum of the effective potential.
In the broken phase,
it can be expanded as
$v_\rmi{min} = [v_0 + v_1 + v_2 + \dots]$,
where $v_0 = g_3 \varphi_0$ is the LO minimum,
\begin{equation}
  \varphi_0 = \frac{1 + \sqrt{1 - (4\pi)^2 xy}}{4\pi x}
  \,.
\end{equation}
The entropy difference,
$\Delta S=\frac{\text{d}}{\text{d}\ln T}\Delta F(y,x)$,
characterizes the released amount of latent heat and therefore
the transition strength.
Defining the quadratic and quartic scalar condensates~\cite{Gould:2019qek}
\begin{align}
    \Delta \langle \phi^\dagger \phi\rangle &\equiv \partial_y \Delta F
    \,,&
    \Delta\langle (\phi^\dagger \phi)^2\rangle &\equiv \partial_x \Delta F 
    \,,
\end{align}
and using the chain-rule,
the entropy difference can be expressed as~\cite{Farakos:1994xh}
\begin{align}
      \Delta S &=
        \biggl(\frac{{\rm d}\,y}{{\rm d} \ln\,T } \biggr) \Delta \langle \phi^\dagger \phi\rangle
      + \biggl(\frac{{\rm d}\,x}{{\rm d} \ln\,T } \biggr) \Delta \langle (\phi^\dagger \phi)^2\rangle
      \,.
\end{align}

Our goal is to determine these thermodynamic parameters at the critical temperature.
Accordingly, we compute
$\yc$, $\Delta \langle \phi^\dagger \phi\rangle_{\rm c}$, and
$\Delta \langle (\phi^\dagger \phi)^2\rangle_{\rm c}$ order by order in $x$, while
including the leading contribution from $c_6$.
The resulting soft-enhanced,
dimension-six corrected expressions for the critical parameters up to N$^2$LO, 
\begin{align}
    \yc\big|_{\rmii{N$^2$LO+}\rmi{dim6}}&=
      \frac{1}{2(3\pi)^2x}\left[
          \mathcal{E}_1^2
        + \frac{9}{2}\mathcal{E}_3\,x\, \text{ln}\,\tilde{\Lambda}_\rmi{3d}
        - \mathcal{E}_1\left(\frac{x}{2}
        \right)^{\frac{3}{2}}\right]
        - \frac{\mathcal{E}_1^4\,c_6}{324\pi^4 x^4}
        \,,\\
    \Delta\langle \phi^\dagger \phi \rangle_{\rm c}\big|_{\rmii{N$^2$LO+}\rmi{dim6}}&=
      \frac{1}{2(3\pi)^2x^2}\left[
          \mathcal{E}_1^2
        - \frac{9}{2}\mathcal{E}_3\,x
        + \frac{25}{2}\mathcal{E}_1\left(\frac{x}{2}\right)^{\frac{3}{2}}\right]
        - \frac{\mathcal{E}_1^4 \,c_6}{81\pi^4x^5}
        \,,\\
    \Delta\langle (\phi^\dagger \phi)^2 \rangle_{\rm c}\big|_{\rmii{N$^2$LO+}\rmi{dim6}}&=
      \frac{1}{2^2(3\pi)^4x^4}\left[
          \mathcal{E}_1^4
        - 9\,\mathcal{E}_1^2\mathcal{E}_3\,x
        + 13\,\mathcal{E}_1^3\left(\frac{x}{2}\right)^{\frac{3}{2}}\right]
        - \frac{\mathcal{E}_1^6\,c_6}{729\pi^6 x^7}
        \,,
\end{align}
depend on the 3d renormalization scale $\Lamd$ through
\begin{equation}
    \tilde{\Lambda}_{\rmi{3d}}\equiv
      e^{\ln\frac{3}{2}+\frac{\mathcal{E}_2}{2\mathcal{E}_3}}\pi x \frac{\Lamd}{\mathcal{E}_1 g_3^2}
      \,.
\end{equation}
Additionally,
setting $\mathcal{E}_1=\mathcal{E}_2=\mathcal{E}_3=1$
recovers the results of~\cite{Ekstedt:2024etx,Gould:2023ovu},
including the dimension-six operator contribution as in~\cite{Bernardo:2025vkz}.
As noted in~\cite{Ekstedt:2024etx,Gould:2023ovu},
the scale dependence of $\yc$ is governed by the $\beta$-function
\begin{equation}
    \beta_y\equiv\partial_{t_3}y=\frac{4}{(4\pi)^2}\Bigl[\mathcal{E}_3-2x+2x^2\Bigr]
    \,,
\end{equation}
with $t_3 = \ln\Lamd$.
We therefore define the renormalization-scale-invariant quantity
\begin{equation}
    \tilde{y}_{\rm c}\equiv \yc-\beta_{y}\,\ln\tilde{\Lambda}_{\rmi{3d}}
    \,.
\end{equation}

Upon the substitution of our matching relations,
the expressions above fully determine
the critical behavior of
the Abelian Higgs model to
$\mathcal{O}(g^6)$ in hard-scale and
$\mathrm{N^2 LO} + \mathrm{dim6}$ in soft-scale
contributions within perturbation theory.

\subsection{Comparing higher-dimensional operators against loop corrections}

In this section, we assess the relative importance of
higher-dimensional operators against higher-loop corrections
in determining the equilibrium thermodynamics of the PT.
This question is particularly relevant in the small-$x$ regime,
which is phenomenologically interesting due to its correlation with
stronger PTs~\cite{Bernardo:2025vkz}.
Specifically, we want to determine the value of $x$
beyond which, at fixed orders in $g$,
the contribution from higher-dimensional terms dominates over
higher-loop hard-scale corrections to super-renormalizable operators.

\begin{figure}[t!]
    \includegraphics[width=0.5\linewidth]{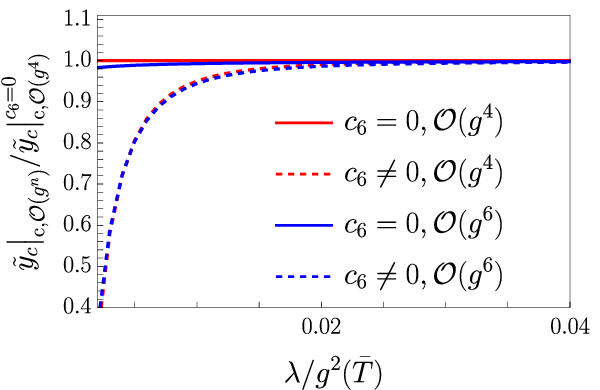}
    \\
    \includegraphics[width=0.5\linewidth]{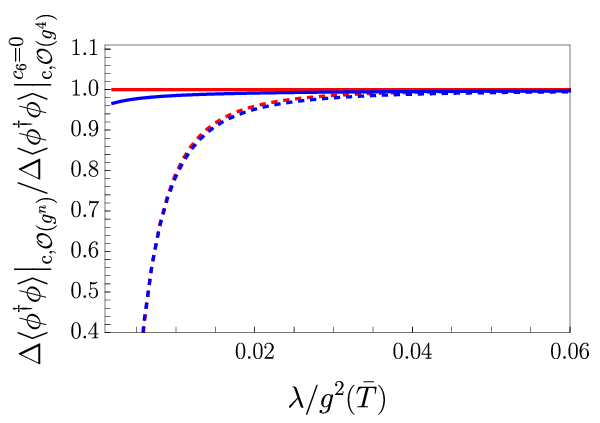}%
    \includegraphics[width=0.5\linewidth]{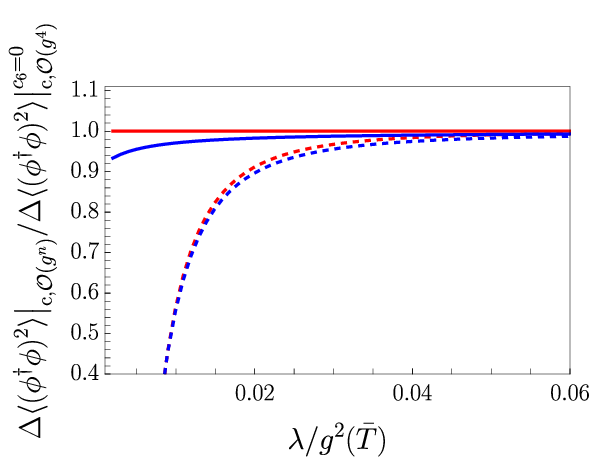}
    \caption{%
    Relative ratios of the thermodynamic quantities $\tilde{y}_{\rm c}$,
    $\Delta\langle\phi^\dagger\phi\rangle_{\rm c}$, and
    $\Delta\langle(\phi^\dagger\phi)^2\rangle_{\rm c}$
    with (dashed) and without (solid) the dimension-six contribution $c_6$,
    computed at
    $\mathcal{O}(g^4)$ (red) and
    $\mathcal{O}(g^6)$ (blue) matching orders.
    For all plots,
    $\Tc=1.2$~GeV,
    $g^2(\overline{T}_{\rm c})=0.3$,
    $\bar{\Lambda}=\overline{T}_{\rm c}\equiv 4\pi e^{\gammaE} T$, and
    $\bar\Lambda_\rmi{3d,opt}=2.85\,\Tc$~\cite{Ghisoiu:2015uza}.
    }
    \label{fig:plots}
\end{figure}

To this aim,
in fig.~\ref{fig:plots}
we plot the thermodynamic quantities
$\tilde{y}_{\rm c}$,
$\Delta\langle\phi^\dagger\phi\rangle_{\rm c}$, and
$\Delta\langle(\phi^\dagger\phi)^2\rangle_{\rm c}$
with and without the leading $c_6$ contribution,
using both
\begin{itemize}
    \item $\mathcal{O}(g^4)$ matching with two-loop masses and one-loop couplings,
    \item $\mathcal{O}(g^6)$ matching with three-loop masses and two-loop couplings.
\end{itemize}
As seen in all panels of fig.~\ref{fig:plots},
for small values of $\lambda/g^2=x_{\rmii{LO}}$,
corrections induced by the $(\phi^\dagger\phi)^3$ operator
exceed those from three-loop hard-scale diagrams.
This small-$x$ regime,
where higher-dimensional operators dominate,
also corresponds to stronger PTs,
as shown in~\cite{Gould:2019qek,Chala:2024xll,Chala:2025oul,Bernardo:2025vkz}.
In the opposite limit of larger $x$,
the situation is reversed,
with higher-loop corrections dominating over
the effects of higher-dimensional operators.
Quantitatively,
this transition occurs only around
$x_{\rmii{LO}}\sim 1$
which is well beyond
the critical endpoint of the theory
which is expected to be 
close to
the value of the critical endpoint in the softer scale EFT~\eqref{eq:DR:A}
located at
$\bar{x}_{\rm c} \approx 0.28$~\cite{Kleinert:1986jp,Mo:2001fi}.%
\footnote{%
    In contrast,
    the exact location of the soft-EFT critical endpoint $\xc$
    is expected to shift slightly in comparison to $\bar{x}_{\rm c}$
    when including dynamical temporal scalars $B_0$
    as seen from the full 4d theory~\cite{Jansen:1985cq,Kleinert:1986te}.
}

\begin{figure}[t]
    \includegraphics[width=0.5\linewidth]{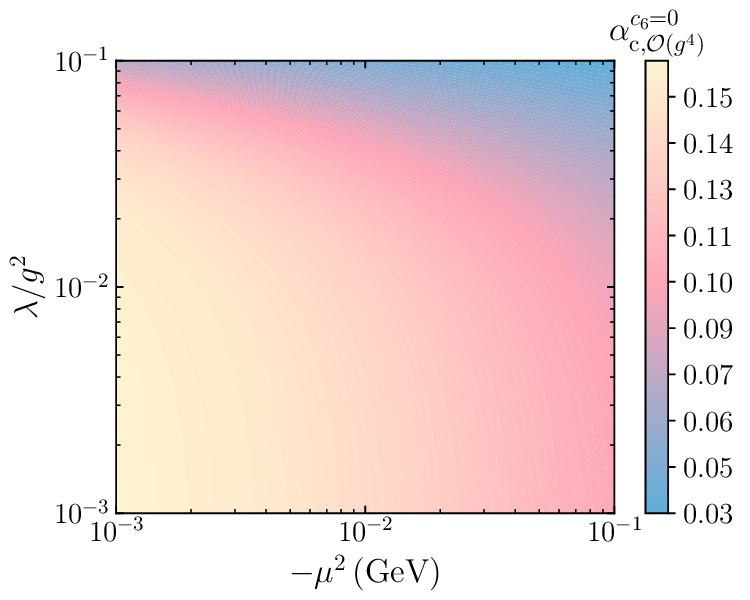}%
    \includegraphics[width=0.5\linewidth]{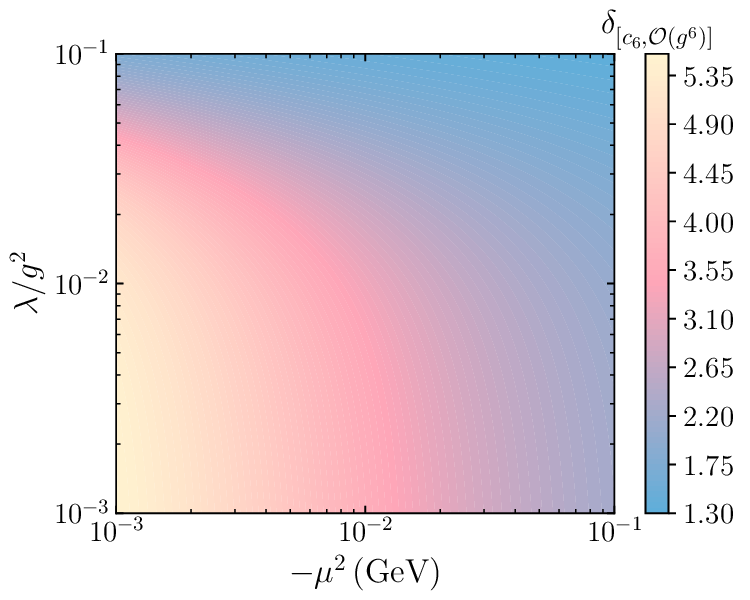}
    \caption{%
    Left:
    Transition strength $\alpha_{\rm c}$ at the critical temperature~\eqref{eq:transition_strength},
    computed using $\mathcal{O}(g^4)$ matching while
    neglecting dimension-six operators.
    Right:
    Ratio of corrections from higher-dimensional operators ($c_6 \neq 0$) to those
    from $\mathcal{O}(g^6)$ higher-loop matching
    as defined in eq.~\eqref{eq:ratio_transition_strength}.
    The four-dimensional couplings are fixed as in~\eqref{eq:BM1}.
    }
    \label{fig:plots_transition_strength}
\end{figure}

Furthermore, fig.~\ref{fig:plots_transition_strength}
displays the transition strength at the critical temperature,
\begin{equation}
\label{eq:transition_strength}
    \alpha_{\rm c}\approx \frac{g_3^6 \Delta S}{3 p_0'} \Big|_{\Tc}
    \,,
\end{equation}
where
the field-independent pressure
$p_0 = \frac{\pi^2}{90}\geff T^4$ is also known as the unit-operator~\cite{Braaten:1995cm},
with
$\geff$ denoting the number of relativistic degrees of freedom in
the Abelian Higgs model,
namely $\geff=4$. 
We extend this analysis across a
wider region of parameter space and compute
the ratio of corrections from dimension-six operators
with $c_6 \neq 0$
to those from $\mathcal{O}(g^6)$ higher-loop matching,
both relative to the $\mathcal{O}(g^4)$ baseline without higher-dimensional operators;
cf.\ fig.~\ref{fig:plots_transition_strength} (right).
To this end, we define
\begin{equation}
\label{eq:ratio_transition_strength}
  \delta_{[c_6,\mathcal{O}(g^6)]} \equiv
  \frac{
    \delta \alpha^{c_6\neq 0}_{{\rm c},\mathcal{O}(g^6)}}{
    \delta \alpha^{c_6= 0}_{{\rm c},\mathcal{O}(g^6)}}\equiv
   \frac{
    \alpha^{c_6\neq 0}_{{\rm c},\mathcal{O}(g^6)}-\alpha^{c_6=0}_{{\rm c},\mathcal{O}(g^4)}}{
    \alpha^{c_6=0}_{{\rm c},\mathcal{O}(g^6)}-\alpha^{c_6=0}_{{\rm c},\mathcal{O}(g^4)}
  }
  \,.
\end{equation}
Fig.~\ref{fig:plots_transition_strength} (left) shows the baseline transition strength
at the critical temperature
computed using $\mathcal{O}(g^4)$ matching while
neglecting dimension-six operators.

As shown in fig.~\ref{fig:plots_transition_strength} (right),
corrections from higher-dimensional operators dominate over those from higher-loop matching
in the parameter space associated with stronger PTs.
This ratio increases as we approach regions of enhanced transition strength,
indicating that the effects of higher-dimensional operators become increasingly significant
in this regime, consistent with the discussion in fig.~\ref{fig:plots}.

While the analysis presented here is specific for
the critical temperature,
we verify that by varying 
the temperature
close to the critical temperature
between $\Tc \to [10^{-2},10^{-1}] \times \Tc$,
similar conclusions should hold at
the nucleation temperature $\Tn < \Tc$.
In fact,
we find that 
$\delta_{[c_6,\mathcal{O}(g^6)]}$ reaches up to $\mathcal{O}(10)$.

Finally,
we comment on the implications of our findings
for classically conformal gauge-Higgs models.
In the conformal limit where the zero-temperature mass parameter $\mu^2$ vanishes
and the PT
scale is generated dynamically by thermal effects,
the Abelian Higgs model~\cite{Coleman:1973jx} with
$\lambda \ll g^2$
exhibits significantly enhanced transition strength $\alpha_{\rm c}$ in the lower left corner
of fig.~\ref{fig:plots_transition_strength}.
However, nucleation occurs at
the escape point of the bubble trajectory,
where field values are typically much smaller than
the broken-phase minimum~\cite{Kierkla:2023von}.
Hence, at $\Tn$,
higher-dimensional operator effects are expected to be
less pronounced and
less dominant over 
higher loop effects in
this class of models.
Nevertheless, such contributions
were never included in
realistic nucleation computations for classically conformal models with strong supercooling and
their study
remains important for accurate GW predictions
and primordial black hole formation from supercooled transitions~\cite{Lewicki:2023ioy,Franciolini:2025ztf,Kierkla:2025vwp,Carr:2026hot}.
We defer a detailed analysis of such a scenario to future work.

%
\section{Outlook}
\label{sec:outlook}

In this work,
we have computed
hard thermal corrections to the equilibrium thermodynamics
of the Abelian Higgs model at high temperatures.
This computation includes
the three-loop corrections to
the Debye and scalar masses, as well as
the four-point gauge-scalar correlators at two-loop level.
These results complete the determination of the effective parameters
of the dimensionally reduced EFT at $\mathcal{O}(g^6)$ accuracy.

For the three-loop scalar mass,
we identify a previously absent contribution in the master integral basis
of~\cite{Ghisoiu:2012kn,Ghisoiu:2015uza}.
While no new sum-integrals emerge at this order and gauge-parameter invariance is preserved,
renormalizability necessitates these new contributions.
As a byproduct, we have carried out a re-evaluation of
the three-loop QCD Debye mass including this missing contribution
yielding a $\sim 7\%$ correction at three-loop level.

Using these results,
we have assessed the relative importance of
higher-loop corrections against higher-dimensional operators
in determining the equilibrium thermodynamics of the PT.
We found that higher-dimensional operators dominate
in the small-$x$ regime,
with $x_\rmii{LO} \sim \lambda/g^2$,
which is phenomenologically interesting due to
the presence of stronger PTs.
As expected, this dominance occurs only for
values of $x$ significantly smaller than those
corresponding to the critical endpoint of the theory
where loop corrections dominate instead.
Our results pave the way for future studies of
the nonequilibrium dynamics of the 
PT
at $\mathcal{O}(g^6)$ accuracy,
including the computation of
the bubble nucleation rate.

Our results are based on the effective potentials of~\cite{Ekstedt:2022zro,Ekstedt:2024etx},
which apply directly to the softer-scale EFT~\eqref{eq:DR:A} up to N$^4$LO accuracy.
A natural next step beyond the soft enhancement of the softer-scale EFT used in
sec.~\ref{sec:dr} is to improve the soft-scale EFT~\eqref{eq:DR:B} by including
temporal gauge-field loop contributions, following~\cite{Ekstedt:2024etx},
again up to N$^4$LO accuracy.
Additionally, our results motivate including
higher-order EFT corrections and further investigating
broad-temperature frameworks~\cite{Laine:2019uua,Curtin:2022ovx,Navarrete:2025yxy}
to study PTs
in BSM extensions in the strongest regimes.

\section*{Acknowledgements}

We thank
Oliver Gould,
Maciej Kierkla,
Pablo Navarrete,
Tuomas V.I.\ Tenkanen,
York Schr\"oder, and
Jorinde van de Vis
for enlightening discussions.
FB and PS are supported by the Swiss National Science Foundation (SNSF) under
grant \href{https://data.snf.ch/grants/grant/215997}{\tt PZ00P2-215997}.
MC is supported by the European Research Council under grant agreement
n.~101230200.
LG is supported by the FPU program under grant number FPU23/02026 and
also acknowledges the support of the European Consortium for Astroparticle Theory in
the form of an Exchange Travel Grant.
This work has received further funding from MICIU/AEI/10.13039/
501100011033 (grants PID2022-139466NB-C21/C22 and PID2024-161668NB-100) as well
as from Junta de Andaluc\'ia (grants FQM 101 and P21-00199).\\

\noindent
{\bf Data availability statement.}
The effective potential expressions used to produce
figs.~\ref{fig:debye_scalar_combined}, \ref{fig:plots}, and \ref{fig:plots_transition_strength}
are publicly available via the software {\tt DRalgo}~\cite{Ekstedt:2024etx}.
Diagrams were generated with {\tt Axodraw}~\cite{Collins:2016aya}.

%
\appendix
\renewcommand{\thesection}{\Alph{section}}
\renewcommand{\thesubsection}{\Alph{section}.\arabic{subsection}}
\renewcommand{\theequation}{\Alph{section}.\arabic{equation}}

%
\section{Renormalization group equations}
\label{app:4d rges}

For completeness,
in this appendix we present the renormalization group (RG) equations
of the 4d Abelian Higgs model, that we compute in the $\overline{\text{MS}}$ scheme.

The 4d bare Lagrangian is defined as
\begin{align}
    \mathcal{L}^{\rmii{$(B)$}}&=
      \frac{1}{4}F^{\rmii{$(B)$}}_{\mu\nu}F^{\rmii{$(B)$}}_{\mu\nu}
    + (D_\mu\phi^{\rmii{$(B)$}})^\dagger (D_\mu\phi^{\rmii{$(B)$}})
    + \mu^{2 \rmii{$(B)$}}(\phi^{\dagger \rmii{$(B)$}}\,\phi^{\rmii{$(B)$}})
    + \lambda^{\rmii{$(B)$}}(\phi^{\dagger \rmii{$(B)$}}\,\phi^{\rmii{$(B)$}})^2
    \nn &
    + \frac{1}{2 \xi^{\rmii{$(B)$}}}(\partial_\mu B^{\rmii{$(B)$}}_\mu)^2
    \,,
\end{align}
with 
$D^{\rmii{$(B)$}}_\mu=\partial_\mu-ig^{\rmii{$(B)$}} B^{\rmii{$(B)$}}_\mu$.
The bare fields and parameters are related to the renormalized ones by
\begin{align}
    \phi^{\rmii{$(B)$}}&=Z_s^{1/2}\phi
    \,,&
    B_\mu^{\rmii{$(B)$}}&=Z_B^{1/2} B_\mu
    \,,&
    \mu^{2\rmii{$(B)$}}&=\mu^2Z_{\mu^2}
    \,,\nn
    \lambda^{\rmii{$(B)$}}&=\lambda\,\Lambda^{2\epsilon}Z_{\lambda}
    \,,&
    g^{2\rmii{$(B)$}}&=g^2\,\Lambda^{2\epsilon}Z_{g^2}
    \,,&
    \xi^{\rmii{$(B)$}}&=\xi Z_{\xi}
    \,,
\end{align}
with $\bar{\Lambda}$ being
the $\overline{\text{MS}}$ scheme scale defined as
$\bar{\Lambda}^2=4\pi e^{-\gammaE}\Lambda^2$, and
$\gammaE$ being the Euler-Mascheroni constant.
At two-loop order,%
\footnote{
  The $Z$-factors presented here remain valid at three-loop level to $\mathcal{O}(g^6)$,
  as $Z_{\mu^2}$ receives no additional running corrections at this order.
}
the normalization factors
are
\begin{align}
    Z_s&=
        1
      + \frac{1}{(4\pi)^2}\frac{1}{\epsilon}(3-\xi)\,g^2
      + \frac{1}{(4\pi)^4}\Big[
          \frac{10-6\xi+\xi^2}{2\epsilon^2}g^4
          + \frac{1}{3\epsilon}(-5g^4-6\lambda^2)
        \Big]
    \,, \nn
    Z_\rmii{$B$}&=
        1
      - \frac{1}{(4\pi)^2}\frac{1}{\epsilon}\frac{1}{3}\,g^2
      + \frac{1}{(4\pi)^4}\Big[-\frac{2}{\epsilon}g^4\Big]
      \,,\nn
    Z_{\xi}&= Z_\rmii{$B$}
      \,,\nn
    Z_{g^2}&=
        1
      + \frac{1}{(4\pi)^2}\frac{1}{\epsilon}\frac{1}{3}\,g^2
      + \frac{1}{(4\pi)^4}\Big[\frac{1}{9\,\epsilon^2}\,g^4+\frac{2}{\epsilon}\,g^4\Big]
    \,,\nn
    \lambda\,Z_{\lambda}&=
        \lambda
      + \frac{1}{(4\pi)^2}\frac{1}{\epsilon}\Big(10\lambda^2-6g^2\lambda+3g^4\Big)
      + \frac{1}{(4\pi)^4}\Big[\frac{1}{\epsilon^2}\,\Big(100\lambda^3-90g^2\lambda^2+47g^4\lambda-8g^6\Big)
    \nn &
      + \frac{1}{\epsilon}\,\Big(-60\lambda^3+28g^2\lambda^2+\frac{79}{3}g^4\lambda-\frac{52}{3}g^6\Big)\Big]
      \,,\nn
    Z_{\mu^2}&=
        1
      + \frac{1}{(4\pi)^2}\frac{1}{\epsilon}\Big(4\lambda-3g^2\Big)
      \nn &
      + \frac{1}{(4\pi)^4}\Big[\frac{1}{\epsilon^2}\Big(28\lambda^2-24g^2\lambda+10g^4\Big)+\frac{1}{\epsilon}\Big(-10\lambda^2+16g^2\lambda+\frac{43}{6}g^4\Big)\Big]
      \,.
\end{align}
As a crosscheck, we verify that
$Z_{g^2}=(Z_\rmii{B})^{-1}$ and $Z_\xi=Z_\rmii{B}$,
according to gauge invariance.
The related $\beta$-functions are
\begin{align}
\label{4d running}
    \beta_{g^2}&\equiv \partial_t g^2=
        \frac{1}{(4\pi)^2}\Big[\frac{2}{3}g^4\Big]+\frac{1}{(4\pi)^4}\Big[8g^6\Big]
        \,,\nn
    \beta_{\lambda}&\equiv \partial_t \lambda =
          \frac{1}{(4\pi)^2}\Big[6g^4-12g^2\lambda+20\lambda^2\Big]
        \nn &
        \hphantom{{}\equiv \partial_t \lambda} 
        + \frac{1}{(4\pi)^4}\Big[-\frac{208}{3}g^6+\frac{316}{3}g^4\lambda+112g^2\lambda^2-240\lambda^3\Big]
        \,,\nn
    \gamma_{\mu^2}&\equiv\frac{\partial_t\mu^2}{\mu^2} =
        \frac{1}{(4\pi)^2}\Big[-6g^2+8\lambda\Big]
      \nn &
      \hphantom{{}\equiv \frac{\partial_t\mu^2}{\mu^2}}
      + \frac{1}{(4\pi)^4}\Big[-40\lambda^2+64g^4\lambda+\frac{86}{3}g^4\Big]
    \,,
\end{align}
with $t\equiv \ln\bar{\Lambda}$.

%
\section{Master integrals}
\label{app:masters}

In the following, we use the same notation as in~\cite{Laine:2016hma}, and
we present all our sum-integral results in
the $\overline{\mathrm{MS}}$ scheme in dimensional regularization,
with $d = 3 - 2\epsilon$.
We adopt the usual notation for sum-integrals
\begin{align}
    \Tint{K} &\equiv T \sum_{n=-\infty}^\infty \int_\vec{k}
    \,,&
    \int_\vec{k} &\equiv \Lambda^{3-d}
      \int \frac{{\rm d}^{d}\vec{k}}{(2\pi)^{d}}
    \,,
\end{align}
where $K = (k_0, \mathbf{k}) = (k_n, \mathbf{k})$ is a loop 4-momentum, and
$n \in \mathbb{Z}$ labels the bosonic Matsubara modes
with zero-momenta $k_n = 2\pi n T$ running in the loop.
We use a basis of sum-integrals of the form
\begin{align}
\label{eq:I:1loop}
    \mathcal{I}_{s}^{\alpha} &= \Tint{K}
        \frac{(k_0)^{\alpha}}{\bigl[ K^2 \bigr]^{s}}
    \;,\\[1mm]
\label{eq:I:2loop}
    \mathcal{I}_{s_1 s_2 s_3}^{\alpha_1 \alpha_2} &= \Tint{K_1 K_2} \frac{
        (k_{1,0})^{\alpha_1}
        (k_{2,0})^{\alpha_2}
        }{
        \bigl[ K_1^2 \bigr]^{s_1}
        \bigl[ K_2^2 \bigr]^{s_2}
        \bigl[ (K_1 - K_2)^2 \bigr]^{s_3}
        }
    \;,\\[1mm]
\label{eq:I:3loop}
    \mathcal{I}_{s_1 s_2 s_3 s_4 s_5 s_6}^{\alpha_1 \alpha_2 \alpha_3} &=
    \Tint{K_1 K_2 K_3}
    \frac{
        (k_{1,0})^{\alpha_1}
        (k_{2,0})^{\alpha_2}
        (k_{3,0})^{\alpha_3}
        }{
        \bigl[ K_1^2\bigr]^{s_1}
        \bigl[ K_2^2\bigr]^{s_2}
        \bigl[ K_3^2\bigr]^{s_3}
        \bigl[ (K_1 - K_2)^2\bigr]^{s_4}
        \bigl[ (K_1 - K_3)^2\bigr]^{s_5}
        \bigl[ (K_2 - K_3)^2\bigr]^{s_6}
        }
    \;.
\end{align}
At one- and two-loop level,
master integrals factorize into one-loop thermal integrals of
the type~\eqref{eq:I:1loop};
see~\cite{Davydychev:2023jto,Davydychev:2022dcw} for a proof.
Full bosonic integrals without Matsubara modes in the numerator
are abbreviated with
$\mathcal{I}_{s_1\dots s_n}^{0\dots0} \equiv \mathcal{I}_{s_1\dots s_n}^{ }$.

At three-loop level,
there exist master integrals which do not factorize into lower-loop cases.
We use an in-house Laporta algorithm~\cite{Laporta:2000dsw}
adapted to finite temperature~\cite{Nishimura:2012ee,Schicho:2020xaf}
to reduce integrals to a finite set of master integrals~\cite{Smirnov:2010hn}
via integration by parts (IBP).%
\footnote{%
  As a crosscheck of all reduction results presented here,
  we also make use of {\tt SIRENA}~\cite{Gil:2026cqz},
  a publicly available \texttt{python}-based tool for sum-integral reduction.
}
Once a basis of master integrals is found, further exploitation of
IBP relations can also allow us to change the basis of master integrals to evaluate.
This becomes a necessity in the evaluation of the three-loop thermal masses,
as discussed in sec.~\ref{app:thermal:masses:U1} and~\cite{Chala:2025oul}.
After these manipulations,
the remaining sum-integrals in eq.~\eqref{eq:masters:3loop} can
be addressed using classic~\cite{Arnold:1994eb,Arnold:1994ps,Ghisoiu:2012yk,Ghisoiu:2013zoj} and
modern methods~\cite{Catani:2008xa,Seppanen:2025owq,Navarrete:2025yxy}.
While the procedure is, in principle, straightforward,
it is technically demanding.
An explicit algorithm for evaluating the basketball-type integrals was developed
in~\cite{Moller:2012chx},
illustrated with the example of $\mathcal{I}^{200}_{310011}$.
For the present calculation,
the dimension-two three-loop basketball master integral
$\mathcal{I}^{ }_{110011}$ is known from~\cite{Arnold:1994ps},
$\mathcal{I}^{ }_{210011}$ from~\cite{Gynther:2007bw,Moller:2010xw}, and
the corresponding spectacles-type integrals were tamed
in~\cite{Ghisoiu:2012yk,Ghisoiu:2013zoj,Ghisoiu:2015uza}.
The different spectacle-type integrals were evaluated
in~\cite{Andersen:2008bz,Schroder:2012hm} for $\mathcal{I}_{111110}^{ }$,
\cite{Ghisoiu:2012kn} for $\mathcal{I}_{211110}^{020}$, and
\cite{Ghisoiu:2012yk} for $\mathcal{I}_{31111-2}^{ }$.
Here, we collect the evaluated expressions for these integrals
that also form the master integral basis~\eqref{eq:masters:3loop},
\begin{align}
  \mathcal{I}_{210011} &=
    \frac{T^2}{(4\pi)^4}
    \Bigl(\frac{1}{4\pi T^2}\Bigr)^{3\epsilon}
    \frac{1}{8\epsilon^2}
  \biggl[
      1
    + \Bigl(
      \frac{17}{6}
      + \gammaE
      + 2 Z_{1}'
      \Bigr)\epsilon
    \\ &
    \hphantom{{}=\frac{T^2}{(4\pi)^4}
    \Bigl(\frac{1}{4\pi T^2}\Bigr)^{3\epsilon}
    \frac{1}{8\epsilon^2}\Bigl[1}
    + \Bigl(
        \frac{131}{12}
      + \frac{31\pi^2}{36}
      + 8\ln(2\pi)
      - \frac{9\gammaE}{2}
      - \frac{15\gammaE^2}{2}
      + (5+2\gammaE) Z_{1}'
    \nn &
    \hphantom{{}=\frac{T^2}{(4\pi)^4}
    \Bigl(\frac{1}{4\pi T^2}\Bigr)^{3\epsilon}
    \frac{1}{8\epsilon^2}\Bigl[1}
      + 2 Z_{1}''
      - 16\gamma_{1}
      + \frac{4\zeta_{3}}{9}
      - 0.145652981107(4)
      \Bigr)\epsilon^2
    + \mathcal{O}(\epsilon^3)
  \biggr]
  \,,\nn[2mm]
  \mathcal{I}^{ }_{111110} &=
  - \frac{T^2}{(4\pi)^4}
    \Bigl(\frac{1}{4\pi T^2}\Bigr)^{3\epsilon}
    \frac{1}{4\epsilon^2}
  \biggl[
      1
    + \Bigl(
      \frac{4}{3}
      + \gammaE
      + 2 Z_{1}'
      \Bigr)\epsilon
    \\ &
    \hphantom{{}=-\frac{T^2}{(4\pi)^4}
    \Bigl(\frac{1}{4\pi T^2}\Bigr)^{3\epsilon}
    \frac{1}{8\epsilon^2}}
    \hspace{-2cm}
    + \Bigl(
      \frac{1}{3}\Bigl[
          46
        + \frac{45\pi^2}{4}
        + 24\ln(2\pi)(\ln(2\pi)-\gammaE)
        - 104\gamma_1
        - 8\gammaE^{ }\Bigl(\frac{5}{16}\gammaE^{ }+1\Bigr)
    \nn &
    \hphantom{{}=-\frac{T^2}{(4\pi)^4}
    \Bigl(\frac{1}{4\pi T^2}\Bigr)^{3\epsilon}
    \frac{1}{8\epsilon^2}}
    \hspace{-2cm}
        + 8(2\gammaE + 3)Z_1'
        + 2Z_1''
      \Bigr]
      - 38.53084275773721000(10)
    \Bigr)\epsilon^2
    + \mathcal{O}(\epsilon^3)
  \biggr]
  \,,\nn[2mm]
  \mathcal{I}^{020}_{211110} &=
  \frac{T^2}{(4\pi)^4}
  \Bigl(\frac{1}{4\pi T^2}\Bigr)^{3\epsilon}
  \frac{1}{96\epsilon^2}
  \Bigl[
      1
    + \Bigl(
      \frac{67}{6}
      + \gammaE
      + 2 Z_{1}'
      \Bigr)\epsilon
    + 93.0894417(2)\epsilon^2
    + \mathcal{O}(\epsilon^3)
  \Bigr]
  \,,\\[2mm]
  \label{eq:I:31111-2}
  \mathcal{I}^{ }_{31111-2} &=
  -\frac{T^2}{(4\pi)^4}
    \Bigl(\frac{1}{4\pi T^2}\Bigr)^{3\epsilon}
    \frac{5}{36\epsilon^2}
  \Bigl[
      1
    + \Bigl(
      \frac{59}{30}
      + \gammaE
      + 2 Z_{1}'
      \Bigr)\epsilon
    + 42.1751477(1)\epsilon^2
    + \mathcal{O}(\epsilon^3)
  \Bigr]\,.
\end{align}
Here, $\gamma_{1}$ is the first of the Stieltjes constants defined via
$\zeta_s = (s-1)^{-1} + \sum_{n=0}^{\infty} (1-s)^n \gamma_n/n!$, and
$\zeta_s \equiv \zeta(s)$ is the Riemann $\zeta$-function.
We also abbreviate derivatives of
the $\zeta$-function by $Z_s^{(n)} = \zeta^{(n)}_{-s} / \zeta_{-s}$.
The divergent piece in
eq.~\eqref{eq:I:31111-2} differs
compared to the original~\cite{Ghisoiu:2013zoj}.
Here, we report the correct result compatible with renormalizability and gauge invariance of the three-loop masses in both the Abelian Higgs and $\mathrm{SU}(2)$ + Higgs theories (see appendix~\ref{app:mismatch}).

%
\section{Details of dimensional reduction}
\label{app:DR:relations}

In this section, we collect all matching relations and expressions of
the effective parameters in the soft-scale EFT~\eqref{eq:DR:B}.
The 3d~EFT bare%
\footnote{%
  Here, by \textit{bare}
  we mean that the matching equations do not include the corresponding counterterms in the 3d EFT.
  4d renormalization of these results is omitted for brevity. To get a finite result, we replace each bare 4d parameter $c_i^\rmii{(B)} \to c_i - \delta c_i$ by its renormalized counterpart perturbatively.}
Lagrangian is given by
\begin{align}
  \label{eq:L:soft:U1}
\mathcal{L}^{\rmii{$(B)$}}_{\rmi{soft}}&=
    \frac{1}{4}F^{\rmii{$(B)$}}_{ij}F^{\rmii{$(B)$}}_{ij}
  + (D_i\phi^{\rmii{$(B)$}})^{\dagger}(D_i\phi^{\rmii{$(B)$}})
  + \widehat{\mu}_3^{2 \rmii{$(B)$}} (\phi^{\dagger \rmii{$(B)$}}\phi^{\rmii{$(B)$}})
  + \frac{1}{2}(\partial_i B_0^{\rmii{$(B)$}})^2
  + \frac{1}{2}\widehat{m}_\rmii{D}^{2 \rmii{$(B)$}} (B_0^{\rmii{$(B)$}})^2
  \nn &
  + \widehat{\lambda}_3^{\rmii{$(B)$}}(\phi^{\dagger \rmii{$(B)$}}\phi^{\rmii{$(B)$}})^2
  + \widehat{h}_3^{\rmii{$(B)$}}(\phi^{\dagger \rmii{$(B)$}}\phi^{\rmii{$(B)$}})(B_0^{\rmii{$(B)$}})^2
  + \widehat{\kappa}_3^{\rmii{$(B)$}}(B_0^{\rmii{$(B)$}})^4
  + \frac{1}{2\xi}(\partial_i B^{\rmii{$(B)$}}_i)^2
  + \mathcal{L}_\text{soft}^{(6)}
  \,,
\end{align}
where $D_i = \partial_i - i\,\widehat{g}_3 B_i^{\rmii{$(B)$}}$.
The $\mathcal{L}_\text{soft}^{(6)}$ piece contains an off-shell basis of dimension-six operators,
listed in tab.~\ref{tab:softop:dim6}.
Hatted effective parameters indicate this off-shell basis;
they 
shift under the field redefinitions to
the on-shell physical basis discussed below.
\begin{table}[t]
\centering
\begin{tabular}[t]{|l|l|}
\hline
\multicolumn{2}{|c|}{dimension-six operator basis}\\
\hline
\hline
$F_{ij}F_{ij} B_0^2$ &
$\alphaF_{B_0^2 F^2}$ \\ \hline
$F_{ij} F_{ij} \phi^\dagger \phi$ &
$\alphaF_{\phi^2 F^2}$ \\ \hline
$(D_i \phi^\dagger D_i \phi) (\phi^\dagger \phi)$ &
$  \alphaF_{D^2\phi^4, 1}$ \\ \hline
$(D_i\phi^\dagger D_i \phi) B_0^2$ &
$\alphaF_{D^2\phi^2 B_0^2, 3}$ \\ \hline
$B_0^6$ & $\alphaF_{B_0^6}$ \\ \hline
$B_0^4 (\phi^\dagger\phi) $ &
$\alphaF_{\phi^2 B_0^4}$ \\ \hline
$B_0^2 (\phi^\dagger\phi)^2 $ &
$\alphaF_{\phi^4 B_0^2}$ \\ \hline
$(\phi^\dagger\phi)^3$ &
$\alphaF_{\phi^6 }$ \\ \hline
\end{tabular}
\hspace{1cm}
\begin{tabular}[t]{|l|l|}
\hline
\multicolumn{2}{|c|}{Redundant operators}\\
\hline
\hline
$(\partial_i F_{ij})^2$ &
$\alphaF_{D^2 F^2}$ \\ \hline
$B_0 \Box^2 B_0$ &
$\alphaF_{D^4B_0^2}$ \\ \hline
$B_0^3 \Box B_0$ &
$\alphaF_{D^2B_0^4}$ \\ \hline
$(D^2 \phi^\dagger) (D^2 \phi)$ &
$\alphaF_{D^4\phi^2}$ \\ \hline
$(\phi^\dagger \phi) (\phi^\dagger D^2 \phi + \textit{h.c.})$ &
$ \alphaF_{D^2\phi^4,2}$ \\ \hline
$(\partial_i F_{ij}) i \phi^\dagger (D_j \phi)$ &
$\alphaF_{D^2\phi^2 F}$ \\ \hline
$(\phi^\dagger \phi) B_0 \Box B_0$ &
$\alphaF_{D^2\phi^2 B_0^2, 1}$ \\ \hline
$\phi^\dagger (D_i ^2\phi) B_0^2 + \textit{h.c.}$ &
$\alphaF_{D^2\phi^2 B_0^2, 2}$ \\ \hline
\end{tabular}
\caption{%
  Off-shell basis of dimension-six operators in the soft-scale 3d effective theory
  in terms of bare fields and couplings;
  cf.\ also~\cite{Bernardo:2025vkz}.
  }
  \label{tab:softop:dim6}
\end{table}

The matching relations for the bare Wilson coefficients
up to $\mathcal{O}(g^6)$,
\begin{align}
    Z_{\phi_3}&=Z_\phi\Big\{1+(\xiNew-3)\, g^2 \,\mathcal{I}_2+\frac{24(d-4)(d-3)}{(d-7)(d-5)(d-2)d} \, \lambda^2 \, (\mathcal{I}_2)^2 + \Big(\frac{10}{3}-2\xiNew\Big) \, g^2\mu^2 \, \mathcal{I}_3 
    \nn
    &\phantom{=} + \Big(\frac{40}{3}-8\xiNew\Big) \, g^2\lambda \, \mathcal{I}_1\mathcal{I}_3 + g^4 \Big[\Big(\frac{2(72-84d-23d^2+5d^3)}{3(d-7)d}-2d\,\xiNew\Big)\,\mathcal{I}_1\mathcal{I}_3
    \nn
    &\phantom{=}+\Big(\frac{192-752d+583d^2-142d^3+11d^4}{2(d-7)(d-5)(d-2)d} - 3\xiNew+\frac{1}{2}\xiNew^2 \Big)(\mathcal{I}_2)^2\Big]\Big\} \,,
    \\
    Z_{\rmii{$B_0$}}&=Z_\rmii{$B$}\Big\{
        1
        + \frac{1}{3}(4-d)g^2\,\mathcal{I}_2
        + \frac{2}{3}(d-6) \, g^2\mu^2 \,\mathcal{I}_3  
    \nn
    &\phantom{=}
        - \frac{4(d-4)(d-3)(d^2-4d-3)}{(d-7)(d-5)(d-2)d}g^4\,(\mathcal{I}_2)^2 
        + \frac{2}{3}(d-6)g^2\Big[dg^2+4\lambda\Big]\,\mathcal{I}_1\,\mathcal{I}_3
    \Big\}
    \,, \\[2mm]
    Z_{\rmii{$B_i$}}&=Z_\rmii{$B$}\Big\{
        1
        + \frac{1}{3}g^2\,\mathcal{I}_2
        -\frac{2}{3}g^2\mu^2\,\mathcal{I}_3        
    \nn
    & \phantom{=}
        + \frac{24(d-4)(d-3)}{(d-7)(d-5)(d-2)d}g^4\,(\mathcal{I}_2)^2
        - \frac{2}{3}g^2\Big[dg^2+4\lambda\Big]\,\mathcal{I}_1\,\mathcal{I}_3
    \Big\}
    \,,\\[2mm]
    \widehat{\mu}_3^{2}&=
        \mu^2
        + \Big[d\,g^2+4\lambda\Big]\,\mathcal{I}_1
        + \Big[3g^2-4\lambda\Big]\,\mu^2\,\mathcal{I}_2
        + \Big[4\,\lambda+\Big(-\frac{10}{3}+\xiNew\Big)g^2\Big] \, \mu^4 \, \mathcal{I}_3
    \nn
    &\phantom{=}
        + \Big[(2+d)g^4-4(d-3)g^2\lambda-16\lambda^2\Big]\mathcal{I}_1\mathcal{I}_2
    \nn
    &\phantom{=}
        + \Big[32\lambda^2+\Big(\frac{8}{3}(3d-10)+8\,\xiNew\Big)g^2\lambda
        - \Big(\frac{4(36-42d-29d^2+5d^3)}{3(d-7)d}-2d\,\xiNew\Big)g^4 \Big]\, \mu^2 \,\mathcal{I}_1\mathcal{I}_3
    \nn
    &\phantom{=}
        + \Big[\frac{-96-254d+89d^2+68d^3-25d^4+2d^5}{(d-7)(d-5)(d-2)d} \, g^4 
    \nn
    &\phantom{=}
        - \frac{8(d-6)(d-1)}{(d-5)(d-2)} \, g^2\lambda
        + \frac{16(d^2-7d-2)(d^2-7d+9)}{(d-7)(d-5)(d-2)d} \, \lambda^2 \Big] \, \mu^2 \, (\mathcal{I}_2)^2
    \nn
    &\phantom{=} - \Big[\frac{96(d-4)(d-3)}{(d-7)(d-5)(d-2)d} \lambda^3
    +\Big(\frac{24(2d^3-27d^2+111d-128)}{(d-7)(d-5)(d-2)}-16\,\xiNew\Big)g^2\lambda^2
    \nn
    &\phantom{=} + \Big(\frac{2(192+508d-899d^2+464d^3-91d^4+6d^5)}{(d-7)(d-5)(d-2)d}-4d\,\xiNew+2\,\xiNew^2\Big)\,g^4\lambda
    \nn
    &\phantom{=} + \Big(\frac{(d-3)(5d^3-55d^2+232d-344)}{2(d-2)(d-5)(d-7)}-2(d-1)\,\xiNew+\frac{1}{2}d\,\xiNew^2\Big)\,g^6\Big]\,\mathcal{I}_1(\mathcal{I}_2)^2
    \nn
    &\phantom{=}
        - \Big[\Big(\frac{160}{3}-32\,\xiNew\Big)g^2\lambda^2
        + \Big(\frac{16(36-42d-29d^2+5d^3)}{3d(d-7)}-16d\,\xiNew\Big)g^4\lambda
    \nn
    &\phantom{=}
        + \Big(\frac{2(72-84d-23d^2+5d^3)}{3(d-7)}-2d^2\,\xiNew\Big)g^6\Big]\,(\mathcal{I}_1)^2\mathcal{I}_3
        +\Pi_{\phi^\dagger \phi}^{3\ell,\rmii{$(B)$}}
    \,, \\[2mm]
    \widehat{m}_\rmii{D}^{2}&=
        2(d-1)g^2\,\mathcal{I}_1
        - 2(d-3)g^2\mu^2\,\mathcal{I}_2
        - \Big[
            \frac{4}{3}(d^2-2d-2)g^4
            + 8(d-3)g^2\lambda
        \Big]\,\mathcal{I}_1\mathcal{I}_2
    \nn
    &\phantom{=}
    - \Big[\frac{2(d-3)(d^2-9d+26)}{3(d-2)}g^4 - 8(d-3)g^2\lambda \Big]\, \mu^2 \, (\mathcal{I}_2)^2+2(d-5)g^2\mu^4\,\mathcal{I}_3
    \nn
    &\phantom{=}
    - \frac{4}{3} (d-6) (d-1) \Big[d g^6 + 4 \lambda g^4 \Big] \,(\mathcal{I}_1)^2\,\mathcal{I}_3
    \nn
    &\phantom{=} 
    - \frac{4}{3} (d-4) \Big[
        2 (d-3)g^4\lambda
      + \frac{162+140d-266d^2-16d^3+67d^4-16d^5+d^6}{3(d-7)(d-5)(d-2)d}g^6
      \Big]\,\mathcal{I}_1(\mathcal{I}_2)^2
    \nn
    &\phantom{=}
    + \Big[\frac{8}{3}(d^2-4d-3)g^4 + 16(d-5)g^2\lambda \Big]\, \mu^2 \,\mathcal{I}_1\mathcal{I}_3 
    + \Pi_{B_0 B_0}^{3\ell,\rmii{$(B)$}} \,,
    \\
    \frac{\widehat{\lambda}_3}{T}&= \lambda + \Big[-dg^4-10\lambda^2+6g^2\lambda\Big]\,\mathcal{I}_2
    +\Big[-\frac{2(3d^3-20d^2+27d+10)}{(d-5)(d-2)}g^6
    \nn
    &\phantom{=}-\frac{2(282-217d+31d^2)}{(d-5)(d-2)}g^2\lambda^2+\frac{4(9d^4-126d^3+487d^2-322d-144)}{(d-7)(d-5)(d-2)d}\lambda^3
    \nn
    &\phantom{=}+\frac{8d^5-95d^4+210d^3+619d^2-1390d-192}{(d-7)(d-5)(d-2)d}g^4\lambda\Big]\,(\mathcal{I}_2)^2 + 20 \Big[\lambda^2 - \frac{1}{3}g^2\lambda\Big]\, \mu^2 \,\mathcal{I}_3
    \nn
    &\phantom{=}
    + \Big[
        4(d-1)g^6+80\lambda^3
        + \frac{20}{3}(3d-4)g^2\lambda^2
        - \frac{4(72-84d-23d^2+5d^3)}{3(d-7)d}g^4\lambda
    \Big]\,\mathcal{I}_1\mathcal{I}_3
    \,, \\[2mm]
    \frac{\widehat{h}_3}{T}&=g^2-\Big[\frac{1}{3}(2d-5)g^4+4(d-3)g^2\lambda\Big]\,\mathcal{I}_2 + \Big(\frac{2}{3}g^4\mu^2+8(d-5)g^2\lambda\mu^2\Big)\,\mathcal{I}_3
    \nn
    &\phantom{=}+\Big[\frac{-2160 - 9274 d + 12805 d^2 - 7372 d^3 + 2191 d^4 - 310 d^5 + 16 d^6}{9(d-7)(d-5)(d-2)d}\,g^6
    \nn
    &\phantom{=}-\frac{4 (d - 4) (d - 3) (28 - d + d^2)}{3(d-5)(d-2)}g^4\lambda
    \nn
    &\phantom{=}+\frac{16 (-3 + d) (6 - 27 d + 45 d^2 - 13 d^3 + d^4)}{d(d-7)(d-5)(d-2)}g^2\lambda^2\Big]\,(\mathcal{I}_2)^2\nn
    &\phantom{=}
    + \Big[
        \frac{2(-72 + 96 d - 31 d^2 + d^3)}{3(d-7)d}g^6
        +\frac{8}{3}(3d^2-15d+1)g^4\lambda
        +32(d-5)g^2\lambda^2
    \Big]\,\mathcal{I}_1\mathcal{I}_3
    \,, \\[2mm]
    \frac{\widehat{g}_3^{2}}{T}&= g^2
    - \frac{1}{3}g^4\,\mathcal{I}_2
    + \frac{d^4-14d^3-157d^2+1442d-2592 }{9(d-7)(d-5)(d-2)d}g^6\,(\mathcal{I}_2)^2
    \nn
    & \phantom{=} + \frac{2}{3}g^4\mu^2\mathcal{I}_3 + \frac{2}{3} \Big[d\,g^6+4 g^4\lambda\Big]\mathcal{I}_1\mathcal{I}_3\,,
    \\[2mm]
    \frac{\widehat{\kappa}_3}{T}&=
    - \frac{1}{6}(d-3)(d-1)g^4\,\mathcal{I}_2
    - \Big[\frac{1}{18}(d-7)(d-3)(2d-5)g^6 - 2(d-3)^2g^4\lambda\Big]\,(\mathcal{I}_2)^2
    \nn & \phantom{=}
    - \frac{1}{3}(d-5)(d-3)g^4\mu^2\,\mathcal{I}_3
    - \frac{1}{3}(d-5)(d-3)g^4\Big[dg^2+4\lambda\Big]\,\mathcal{I}_1\mathcal{I}_3
    \,,
\end{align}
are given in terms of unevaluated master sum-integrals, and
are explicitly gauge parameter ($\xiNew$) dependent.
We omitted the superscript $(B)$ on the left-hand side and
collected bare three-loop contributions to the
scalar and Debye masses into
$\Pi_{\phi^\dagger \phi}^{3\ell,\rmii{$(B)$}}$ and 
$\Pi_{B_0 B_0}^{3\ell,\rmii{$(B)$}}$,
which are discussed separately in eqs.~\eqref{eq:Pi:phiphi} and~\eqref{eq:Pi:B0B0}.

The dimension-six effective parameters and
the field redefinitions required to change from
the off-shell basis to the on-shell one are given in~\cite{Bernardo:2025vkz}.
After these redefinitions, and truncating at $\mathcal{O}(g^6)$,
we find the following expressions for the bare, on-shell parameters
\begin{align}
    \mu_3^{2}&= \mu^2 + \Big[d\,g^2+4\lambda\Big]\,\mathcal{I}_1 +\Big[(2+d)g^4-4(d-3)g^2\lambda-16\lambda^2\Big]\mathcal{I}_1\mathcal{I}_2
    \nn
    &\phantom{=}+ \Big[3g^2-4\lambda\Big]\,\mu^2\,\mathcal{I}_2 + \Big[\frac{-96-254d+89d^2+68d^3-25d^4+2d^5}{(d-7)(d-5)(d-2)d}\,g^4
    \nn
    &\phantom{=} -\frac{8(d-6)(d-1)}{(d-5)(d-2)}\,g^2\lambda +\frac{16(d^2-7d-2)(d^2-7d+9)}{(d-7)(d-5)(d-2)d}\,\lambda^2\Big] \, \mu^2 \, (\mathcal{I}_2)^2
    \nn
    &\phantom{=} -\Big[\frac{96(d-4)(d-3)}{(d-7)(d-5)(d-2)d}\lambda^3 +\frac{24(2d^3-27d^2+111d-128)}{(d-7)(d-5)(d-2)}g^2\lambda^2
    \nn
    &\phantom{=} +\frac{2(192+508d-899d^2+464d^3-91d^4+6d^5)}{(d-7)(d-5)(d-2)d}\,g^4\lambda 
    \nn
    &\phantom{=}+\frac{(d-3)(5d^3-55d^2+232d-344)}{2(d-2)(d-5)(d-7)}\,g^6\Big]\,\mathcal{I}_1(\mathcal{I}_2)^2+\Big[4\,\lambda-\frac{5}{3}g^2\Big]\, \mu^4 \,\mathcal{I}_3
    \nn
    &\phantom{=}+\Big[32\lambda^2 + \frac{8}{3}(3d-5)g^2\lambda -\frac{2(5d^3-23d^2-84d+72)}{3(d-7)d}g^4\Big] \, \mu^2 \,\mathcal{I}_1\mathcal{I}_3
    \nn
    &\phantom{=}-\Big[\frac{80}{3}g^2\lambda^2 + \frac{8(72-84d-23d^2+5d^3)}{3d(d-7)}g^4\lambda + \frac{5d^3-11d^2-168+144}{3(d-7)}g^6\Big]\,(\mathcal{I}_1)^2\mathcal{I}_3
    \nn
    &\phantom{=} -\frac{1}{2}(dg^2+4\lambda)\,g^4\,\mathcal{I}_1(\mathcal{I}_2)^2 \xiNew^2+\Big\{2\Big[(d-1)g^4+2d\,g^2\lambda+8\lambda^2\Big]g^2\,\mathcal{I}_1(\mathcal{I}_2)^2
    \nn
    &\phantom{=} +\Big(d\,g^2+4\lambda\Big)^2g^2\,(\mathcal{I}_1)^2\mathcal{I}_3\Big\}\xiNew
    + \Pi_{\phi^\dagger \phi}^{3\ell,\rmii{$(B)$}} \,,
    \\
    \mD^{2}&= 2(d-1)g^2\,\mathcal{I}_1 -2(d-3)g^2\mu^2\,\mathcal{I}_2 - \Big[\frac{4}{3}(d^2-2d-2)g^4 + 8(d-3)g^2\lambda\Big]\,\mathcal{I}_1\mathcal{I}_2
    \nn
    &\phantom{=} - \Big[\frac{2(d-3)(d^2-9d+26)}{3(d-2)}g^4 - 8(d-3)g^2\lambda\Big]\, \mu^2 \,(\mathcal{I}_2)^2 + 2(d-5)g^2\mu^4\,\mathcal{I}_3
    \nn
    &\phantom{=}- \frac{4}{3} (d-6)(d-1) \Big[\frac{1}{5}(4d - 1) g^6 + 4 g^4\lambda\Big]\,(\mathcal{I}_1)^2\,\mathcal{I}_3
    \nn
    &\phantom{=}
    - \frac{4}{3} (d-4) \Big[
        2(d-3)g^4\lambda
      + \frac{162+140d-266d^2-16d^3+67d^4-16d^5+d^6}{3(d-7)(d-5)(d-2)d}g^6
      \Big]\,\mathcal{I}_1(\mathcal{I}_2)^2
    \nn
    &\phantom{=}+\Big[\frac{8}{3}(d^2-4d-3)g^4+16(d-5)g^2\lambda\Big]\, \mu^2 \,\mathcal{I}_1\mathcal{I}_3
    + \Pi_{B_0 B_0}^{3\ell,\rmii{$(B)$}} \,,
    \\
    \frac{\lambda_3}{T}&=\lambda+\Big[-dg^4-10\lambda^2+6g^2\lambda\Big]\,\mathcal{I}_2 + \Big[-\frac{2(3d^3-20d^2+27d+10)}{(d-5)(d-2)}g^6
    \nn
    &\phantom{=}-\frac{2(282-217d+31d^2)}{(d-5)(d-2)}g^2\lambda^2+\frac{4(9d^4-126d^3+487d^2-322d-144)}{(d-7)(d-5)(d-2)d}\lambda^3
    \nn
    &\phantom{=}+\frac{8d^5-95d^4+210d^3+619d^2-1390d-192}{(d-7)(d-5)(d-2)d}g^4\lambda\Big]\,(\mathcal{I}_2)^2+\Big[-\frac{2}{15}(13+5d)g^4
    \nn
    &\phantom{=}+\frac{44}{3}\lambda^2-\frac{20}{3}g^2\lambda\Big]\, \mu^2\,\mathcal{I}_3+\Big[-\frac{2}{15}(30-17d+5d^2)g^6+\frac{176}{3}\lambda^3+\frac{4}{3}(11d-20)g^2\lambda^2
    \nn
    &\phantom{=}-\frac{4(360-602d-159d^2+35d^3)}{15(d-7)d}g^4\lambda\Big]\,\mathcal{I}_1\mathcal{I}_3 \,,
    \label{eq:lambda3 matching}
    \\
    \frac{h_3}{T}&=g^2-\Big[\frac{1}{3}(2d-5)g^4+4(d-3)g^2\lambda\Big]\,\mathcal{I}_2 + \Big[-\frac{2}{3}(2d-11)g^4\mu^2+8(d-5)g^2\lambda\mu^2\Big]\,\mathcal{I}_3
    \nn
    &\phantom{=}+ (d-3) \Big[\frac{ -2160 - 9274 d + 12805 d^2 - 7372 d^3 + 2191 d^4 - 310 d^5 + 16 d^6}{9(d-7)(d-5)(d-3)(d-2)d}\,g^6
    \nn
    &\phantom{=}-\frac{4 (-4 + d) (28 - d + d^2)}{3(d-5)(d-2)}g^4\lambda +\frac{16 (6 - 27 d + 45 d^2 - 13 d^3 + d^4)}{d(d-7)(d-5)(d-2)}g^2\lambda^2\Big]\,(\mathcal{I}_2)^2
    \nn
    &\phantom{=}+ \Big[\frac{2(-360 + 514 d +855 d^2 -273d^3+22d^4)}{15(d-7)d}g^6 +\frac{8}{3}(d^2-5d+1)g^4\lambda
    \nn
    &\phantom{=}+32(d-5)g^2\lambda^2\Big]\,\mathcal{I}_1\mathcal{I}_3 \,,
    \\
    \frac{g_3^{2}}{T}&=g^2-\frac{1}{3}g^4\,\mathcal{I}_2+\frac{d^4-14d^3-157d^2+1442d-2592 }{9(d-7)(d-5)(d-2)d}g^6\,(\mathcal{I}_2)^2+\frac{2}{3}g^4\mu^2\mathcal{I}_3
    \nn
    &\phantom{=} +\frac{2}{3}\Big[d\,g^6+4g^4\lambda\Big]\mathcal{I}_1\mathcal{I}_3\Big\} \,,
    \\
    \frac{\kappa_3}{T}&=-\frac{1}{6}(d-3)(d-1)g^4\,\mathcal{I}_2 - \Big[\frac{1}{18}(d-7)(d-3)(2d-5)g^6 - 2(d-3)^2g^4\lambda\Big]\,(\mathcal{I}_2)^2
    \nn
    &\phantom{=}+\frac{1}{3}(d-5)(d-3)g^4\mu^2\,\mathcal{I}_3+(d-5)\Big[\frac{1}{9}(d^2+d-8)g^6+\frac{4}{3}(d-3)g^4\lambda\Big]\,\mathcal{I}_1\mathcal{I}_3 \,,
\end{align}
while the physical dimension-six effective parameters are
\begin{eqnarray}
\label{matching:B0^6}
    \alphaFR_{B_0^6} &=&\frac{1}{45}(d-5)(d-3)(d-1)g^{6}\mathcal{I}_{3}^{ }\,T^{2}
    \;, \\[2mm] 
    \alphaFR_{\phi^2B_0^4} &=&\frac{1}{9}\Bigl[(d-5)(d-1)g^{6}+12(d-5)(d-3)g^{4}\lambda\Bigr]\mathcal{I}_{3}^{ }\,T^{2}
    \;, \\[2mm] 
    \alphaFR_{\phi^4B_0^2} &=&\frac{2}{15}\Bigl[(36d-139)g^{6}+10(35-3d)g^{4}\lambda+10(13d-67)g^{2}\lambda^2\Bigr]\mathcal{I}_{3}^{ }\,T^{2}
    \;, \\[2mm] 
    \alphaFR_{\phi^6} &=&\frac{4}{15}\Bigl[5d g^{6}-(5d+3)g^{4}\lambda+75g^{2}\lambda^2+100\lambda^3\Bigr]\mathcal{I}_{3}^{ }\,T^{2}
    \;, \\[2mm] 
    \alphaFR_{D^2\phi^2B_0^2} &=&\frac{1}{3}(d-4)\Bigl[7g^{4}-4g^{2}\,\lambda\Bigr]\mathcal{I}_{3}^{ }\,T^{}
    \;, \\[2mm] 
    \alphaFR_{B_0^2F^2} &=&-\frac{1}{6}(d-5)g^{4}\,\mathcal{I}_{3}^{ }\,T^{}
    \;, \\[2mm] 
    \alphaFR_{\phi^2F^2} &=&\frac{1}{6}\Bigl[7g^{4}-4g^{2}\,\lambda\Bigr]\mathcal{I}_{3}^{ }\,T^{}
    \;, \\[2mm] 
\label{matching:D^2phi^4}
    \alphaFR_{D^2\phi^4} &=&\frac{2}{15}\Bigl[(66-5d)g^{4}+200g^{2}\,\lambda-20\lambda^2\Bigr]\mathcal{I}_{3}^{ }\,T^{}
    \;.
\end{eqnarray}
These results are consistent with (A.24)--(A.29) in~\cite{Hirvonen:2021zej}.
The remaining
residual gauge dependence in the scalar mass,
cancels with the gauge-dependent contribution coming from
the three-loop matching contributions. 

The bare fields and parameters of the super-renormalizable part of the 3d Lagrangian are
related to the renormalized ones by
\begin{align}
    \phi^{\rmii{$(B)$}}_3&=Z_{\phi_3}^{1/2}\,\phi_3
    \,,&
    B_{i}^{\rmii{$(B)$}}&=Z_{B_i}^{1/2}\, B_{i}
    \,,&
    B_{0}^{\rmii{$(B)$}}&=Z_{B_0}^{1/2}\, B_{0}
    \,,\nn
    \mu^{2\rmii{$(B)$}}_3&=\mu^2_3+\delta\mu^2_3
    \,,&
    \mD^{2\rmii{$(B)$}}&=\mD^2+\delta\mD^2
    \,,&
    g^{2\rmii{$(B)$}}_3&=Z_{g_3^2}\,(\Lambda_\rmi{3d})^{2\epsilon}\,g^2_3
    \,,\nn
    \lambda^{\rmii{$(B)$}}_3&=Z_{\lambda_3}\,(\Lambda_\rmi{3d})^{2\epsilon}\,\lambda_3
    \,,&
    h^{\rmii{$(B)$}}_3&=Z_{h_3}\,(\Lambda_\rmi{3d})^{2\epsilon}\,h_3
    \,,&
    \kappa^{\rmii{$(B)$}}_3&=Z_{\kappa_3}\,(\Lambda_\rmi{3d})^{2\epsilon}\,\kappa_3
    \,.
\end{align}
At super-renormalizable level,
the expression for the counterterms
is exact at the two-loop level.
We verify that the contribution to
the counterterms 
from
higher-dimensional operators starts only at $\mathcal{O}(g^{8})$~\cite{Chala:2025cya}.
The expressions for the 3d counterterms 
are~\cite{Hirvonen:2021zej,Farakos:1994kx,Laine:1995np}
\begin{eqnarray}
  Z_{g_3^2}=Z_{\lambda_3}=Z_{h_3}=Z_{\kappa_3}&=&1
  \,,\\
  \delta\mu_3^2&=&
    \frac{1}{(4\pi)^2}\frac{1}{\epsilon}\Bigl[g^4_3+\frac{1}{2}h_3^2-2g_3^2\,\lambda_3^{ }+2\lambda^2_3\Bigr]
  \,,\\
  \delta\mD^2&=&
    -\frac{1}{(4\pi)^2}\frac{1}{\epsilon}\Bigl[g^2_3h_3^{ }-h_3^2-24\kappa^2_3\Bigr]
  \,.
\end{eqnarray}

Following the procedure in~\cite{Ghisoiu:2015uza},
we substitute the matching relations
eqs.~\eqref{matching:B0^6}--\eqref{matching:D^2phi^4}
into the expressions for the 3d couplings: 
\begin{align}
  \label{eq:mu32 3d CT}
  \delta\mu_3^2&=
  \frac{T^2}{(4\pi)^2}\frac{1}{\epsilon}\left[\frac{3}{2}g^2-4g^2\lambda+4\lambda^2\right]\left(\frac{\Lambda}{\Lamd}\right)^{4\epsilon}
  \nn &
  + \frac{T^2}{(4\pi)^2}\frac{1}{\epsilon}\left[\left(1+\frac{4}{3}d\right)g^6+\left(\frac{2}{3}-8d\right)g^4\lambda+44g^2\lambda^2-40\lambda^3\right]\left(\frac{\Lambda}{\Lamd}\right)^{4\epsilon}\Lambda^{2\epsilon}\,\mathcal{I}_2
  \nn &
  - \frac{T^2}{(4\pi)^4}\frac{1}{\epsilon^2}\left[5g^6-\frac{70}{3}g^4\lambda+44g^2\lambda^2-40\lambda^3\right]\left(\frac{\Lambda}{\Lamd}\right)^{4\epsilon}+\mathcal{O}(g^8)
  \,,
  \\
  \label{eq:mD2 3d CT}
  \delta\mD^2&=-\frac{T^2}{(4\pi)^2}\frac{1}{\epsilon}(d-3) g^2\left(g^2+6\lambda\right)\left(\frac{\Lambda}{\Lamd}\right)^{4\epsilon}\Lambda^{2\epsilon}\mathcal{I}_2+\mathcal{O}(g^8)
  \,.
\end{align}
Upon expanding about $d = 3 - 2\epsilon$,
the double pole $1/\epsilon^{2}$ of $\delta\mu_{3}^{2}$ in eq.~\eqref{eq:mu32 3d CT} cancels.
The single pole removes the remaining divergence in $\mu_3^2$ after 4d renormalization,
rendering the effective scalar mass finite.

As one of the main results of this work,
we now present the computation of
the three-loop contributions to the scalar and Debye masses.
The final renormalized expressions
are given in
eqs.~\eqref{eq:scalar mass 3 loop} and~\eqref{eq:debye mass 3 loop}.

\subsection{Three-loop thermal masses in the Abelian Higgs model}
\label{app:thermal:masses:U1}

The three-loop two-point correlators for the temporal and Lorentz scalars
contribute to the corresponding
Debye mass~\eqref{eq:Debye:mass:def} and
scalar thermal mass~\eqref{eq:scalar:mass:def}.
In the initial form of this IBP-reduced result, some of our expressions exhibit $1/(d-3)^2$ singularities. These divergences imply that the corresponding master integrals
would be required up to $\mathcal{O}(\epsilon^2)$.
A possible way to circumvent this is to perform a basis transformation within
the IBP reduction.
The idea is to exploit additional IBP relations that modify the coefficients of the master sum-integrals in such a way that the explicit $(d-3)$ factors in the denominators are removed.
In practice, this corresponds to shifting from the basketball-type basis to one involving spectacle diagrams.
Although these may be somewhat more cumbersome to evaluate individually,
such a transformation can simplify the overall structure of
the reduction and avoid the need for higher-order $\epsilon$-expansions.
This procedure closely mirrors the treatment in appendix~D of~\cite{Ghisoiu:2015uza}.

To this end, we list the corresponding IBP basis transformation that we obtained with
an in-house {\tt FORM}~\cite{Davies:2026cci}
Laporta-type algorithm~\cite{Laporta:2000dsw} for
finite-temperature sum-integrals~\cite{Nishimura:2012ee} (cf.\cite{Ghisoiu:2015uza} for a crosscheck):
\begin{align}
  \mathcal{I}^{ }_{111110} &=
  \frac{2}{3(d-3)^2}\Bigl[
    \frac{3d^2 -24d + 47}{(d-4)}
    \mathcal{I}_{210011}^{ }
  + 8\, \mathcal{I}_{310011}^{200}
  \Bigr]
  \,,
  \\[2mm]
  \mathcal{I}^{020}_{211110} &=
    \frac{(d-9)(d-7)(d-2)}{2(d-6)(d-5)(d-4)(d-3)^2}\,
    (\mathcal{I}_{1}^{ })^2\mathcal{I}_{3}^{ }
    +\frac{519-312d+61d^2-4d^3}{2(d-6)(d-5)^2(d-4)(d-3)}\,
    (\mathcal{I}_{2}^{ })^2\mathcal{I}_{1}^{ }
    \nn&-
    \frac{(3d-10)(10791-9060d+2806d^2-380d^3+19d^4)}
      {12(d-6)(d-5)(d-4)^2(d-3)^2}\,
      \mathcal{I}_{210011}^{ }
    +\nn&+
      \frac{3(d-7)}{2(d-6)(d-4)(d-3)}\,
      \mathcal{I}_{220011}^{002}
     -\frac{(d-9)(d-7)}{2(d-6)(d-5)(d-4)(d-3)}\,
      \mathcal{I}_{310011}^{020}
    +\nn&+
    \frac{31401-16707d+2951d^2-173d^3}{6(d-6)(d-5)(d-4)(d-3)^2}\,
      \mathcal{I}_{310011}^{200}
    +\frac{512}{(d-5)(d-4)(d-3)^2}\,
      \mathcal{I}_{510011}^{600}
  \,,
  \\[2mm]
  \mathcal{I}^{ }_{31111-2} &=
  - \frac{60 d^{6} - 1381 d^{5} + 12352 d^{4} - 52890 d^{3} + 103142 d^{2} - 49577 d - 60810}
       {3(d-6)(d-5)(d-4)(d-3)^{2}(d-2)(d-1)}
  \,\mathcal{I}_{210011}^{ }
  \nn[2mm]
  &\quad
  + \frac{2(7 d^{2} - 58 d + 75)}
         {(d-6)(d-3)(d-2)(d-1)}
  \,\mathcal{I}_{220011}^{002}
  \nn[2mm]
  &\quad
  - \frac{2\bigl(8 d^{6} - 207 d^{5} + 2195 d^{4} - 12246 d^{3} + 38222 d^{2} - 64347 d + 46743\bigr)}
         {(d-7)(d-6)(d-5)^{2}(d-3)(d-2)(d-1)}
  \,\mathcal{I}_{221000}^{ }
  \nn[2mm]
  &\quad
  - \frac{2(d-9)(d-7)(d+1)}
         {(d-6)(d-5)(d-3)(d-2)(d-1)}
  \,\mathcal{I}_{310011}^{020}
  \nn[2mm]
  &\quad
  - \frac{2\bigl(185 d^{4} - 3002 d^{3} + 15280 d^{2} - 19126 d - 26841\bigr)}
         {3(d-6)(d-5)(d-3)^{2}(d-2)(d-1)}
  \,\mathcal{I}_{310011}^{200}
  \nn[2mm]
  &\quad
  \hspace{-1.3cm}
  - \frac{2\bigl(6 d^{7} - 166 d^{6} + 1971 d^{5} - 13028 d^{4} + 51730 d^{3} - 123276 d^{2} + 163349 d - 92682\bigr)}
         {(d-7)(d-6)(d-5)(d-3)^{2}(d-2)(d-1)}
  \,\mathcal{I}_{311000}^{ }
  \nn[2mm]
  &\quad
  + \frac{16 (d-5)(d-4)}{(d-2)(d-1)}\,
  \mathcal{I}_{311110}^{022}
  + \frac{2048 (d+1)}{(d-5)(d-3)^{2}(d-2)(d-1)}\,
  \mathcal{I}_{510011}^{600}
  \,.
\end{align}
After applying these basis transformations,
the bare correlators are compactly written as follows
\begin{align}
\label{eq:Pi:B0B0}
  \Pi_{B_0 B_0}^{3\ell,\rmii{$(B)$}}&=
    2\,g^{6}
    \biggl[
      (d-5)d^2(\mathcal{I}_{1})^{2} \mathcal{I}_{3}
    + \frac{(d-3)(3d^3 - 27d^2 + 52d + 16)}{(d-5)(d-2)}\mathcal{I}_{1} (\mathcal{I}_{2})^{2}
  \nn & \qquad
    + 2(d-3)\biggl(
        2(d-1)\mathcal{I}_{210011}
      - (d-3)(d-1)\mathcal{I}^{ }_{111110}
      - 8(d-4)\mathcal{I}^{020}_{211110}
      \biggr)
    \biggr]
  \nn &
    + 8\,g^{4} \lambda
    \biggl[
        2d(d-5)\,(\mathcal{I}_{1})^{2} \mathcal{I}_{3}
      + \frac{(d+2)(d-3)^{2}}{d-2} \mathcal{I}_{1} (\mathcal{I}_{2})^{2}
      + 2(d-3)(3d-7)\,\mathcal{I}^{ }_{111110}
    \biggr]
  \nn &
    + 32(d-3)\,g^{2} \lambda^{2}
    \biggl[
        \frac{d-5}{d-3}\,(\mathcal{I}_{1})^{2} \mathcal{I}_{3}
      + \mathcal{I}_{1} (\mathcal{I}_{2})^{2}
      + \mathcal{I}^{ }_{210011}
      - \frac{d-3}{2}\,\mathcal{I}^{ }_{111110}
    \biggr]
  \,,
  \\[3mm]
\label{eq:Pi:phiphi}
   \Pi_{\phi^\dagger \phi}^{3\ell,\rmii{$(B)$}} &=
    g^{6}
    \biggl[
        \bigl(
          4(d-5)
        - d^2\xiNew
        \bigr)
        (\mathcal{I}^{ }_{1})^{2}
        \mathcal{I}^{ }_{3}
  \nn & \qquad
    + \biggl(
          \frac{4d^4 -29d^3 + 43d^2 - 24d + 32}{2(d-5)(d-2)}
        + \frac{d \xiNew^2}{2}
        - 2(d-1)\xiNew
      \biggr)\mathcal{I}^{ }_{1}
      (\mathcal{I}^{ }_{2})^{2}
  \nn & \qquad
      + 6(d-1)\mathcal{I}^{ }_{210011}
      + 4(d-2)\mathcal{I}^{ }_{111110}
      - 8(d-4)\mathcal{I}^{020}_{211110}
      + 4\mathcal{I}^{ }_{31111-2}
    \biggr]
  \nn[2mm] &
    + 4 g^{4} \lambda
    \biggl[
        d(d-2\xiNew)(\mathcal{I}^{ }_{1})^{2}
        \mathcal{I}^{ }_{3}
      + \biggl(
          \frac{10d^3 - 75d^2 + 139d - 8}{2(d-5)(d-2)}
          - \frac{(2d-\xiNew)\xiNew}{2}
        \biggr)
        \mathcal{I}^{ }_{1} (\mathcal{I}^{ }_{2})^{2}
  \nn & \qquad
      + 2(d-1)\mathcal{I}^{ }_{210011}
      + 4(d-1)\mathcal{I}^{ }_{111110}
      - 8(d-4)\mathcal{I}^{020}_{211110}
    \biggr]
  \nn[2mm] &
    + 8\,g^{2} \lambda^{2}
    \biggl[
      2 (2 d - \xiNew)\,
      (\mathcal{I}^{ }_{1})^{2}
      \mathcal{I}^{ }_{3}
    + \mathcal{I}^{ }_{1}
    (\mathcal{I}^{ }_{2})^{2}\biggl(
        \frac{2 d^3-12 d^2+3 d+36}{(d-5) (d-2)}
      - 2\xiNew
      \biggr)
  \nn & \qquad
    + \mathcal{I}^{ }_{210011}
    - 7\,\mathcal{I}^{ }_{111110}
    \biggr]
  \nn &
    + 32\lambda^{3}
    \biggl[
        2\,(\mathcal{I}^{ }_{1})^{2} \mathcal{I}^{ }_{3}
      + \frac{2 d^{2} - 14 d + 17}{(d-2)(d-5)}
        \mathcal{I}^{ }_{1} (\mathcal{I}^{ }_{2})^{2}
      + \mathcal{I}^{ }_{210011}
      + \frac{5}{2}\,\mathcal{I}^{ }_{111110}
    \biggr]
  \,.
\end{align}
The remaining master integrals of mass-dimension $\mbox{dim} = 2$ in
eqs.~\eqref{eq:Pi:B0B0} and~\eqref{eq:Pi:phiphi} are
\begin{align}
\label{eq:masters:3loop}
    \mbox{dim}=2\,, & &&
    \mathcal{I}^{ }_{111110}\,,\;
    \mathcal{I}^{020}_{211110}\,,\;
    \mathcal{I}^{000}_{31111-2}\,,\;
    \mathcal{I}^{000}_{210011}\,,
\end{align}
and require case-by-case treatment,
as done in appendix~\ref{app:masters}.

\subsection{Three-loop thermal masses in the ${\rm SU}(N)$ + fundamental scalar model}
\label{app:thermal:masses:SU2}

Our computations are straightforwardly generalized
to obtain the three-loop contributions to thermal masses in
a ${\rm SU}(N)$ + fundamental scalar theory.
The model in 4d Euclidean spacetime is
\begin{equation}
    \mathcal{L} = \frac{1}{4} F_{\mu\nu}^a F_{\mu\nu}^a 
    + (D_\mu \phi)^\dagger (D_\mu \phi)
    + \mu^2 \phi^\dagger \phi
    + \lambda (\phi^\dagger \phi)^2
    + \mathcal{L}_\rmi{gh}
    + \mathcal{L}_\rmii{GF}
  \,,
\end{equation}
where $\phi$ is a scalar in the fundamental representation of ${\rm SU}(N)$,
$F_{\mu\nu}^a =
    \partial_\mu A_\nu^a
  - \partial_\nu A_\mu^a
  + g f^{a b c} A_\mu^b A_\nu^c$
the field-strength tensor,
and
$D_\mu \phi =
    \partial_\mu \phi
  - i g\, T^a\!A_\mu^a \phi$
the covariant derivative
where $g$ is the gauge coupling.
$f^{a b c}$ are the structure constants, and
$[T^a]_{ij}$ are the generators of the algebra.
Finally,
$\mathcal{L}_\rmi{gh}$ 
is the ghost, and 
$\mathcal{L}_\rmii{GF}$ the $R_\xi$ gauge-fixing Lagrangian,
which we do not display here.
The latter is similar to the U(1) case~\eqref{eq:gf:U1},
but in the background field gauge receives further contributions~\cite{Abbott:1980hw,Laine:2005ai}.
The corresponding soft Lagrangian is similar to ${\rm U}(1)$ case~\eqref{eq:soft lag},
with additional group invariants appearing for general $N$~\cite{Laine:2005ai}.

The dimensional reduction of subsectors of this model
up to $\mathcal{O}(g^4)$ has been presented
in~\cite{Gynther:2005dj,Gynther:2005av,Ghisoiu:2013zoj,Ghisoiu:2015uza} and
in {\tt DRalgo}~\cite{Ekstedt:2022bff}.
The relevant RG equations are also given in~\cite{Born:2024mgz,Ekstedt:2022bff},
specialized to ${\rm SU}(2)$.
Here, we focus on the three-loop two-point correlators for
the temporal and Lorentz scalars, which contribute to the corresponding
Debye mass~\eqref{eq:Debye:mass:def} and
scalar thermal mass~\eqref{eq:scalar:mass:def}.
In the following, the results are given for general ${\rm SU}(N)$;
by setting $N=2$, one recovers the results for the ${\rm SU}(2)$ + Higgs sector of the SM.
Omitting now the $(B)$ superscript to avoid the clutter of notation,
the corresponding scalar and Debye masses at one-loop level are given by the bare correlators
\begin{align}
  \Pi_{A_0 A_0}^{1\ell} &=
    g^{2}(d-1)\Bigl(
      (d-1)\CA
      + 1
    \Bigr) \mathcal{I}_{1}
  \,,
  \\[3mm]
  \Pi_{A_0 A_0}^{1\ell,\rmii{(1)}} &=
    - g^2 \biggl[
        \biggl(
          \frac{d^2+d+10}{6} - (d-3)\xi
        \biggr)\CA
      + \frac{d-4}{6}
    \biggr] \mathcal{I}_{2}
  \,,
  \\[3mm]
  \Pi_{A_0 A_0}^{1\ell,\rmii{(2)}} &=
      g^2 \biggl[
        \biggl(
            \frac{2d^2+11d+2}{60}
          - \frac{d-4}{2}\xi
          + \frac{d-6}{12}\xi^2
        \biggr)\CA
      + \frac{d-6}{30}
    \biggr] \mathcal{I}_{3}
  \,,
  \\[3mm]
  \Pi_{\phi^\dagger \phi}^{1\ell} &=
    \biggl[
        dg^{2}\CF
      + 2\lambda(\CA + 1)
    \biggr]
    \mathcal{I}_{1}
  \,,
  \\[3mm]
  \Pi_{\phi^\dagger \phi}^{1\ell,\rmii{(1)}} &=
    - g^{2} \CF \bigl(
      3 - \xi
    \bigr) \mathcal{I}_{2}
  \,,
  \\[3mm]
  \Pi_{\phi^\dagger \phi}^{1\ell,\rmii{(2)}} &=
    g^{2} \CF \frac{5 - 3\xi}{3} \mathcal{I}_{3}
  \,.
\end{align}
At two-loop level, the bare correlators read
\begin{align}
  \Pi_{A_0 A_0}^{2\ell} &=
      g^4 (d-3)\biggl[
          \CA^2 (d-1)^2 (\xi-2)
        + \frac{\CA - 2\CF}{2} \bigl(d + \CA^2 (4 - 5 d + 2 (d-1) \xi)\bigr)
      \biggr]
      \mathcal{I}_{1} \mathcal{I}_{2}
    \nn[2mm] &
    - 2g^2 \lambda
      (d-3)
      (\CA + 1) 
      \mathcal{I}_{1} \mathcal{I}_{2}
  \,,
  \\[3mm]
  \Pi_{A_0 A_0}^{2\ell , \rmii{(1)}} &=
    g^4\biggl[
        \frac{(d-3)}{(d-7) (d-5) (d-2) d}
        \biggl(
            \CA^{2} \frac{4 d^6 - 67 d^5 + 385 d^4 - 1033 d^3 + 1459 d^2 - 700 d + 672}{24}
        \nn[2mm] &\hphantom{{}=g^4\biggl[+\biggl(}
          + \CA \frac{d^5 - 18 d^4 + 106 d^3 - 234 d^2 + 169 d -96}{6}
        \nn[2mm] &\hphantom{{}=g^4\biggl[+\biggl(}
          +(\CA - 2\CF)(d-4) (d^2 -4 d -3)
        \biggr)
        \mathcal{I}_{2}^2
      \nn[2mm] &\hphantom{{}=g^4\biggl[+\biggl(}
      - \CA(d-3)\biggl(
          \biggl(
              \frac{d-4}{6}
            + \CA^{ } \frac{2 d^3 - 5 d^2 + 19 d -40}{12 (d-2)}
        \biggr) \xi
        \nn[2mm] &\hphantom{{}=g^4\biggl[+\biggl(-\CA(d-3)\biggl(}
        - \CA^{ } \frac{3 d^2 - 13 d + 16}{8 (d-2)} \xi^2
      \biggr)\, \mathcal{I}_{2}^2
    \nn[2mm] &\hphantom{{}=g^4\biggl[}
      + \biggl(
          \CA^{2} \frac{(d-1)^2 (5 d^3 - 53 d^2 + 126 d -144)}{6 (d-7) d}
      \nn[2mm] &\hphantom{{}=g^4\biggl[+\biggl(}
        + \CA^{ } \frac{6 d^4 - 71 d^3 + 221 d^2 - 270 d + 144}{6 (d-7) d}
        - \frac{(d-6) d}{6} (\CA - 2\CF)
      \nn[2mm] &\hphantom{{}=g^4\biggl[+\biggl(}
        - \frac{\CA}{3} \bigl(\CA(d-1) + 1\bigr)(d-1)\Bigl(
            2(d-3)\xi
          - \frac{d-6}{2}\xi^2
        \Bigr)
      \biggr) \mathcal{I}_{1} \mathcal{I}_{3}
    \biggr]
    \nn[2mm] &
    + \frac{2}{3} g^2 \lambda
      (d-6)
      (\CA + 1) 
      \mathcal{I}_{1} \mathcal{I}_{3}
  \,,
  \\[3mm]
  \Pi_{\phi^\dagger \phi}^{2\ell} &=
      g^4 \CF
        \Bigl(
          -(d-1)(\CA (d-1) + 1)
          + \CF d\xi
        \Bigr) 
      \mathcal{I}_{1} \mathcal{I}_{2}
    \nn[2mm] &
    - 2g^2 \lambda
        \CF\bigl(\CA + 1\bigr)
        \bigl(d - \xi\bigr)
      \mathcal{I}_{1} \mathcal{I}_{2}
    - 4\lambda^2
      (\CA + 1)^2
      \mathcal{I}_{1} \mathcal{I}_{2}
  \,,
  \\[3mm]
  \Pi_{\phi^\dagger \phi}^{2\ell,\rmii{(1)}} &=
      g^4\CF\biggl[
        \frac{\CA - 2\CF}{(d-7) (d-5) (d-2) d}
        \biggl(
            \frac{\CA^2 (19d^4 - 238 d^3 + 995 d^2 - 1436 d -384)}{8}
        \nn[2mm] &
        \hphantom{{}=g^4\CF\biggl[}
          + 2 \CA (d^2 - 7d + 15)
          - \frac{11 d^4 - 142 d^3 + 575 d^2 - 696 d + 72}{4}
        \biggr) \mathcal{I}_{2}^2
        \nn[2mm] &
        \hphantom{{}=g^4\CF\biggl[}
        +\frac{\CA-2\CF}{4}\biggl(
          6 - \frac{\CA^2 (78 + 7 (d-7) d)}{(d-5) (d-2)}
        \biggr)\xi
        \mathcal{I}_{2}^2
        \nn[2mm] &
        \hphantom{{}=g^4\CF\biggl[}
        -\frac{\CA-2\CF}{8}\biggl(
          2 - \frac{\CA^2 (28 + 3 (d-7) d)}{(d-5) (d-2)}
        \biggr)\xi^2
        \mathcal{I}_{2}^2
      \nn[2mm] &
      \hphantom{{}=g^4\CF\biggl[}
      +\biggl(
          \frac{4(d-6)(d-1)}{(d-7)d}(\CA(d-1) + 1)
        + \frac{2d}{3}(5-3\xi)\CF
      \biggr)
      \mathcal{I}_{1} \mathcal{I}_{3}
    \biggr]    
    \nn[2mm] &
    + \frac{4}{3}g^2 \lambda 
        \CF(\CA + 1)(5-3\xi)
      \mathcal{I}_{1} \mathcal{I}_{3}
    + 12\lambda^2 (\CA + 1) 
        \frac{(d-3)(d-4)}{(d-7)(d-5)(d-2)d}
      \mathcal{I}_{2}^2
  \,.
\end{align}
The Debye mass is computed in
the background field gauge~\cite{Abbott:1981ke}.
We have crosschecked that the pure gauge contributions
agree with~\cite{Ghisoiu:2015uza} when adapting our gauge parameter convention via 
$\xi \to 1 - \xi_{\!\!\text{\cite{Ghisoiu:2015uza}}}$.

The three-loop part of two-point bare correlators for
the temporal and Lorentz scalars contribute to
the corresponding Debye mass~\eqref{eq:Debye:mass:def} and scalar thermal mass~\eqref{eq:scalar:mass:def}.
They are given by
\begin{align}
\label{eq:Pi:B0B0:SUN}
  \Pi_{A_0 A_0}^{3\ell} &=
    g^{6}
    \biggl[
      \biggl(
          \frac{(d-1)^2 (d^3+40 d^2-347 d+594)}{12} \CA^{3}
        + \frac{2 d^4 + 81 d^3 - 789 d^2 + 1882 d - 1188}{12} \CA^{2}
      \nn & \qquad\quad
        + \frac{d^3 + 40 d^2 - 347 d + 594}{12} \CA^{} 
        - \frac{(2\CA^2-1) (d-5) d^2}{4\CA^2}
      \nn & \qquad\quad
        + \frac{(d-1)^2}{12} \CA (\CA (d-1) + 1)^2 ((d-6)\xi - 2d)\xi
      \biggr)
      (\mathcal{I}_{1})^{2} \mathcal{I}_{3}
    \nn & \qquad
    + (d-3)\biggl(
        \frac{(d-1)^2 (19 d^3-180 d^2+421 d-60)}{8 (d-5) (d-2)} \CA^{3}
      \nn & \qquad\quad
      + \frac{31 d^4-301 d^3+827 d^2-589 d+72}{8 (d-5) (d-2)} \CA^{2}
      + \frac{d^3 - 7 d^2 + 9 d - 3}{2 (d - 5) (d - 2)} \CA
      \nn & \qquad\quad
      - \frac{3 d^4 - 26 d^3 + 54 d^2 - 3 d - 8}{2 (d - 5) (d - 2)}
      - \frac{d^3 - 8 d^2 + 11 d + 8}{2 (d - 5) (d - 2) \CA}
      + \frac{(d - 6) d}{4 (d - 2)\CA^{2}}
      \nn & \qquad\quad
      + \frac{(d-1)^2}{d-2}\biggl(
          \frac{3 d^2-13 d+16}{8}\xi^2
        - \frac{7 d^2 - 39 d + 48}{4}\xi
      \biggr)\CA^3
      \nn & \qquad\quad
      + \frac{1}{d-2}\biggl(
          \frac{3 d^3 - 16 d^2 + 29 d - 16}{8}\xi^2
        - \frac{9 d^3-56 d^2+99 d-48}{4}\xi
      \biggr)\CA^2
      \nn & \qquad\quad
      + \frac{(d - 3) d}{2}\xi
    \biggr)\mathcal{I}_{1} (\mathcal{I}_{2})^{2}
    \nn & \qquad
    - (d-3)\biggl(
        3 (d-1) \CA^{2}
      + \frac{3}{2} \CA^{ } 
      + \frac{4 d - 1}{2}
      - \frac{3}{2} \frac{1}{\CA^{ }}
      - \frac{2 d + 1}{2\CA^{2}}
      \biggr)\mathcal{I}_{210011}
    \nn & \qquad
    - (d-3)\biggl(
        \frac{(d - 1)^2 (7 d - 13)}{4} \CA^{3}
      + \frac{(d-1)(d+3)}{2} \CA^{2}
      + \frac{d + 3}{4} \CA^{ }
    \nn & \qquad\quad
      - \frac{5(d^2 - 3 d + 1)}{2}
      - \frac{3 d - 5}{4\CA^{ }}
      + \frac{2 d^2 - 11 d + 11}{4\CA^{2}}
    \biggr)\mathcal{I}^{ }_{111110}
    \nn & \qquad
      - 8 (d - 4) (d - 3) (\CA(d-1) + 1) (\CA^{2}(d-1) + \CF)\mathcal{I}^{020}_{211110}
    \nn[2mm] & \qquad
      - \frac{(d - 7) (d - 3)}{2}\CA^{ }(\CA(d-1) + 1)^2\mathcal{I}^{ }_{31111-2}
    \biggr]
  \nn &
    + (\CA + 1)(\CA - 2\CF)\, g^{4} \lambda
    \biggl[
        2d(d-5)
        \dA
        \,(\mathcal{I}_{1})^{2} \mathcal{I}_{3}
      \nn & \qquad
      + \biggl(
        \frac{(d-3)^2 (\CA^2 (5d-2) - (d+2))}{d-2}
        - 2 \CA^{2} (d-3)^2 \xi
      \biggr)
        \mathcal{I}_{1} (\mathcal{I}_{2})^{2}
      \nn & \qquad
      + (d-3)(3d-7)
        \frac{\CA^2 + 2(\CA - 1)}{\CA + 1}
      \,\mathcal{I}^{ }_{111110}
    \biggr]
  \nn &
    + 4(d-3)\,g^{2} \lambda^{2}
      (\CA + 1)
    \biggl[
        (\CA + 1)\frac{d-5}{d-3}\,(\mathcal{I}_{1})^{2} \mathcal{I}_{3}
      + (\CA + 1)\mathcal{I}_{1} (\mathcal{I}_{2})^{2}
    \nn & \qquad
      + 2\mathcal{I}^{ }_{210011}
      - (d-3)\,\mathcal{I}^{ }_{111110}
    \biggr]
  \,,
  \\[3mm]
\label{eq:Pi:phiphi:SUN}
  \Pi_{\phi^\dagger \phi}^{3\ell} &=
    g^{6}\CF
    \biggl[
        \bigl(
          (\CA (d-1) + 1)^2 (d-5)
        - d^2\CF^{2}\xi
        \bigr)
        (\mathcal{I}^{ }_{1})^{2}
        \mathcal{I}^{ }_{3}
    \nn & \qquad
    + \biggl(
          \frac{d (3 d^2 - 25 d + 36)}{8 \CA^2 (d-5) (d-2)}
        - \frac{9 d^3 - 123 d^2 + 252 d - 64}{16 (d-5) (d-2)}
      \nn & \qquad\quad
        - \frac{d^4 - 8 d^3 + 17 d^2 - 15 d + 12}{2 \CA (d-5) (d-2)}
        + \frac{\CA (3 d^4 - 26 d^3 + 63 d^2 - 37 d + 4)}{2 (d-5) (d-2)}
      \nn & \qquad\quad
        + \frac{\CA^2 d (16 d^4-160 d^3+515 d^2-585 d+228)}{16 (d-5) (d-2)}
      \nn & \qquad\quad
        - \CF \biggl(
          (d-1)
          + \CA\frac{4 d^4 - 35 d^3 + 93 d^2 - 90 d + 40}{4(d-5) (d-2)}
        \biggr)\xi
      \nn & \qquad\quad
        - d\CF \biggl(
            \frac{1}{4\CA}
          - \CA\frac{3 d^2 - 21 d + 28}{8 (d-5) (d-2)}
        \biggr)\xi^2
      \biggr)\mathcal{I}^{ }_{1}
      (\mathcal{I}^{ }_{2})^{2}
    \nn & \qquad
      - \biggl(
          \frac{d - 1}{2}\CA^{2}
        + \frac{4d-1}{4} \CA^{ }
        + \frac{4d-7}{4\CA}
        - \frac{2d+1}{4\CA^{2}}
        + \frac{3}{4}
      \biggr)
      \mathcal{I}^{ }_{210011}
    \nn & \qquad
      - \biggl(
          \frac{2 d^2 - 11 d + 1}{2} \CA^2
        + \frac{6 d + 5}{4} \CA
        + \frac{7 - 4d}{4\CA^{2}}
        - 5d
        + 4
      \biggr)\mathcal{I}^{ }_{111110}
    \nn & \qquad
      - 2 (d-4) (\CA - 2\CF) (\CA^2 (2d - 1) - 1) (\CA (d-1) + 1)
        \mathcal{I}^{020}_{211110}
    \nn & \qquad
      + (\CA(d-1) + 1)^2
        \mathcal{I}^{ }_{31111-2}
    \biggr]
  \nn[2mm] &
    + g^{4} \lambda \,
    \dA (\CA + 1)
    \biggl[
        2 d(d-2\xi)
        \frac{\CF^{2}}{\dA}
        (\mathcal{I}^{ }_{1})^{2}
        \mathcal{I}^{ }_{3}
      \nn & \qquad
      + \biggl(
            \frac{8 d^4 - 68 d^3 + 159 d^2 - 89 d + 52}{8 (d-5) (d-2)}
        \nn & \qquad \quad
          + \frac{2 d^3 - 16 d^2 + 32 d - 11}{\CA (d-5) (d-2)}
          - \frac{4 d^3 - 22 d^2 + 22 d + 72}{8\CA^2 (d-5) (d-2)}
        \nn & \qquad \quad
          + \biggl(
            \frac{2d}{\CA^2}
            - \frac{2 d^3 - 13 d^2 + 13 d + 18}{(d-5) (d-2)}
          \biggr)\frac{\xi}{4}
          - \biggl(
            \frac{1}{\CA^2}
            - \frac{3 d^2 - 21 d + 28}{2 (d-5) (d-2)}
          \biggr)\frac{\xi^2}{4}
        \biggr)
        \mathcal{I}^{ }_{1} (\mathcal{I}^{ }_{2})^{2}
      \nn & \qquad
      - \frac{2d + 1 + 3\CA +2 \CA^2 (d-1)}{2\CA^{2}}\mathcal{I}^{ }_{210011}
      \nn & \qquad
      - \frac{\CA-2\CF}{2\CA (\CA + 1)}
        \bigl(2(2d - 1) - \CA (4(d-3) + 13\CA^{} - 2 \CA^{2} (d-9))\bigr)
        \mathcal{I}^{ }_{111110}
      \nn & \qquad
      - 4(d-4)(\CA - 2\CF)(\CA(d-1) + 1)\mathcal{I}^{020}_{211110}
    \biggr]
  \nn[2mm] &
    + 2\,g^{2} \lambda^{2}
      \,
      \dA(\CA + 1)(\CA - 2\CF)
    \biggl[
      (\CA + 1) (2 d - \xi)\,
      (\mathcal{I}^{ }_{1})^{2}
      \mathcal{I}^{ }_{3}
  \nn & \qquad
    + \biggl(
        \frac{d^3 - 6 d^2 + 18 + \CA (d-3) (d^2 - 3d -6)}{(d-5) (d-2)}
      - (\CA + 1)\xi
      \biggr)
      \mathcal{I}^{ }_{1}
      (\mathcal{I}^{ }_{2})^{2}
  \nn & \qquad
    + \mathcal{I}^{ }_{210011}
    - 7\,\mathcal{I}^{ }_{111110}
    \biggr]
  \nn &
    + 8\lambda^{3}
    (\CA + 1)^{2}
    \biggl[
        (\CA + 1)^{}(\mathcal{I}^{ }_{1})^{2} \mathcal{I}^{ }_{3}
      + \Bigl(
            \CA
          + \frac{d^2 -7d + 7}{(d-5)(d-2)}
        \Bigr)
        \mathcal{I}^{ }_{1} (\mathcal{I}^{ }_{2})^{2}
  \nn & \qquad
      + \mathcal{I}^{ }_{210011}
      + \frac{\CA + 4}{\CA + 1}\mathcal{I}^{ }_{111110}
    \biggr]
  \,.
\end{align}
In the expressions above,
$\CA = N$ and
$\CF = (N^2-1)/(2N)$
are the Casimirs of the adjoint and fundamental representations of $\mathrm{SU}(N)$, respectively, and
$\dA = N^2-1$ is the dimension of the adjoint representation.

\subsection{On the mismatch in the master integral $\mathcal{I}_{31111-2}$}
\label{app:mismatch}

In our original derivation of the full $\mathcal{O}(g^6)$ dimensional reduction of
the Abelian Higgs model,
using the evaluation of $\mathcal{I}_{31111-2}$ from~\cite{Ghisoiu:2015uza},
we found that the renormalized scalar mass $\mu_3^2$ contains
a residual $1/\epsilon$ pole that is not canceled by
the corresponding 3d counterterm $\delta \mu_3^2$.
In this appendix,
we discuss our initial finding of this mismatch, and how we resolved it.

To further examine this issue,
we explore
the $\mathrm{SU}(2)$ + fundamental scalar theory,
employing the results in appendix~\ref{app:thermal:masses:SU2} for $N=2$.
In this model, the two-loop counterterm for
the 3d scalar mass is known~\cite{Kajantie:1995dw, Laine:1995np},
{\em viz.}
\begin{equation}
    \delta \mu_3^2 = -\frac{1}{4 (4\pi)^2}\frac{1}{\epsilon} \left(
        \frac{39}{16} g_3^4
      + 12 h_3^{ } g_3^2
      - 6 h_3^2
      + 9 \lambda_3^{ } g_3^2
      - 12 \lambda_3^2
    \right)
    \,,
\end{equation}
where the 3d soft scale couplings are the same as those in
eq.~\eqref{eq:soft lag}, upon replacing the fields
in $\mathrm{U}(1)$ representations with the corresponding ones in the $\mathrm{SU}(2)$ case.
Its ${\rm SU}(N)$ generalization,
\begin{align}
    \delta \mu_3^2 &=
      - \frac{1}{(4\pi)^2}\frac{1}{\epsilon} \biggl[
          \CF\frac{3\CA^2 - \CA + 3}{8\CA} g_3^4
        + (\CA + 1)(\CF\lambda_3^{ } g_3^2 - \lambda_3^2) 
        \\ &
        \hphantom{{}= - \frac{4}{(4\pi)^2 \epsilon} \biggl(}
        + \CF\CA(\CA h_3^{ } + 2 h_{3,2}^{ }) g_3^2
        - \CF\Bigl(
            \CA h_{3}^{2}
          + 4 h_{3}^{ } h_{3,2}^{ }
          - 2(\CA - 4\CF) h_{3,2}^{2}
        \Bigr)
      \biggr]
    \,,
    \nonumber
\end{align}
features a second scalar-gauge coupling, $h_{3,2}$,
that appears in the $\mathrm{SU}(N)$ case
besides $h_3$.%
\footnote{%
  Similarly, the temporal quartic couplings
  are linearly dependent only for $N \leq 3$~\cite{Laine:2005ai}.
}
To see this, we decompose the product of generators as
$T^a T^b = \frac{1}{2} \{T^a, T^b\} + \frac{1}{2} [T^a, T^b]$,
and note that the antisymmetric part
vanishes since $A_0^a A_0^b$ is symmetric under interchange of indices.
Then, by using
$\{T^a, T^b\} = \frac{1}{N} \delta^{ab} + d^{abc} T^c$,
we see that in general
the ${\rm SU}(N)$ counterpart of the
3d soft-scale Lagrangian~\eqref{eq:L:soft:U1} contains
the two distinct scalar-gauge interaction terms
\begin{align}
\mathcal{L}_\rmi{soft} &\supset
    h_{3}^{ }\, \phi^\dagger \phi A_0^a A_0^a
  + h_{3,2}^{ }\, \phi^\dagger T^c \phi\, d^{abc} A_0^a A_0^b
  \,.
\end{align}
While in general ${\rm SU}(N)$
the symmetric structure constants $d^{abc}$ are non-vanishing,
for ${\rm SU}(2)$ they identically vanish
and therefore $h_{3,2}$ does not contribute.

The mismatch in the $\mathcal{O}(\epsilon^{-1})$ coefficient of $\mathcal{I}_{31111-2}$
affects only scalar mass renormalization.
While this integral appears in
the pure gauge Debye mass in eq.~\eqref{eq:Pi:B0B0:SUN} (cf.~\cite{Ghisoiu:2015uza}),
the corresponding term is rendered finite by an explicit $(d-3)$ factor.
In contrast, for scalar mass renormalization in eq.~\eqref{eq:Pi:phiphi:SUN},
the prefactor is not lifted to $\mathcal{O}(\epsilon^0)$,
leading to the leftover pole
\begin{equation}
    \mu_3^2 \Big|_{\mathrm{SU}(2)} = -\frac{25}{24 (4 \pi)^4 \epsilon} g^6 T^2 + \mathcal{O}(\epsilon^0)\,,
\end{equation}
which is neatly canceled if $\mathcal{I}_{31111-2}$ is evaluated to eq.~\eqref{eq:I:31111-2}, the same as for the Abelian Higgs case.

In the original derivation of the three-loop Debye mass~\cite{Ghisoiu:2015uza},
errors in the $1/\epsilon$ coefficient of $\mathcal{I}_{31111-2}$
were masked as finite contributions due to the multiplicative $(d-3)$ factor
and could not be verified by renormalization.
This finding has led to a re-evaluation of this standard result (see \texttt{v2}
of~\cite{Ghisoiu:2012yk} on \texttt{arXiv}),
which has confirmed our correct determination of the divergent part.
This introduces a new finite contribution to
the three-loop Debye mass of hot Yang-Mills theories~\cite{Ghisoiu:2015uza},
as reported in eq.~\eqref{eq:new debye} in the main body.

{\small
%

\begin{thebibliography}{100}

\bibitem{Harry:2006fi}
G.~M. Harry, P.~Fritschel, D.~A. Shaddock, W.~Folkner, and E.~S. Phinney,
  {\em{Laser interferometry for the big bang observer},}
  \href{http://dx.doi.org/10.1088/0264-9381/23/15/008}{Class. Quant. Grav. {\bf
  23} (2006) 4887}.

\bibitem{Kawamura:2006up}
S.~Kawamura {\em et~al.}, {\em{The Japanese space gravitational wave antenna
  DECIGO},} \href{http://dx.doi.org/10.1088/0264-9381/23/8/S17}{Class. Quant.
  Grav. {\bf 23} (2006) S125}.

\bibitem{Ruan:2018tsw}
W.-H. Ruan, Z.-K. Guo, R.-G. Cai, and Y.-Z. Zhang, {\em{Taiji program:
  Gravitational-wave sources},}
  \href{http://dx.doi.org/10.1142/S0217751X2050075X}{Int. J. Mod. Phys. A {\bf
  35} (2020) 2050075} [\href{http://arxiv.org/abs/1807.09495}{{\ttfamily
  1807.09495}}].

\bibitem{LIGOScientific:2014pky}
{\bfseries LIGO Scientific} Collaboration, J.~Aasi {\em et~al.}, {\em{Advanced
  LIGO},} \href{http://dx.doi.org/10.1088/0264-9381/32/7/074001}{Class. Quant.
  Grav. {\bf 32} (2015) 074001}
  [\href{http://arxiv.org/abs/1411.4547}{{\ttfamily 1411.4547}}].

\bibitem{Caprini:2019egz}
C.~Caprini {\em et~al.}, {\em{Detecting gravitational waves from cosmological
  phase transitions with LISA: an update},}
  \href{http://dx.doi.org/10.1088/1475-7516/2020/03/024}{JCAP {\bf 03} (2020)
  024} [\href{http://arxiv.org/abs/1910.13125}{{\ttfamily 1910.13125}}].

\bibitem{NANOGrav:2020bcs}
{\bfseries NANOGrav} Collaboration, Z.~Arzoumanian {\em et~al.}, {\em{The
  NANOGrav 12.5 yr Data Set: Search for an Isotropic Stochastic
  Gravitational-wave Background},}
  \href{http://dx.doi.org/10.3847/2041-8213/abd401}{Astrophys. J. Lett. {\bf
  905} (2020) L34} [\href{http://arxiv.org/abs/2009.04496}{{\ttfamily
  2009.04496}}].

\bibitem{Kajantie:1996mn}
K.~Kajantie, M.~Laine, K.~Rummukainen, and M.~E. Shaposhnikov, {\em{Is there a~
  hot electroweak phase transition at $m_H \gtrsim m_W$?},}
  \href{http://dx.doi.org/10.1103/PhysRevLett.77.2887}{Phys. Rev. Lett. {\bf
  77} (1996) 2887} [\href{http://arxiv.org/abs/hep-ph/9605288}{{\ttfamily
  hep-ph/9605288}}].

\bibitem{Gurtler:1997hr}
M.~Gurtler, E.-M. Ilgenfritz, and A.~Schiller, {\em{Where the electroweak phase
  transition ends},} \href{http://dx.doi.org/10.1103/PhysRevD.56.3888}{Phys.
  Rev. D {\bf 56} (1997) 3888}
  [\href{http://arxiv.org/abs/hep-lat/9704013}{{\ttfamily hep-lat/9704013}}].

\bibitem{Csikor:1998eu}
F.~Csikor, Z.~Fodor, and J.~Heitger, {\em{Endpoint of the hot electroweak phase
  transition},} \href{http://dx.doi.org/10.1103/PhysRevLett.82.21}{Phys. Rev.
  Lett. {\bf 82} (1999) 21}
  [\href{http://arxiv.org/abs/hep-ph/9809291}{{\ttfamily hep-ph/9809291}}].

\bibitem{Braaten:1995cm}
E.~Braaten and A.~Nieto, {\em{Effective field theory approach to high
  temperature thermodynamics},}
  \href{http://dx.doi.org/10.1103/PhysRevD.51.6990}{Phys. Rev. D {\bf 51}
  (1995) 6990} [\href{http://arxiv.org/abs/hep-ph/9501375}{{\ttfamily
  hep-ph/9501375}}].

\bibitem{Braaten:1994na}
E.~Braaten, {\em{Solution to the perturbative infrared catastrophe of hot gauge
  theories},} \href{http://dx.doi.org/10.1103/PhysRevLett.74.2164}{Phys. Rev.
  Lett. {\bf 74} (1995) 2164}
  [\href{http://arxiv.org/abs/hep-ph/9409434}{{\ttfamily hep-ph/9409434}}].

\bibitem{Braaten:1995jr}
E.~Braaten and A.~Nieto, {\em{Free energy of QCD at high temperature},}
  \href{http://dx.doi.org/10.1103/PhysRevD.53.3421}{Phys. Rev. D {\bf 53}
  (1996) 3421} [\href{http://arxiv.org/abs/hep-ph/9510408}{{\ttfamily
  hep-ph/9510408}}].

\bibitem{Kajantie:1997tt}
K.~Kajantie, M.~Laine, K.~Rummukainen, and M.~E. Shaposhnikov, {\em{3-D SU(N) +
  adjoint Higgs theory and finite temperature QCD},}
  \href{http://dx.doi.org/10.1016/S0550-3213(97)00425-2}{Nucl. Phys. B {\bf
  503} (1997) 357} [\href{http://arxiv.org/abs/hep-ph/9704416}{{\ttfamily
  hep-ph/9704416}}].

\bibitem{Laine:2019uua}
M.~Laine, P.~Schicho, and Y.~Schr{\"o}der, {\em{A QCD Debye mass in a broad
  temperature range},}
  \href{http://dx.doi.org/10.1103/PhysRevD.101.023532}{Phys. Rev. D {\bf 101}
  (2020) 023532} [\href{http://arxiv.org/abs/1911.09123}{{\ttfamily
  1911.09123}}].

\bibitem{Laine:2018lgj}
M.~Laine, P.~Schicho, and Y.~Schr{\"o}der, {\em{Soft thermal contributions to
  3-loop gauge coupling},}
  \href{http://dx.doi.org/10.1007/JHEP05(2018)037}{JHEP {\bf 05} (2018) 037}
  [\href{http://arxiv.org/abs/1803.08689}{{\ttfamily 1803.08689}}].

\bibitem{Ghiglieri:2021bom}
J.~Ghiglieri, G.~D. Moore, P.~Schicho, and N.~Schlusser, {\em{The
  force-force-correlator in hot QCD perturbatively and from the lattice},}
  \href{http://dx.doi.org/10.1007/JHEP02(2022)058}{JHEP {\bf 02} (2022) 058}
  [\href{http://arxiv.org/abs/2112.01407}{{\ttfamily 2112.01407}}].

\bibitem{Navarrete:2024ruu}
P.~Navarrete and Y.~Schr{\"o}der, {\em{The g$^{6}$ pressure of hot Yang-Mills
  theory: canonical form of the integrand},}
  \href{http://dx.doi.org/10.1007/JHEP11(2024)037}{JHEP {\bf 11} (2024) 037}
  [\href{http://arxiv.org/abs/2408.15830}{{\ttfamily 2408.15830}}].

\bibitem{Gorda:2025cwu}
T.~Gorda, P.~Navarrete, R.~Paatelainen, L.~Sandbote, and K.~Sepp{\"a}nen,
  {\em{A new approach to determine the thermodynamics of deconfined matter to
  high accuracy},} [\href{http://arxiv.org/abs/2511.09627}{{\ttfamily
  2511.09627}}].

\bibitem{Matsubara:1955ws}
T.~Matsubara, {\em{A New approach to quantum statistical mechanics},}
  \href{http://dx.doi.org/10.1143/PTP.14.351}{Prog. Theor. Phys. {\bf 14}
  (1955) 351}.

\bibitem{Ginsparg:1980ef}
P.~H. Ginsparg, {\em{First Order and Second Order Phase Transitions in Gauge
  Theories at Finite Temperature},}
  \href{http://dx.doi.org/10.1016/0550-3213(80)90418-6}{Nucl. Phys. B {\bf 170}
  (1980) 388}.

\bibitem{Appelquist:1981vg}
T.~Appelquist and R.~D. Pisarski, {\em{High-Temperature Yang-Mills Theories and
  Three-Dimensional Quantum Chromodynamics},}
  \href{http://dx.doi.org/10.1103/PhysRevD.23.2305}{Phys. Rev. D {\bf 23}
  (1981) 2305}.

\bibitem{Brauner:2016fla}
T.~Brauner, T.~V.~I. Tenkanen, A.~Tranberg, A.~Vuorinen, and D.~J. Weir,
  {\em{Dimensional reduction of the Standard Model coupled to a new singlet
  scalar field},} \href{http://dx.doi.org/10.1007/JHEP03(2017)007}{JHEP {\bf
  03} (2017) 007} [\href{http://arxiv.org/abs/1609.06230}{{\ttfamily
  1609.06230}}].

\bibitem{Andersen:2017ika}
J.~O. Andersen, T.~Gorda, A.~Helset, {\em et~al.}, {\em{Nonperturbative
  Analysis of the Electroweak Phase Transition in the Two Higgs Doublet
  Model},} \href{http://dx.doi.org/10.1103/PhysRevLett.121.191802}{Phys. Rev.
  Lett. {\bf 121} (2018) 191802}
  [\href{http://arxiv.org/abs/1711.09849}{{\ttfamily 1711.09849}}].

\bibitem{Niemi:2018asa}
L.~Niemi, H.~H. Patel, M.~J. Ramsey-Musolf, T.~V.~I. Tenkanen, and D.~J. Weir,
  {\em{Electroweak phase transition in the real triplet extension of the SM:
  Dimensional reduction},}
  \href{http://dx.doi.org/10.1103/PhysRevD.100.035002}{Phys. Rev. D {\bf 100}
  (2019) 035002} [\href{http://arxiv.org/abs/1802.10500}{{\ttfamily
  1802.10500}}].

\bibitem{Gorda:2018hvi}
T.~Gorda, A.~Helset, L.~Niemi, T.~V.~I. Tenkanen, and D.~J. Weir,
  {\em{Three-dimensional effective theories for the two Higgs doublet model at
  high temperature},} \href{http://dx.doi.org/10.1007/JHEP02(2019)081}{JHEP
  {\bf 02} (2019) 081} [\href{http://arxiv.org/abs/1802.05056}{{\ttfamily
  1802.05056}}].

\bibitem{Kainulainen:2019kyp}
K.~Kainulainen, V.~Keus, L.~Niemi, K.~Rummukainen, T.~V.~I. Tenkanen, and
  V.~Vaskonen, {\em{On the validity of perturbative studies of the electroweak
  phase transition in the Two Higgs Doublet model},}
  \href{http://dx.doi.org/10.1007/JHEP06(2019)075}{JHEP {\bf 06} (2019) 075}
  [\href{http://arxiv.org/abs/1904.01329}{{\ttfamily 1904.01329}}].

\bibitem{Croon:2020cgk}
D.~Croon, O.~Gould, P.~Schicho, T.~V.~I. Tenkanen, and G.~White,
  {\em{Theoretical uncertainties for cosmological first-order phase
  transitions},} \href{http://dx.doi.org/10.1007/JHEP04(2021)055}{JHEP {\bf 04}
  (2021) 055} [\href{http://arxiv.org/abs/2009.10080}{{\ttfamily 2009.10080}}].

\bibitem{Gould:2019qek}
O.~Gould, J.~Kozaczuk, L.~Niemi, M.~J. Ramsey-Musolf, T.~V.~I. Tenkanen, and
  D.~J. Weir, {\em{Nonperturbative analysis of the gravitational waves from a
  first-order electroweak phase transition},}
  \href{http://dx.doi.org/10.1103/PhysRevD.100.115024}{Phys. Rev. D {\bf 100}
  (2019) 115024} [\href{http://arxiv.org/abs/1903.11604}{{\ttfamily
  1903.11604}}].

\bibitem{Niemi:2020hto}
L.~Niemi, M.~J. Ramsey-Musolf, T.~V.~I. Tenkanen, and D.~J. Weir,
  {\em{Thermodynamics of a Two-Step Electroweak Phase Transition},}
  \href{http://dx.doi.org/10.1103/PhysRevLett.126.171802}{Phys. Rev. Lett. {\bf
  126} (2021) 171802} [\href{http://arxiv.org/abs/2005.11332}{{\ttfamily
  2005.11332}}].

\bibitem{Gould:2021ccf}
O.~Gould and J.~Hirvonen, {\em{Effective field theory approach to thermal
  bubble nucleation},}
  \href{http://dx.doi.org/10.1103/PhysRevD.104.096015}{Phys. Rev. D {\bf 104}
  (2021) 096015} [\href{http://arxiv.org/abs/2108.04377}{{\ttfamily
  2108.04377}}].

\bibitem{Gould:2021dzl}
O.~Gould, {\em{Real scalar phase transitions: a nonperturbative analysis},}
  \href{http://dx.doi.org/10.1007/JHEP04(2021)057}{JHEP {\bf 04} (2021) 057}
  [\href{http://arxiv.org/abs/2101.05528}{{\ttfamily 2101.05528}}].

\bibitem{Schicho:2021gca}
P.~M. Schicho, T.~V.~I. Tenkanen, and J.~\"Osterman, {\em{Robust approach to
  thermal resummation: Standard Model meets a singlet},}
  \href{http://dx.doi.org/10.1007/JHEP06(2021)130}{JHEP {\bf 06} (2021) 130}
  [\href{http://arxiv.org/abs/2102.11145}{{\ttfamily 2102.11145}}].

\bibitem{Niemi:2021qvp}
L.~Niemi, P.~Schicho, and T.~V.~I. Tenkanen, {\em{Singlet-assisted electroweak
  phase transition at two loops},}
  \href{http://dx.doi.org/10.1103/PhysRevD.103.115035}{Phys. Rev. D {\bf 103}
  (2021) 115035} [\href{http://arxiv.org/abs/2103.07467}{{\ttfamily
  2103.07467}}].

\bibitem{Camargo-Molina:2021zgz}
J.~E. Camargo-Molina, R.~Enberg, and J.~L\"ofgren, {\em{A new perspective on
  the electroweak phase transition in the Standard Model Effective Field
  Theory},} \href{http://dx.doi.org/10.1007/JHEP10(2021)127}{JHEP {\bf 10}
  (2021) 127} [\href{http://arxiv.org/abs/2103.14022}{{\ttfamily 2103.14022}}].

\bibitem{Niemi:2022bjg}
L.~Niemi, K.~Rummukainen, R.~Sepp\"a, and D.~J. Weir, {\em{Infrared physics of
  the 3D SU(2) adjoint Higgs model at the crossover transition},}
  \href{http://dx.doi.org/10.1007/JHEP02(2023)212}{JHEP {\bf 02} (2023) 212}
  [\href{http://arxiv.org/abs/2206.14487}{{\ttfamily 2206.14487}}].

\bibitem{Ekstedt:2022ceo}
A.~Ekstedt, {\em{Convergence of the nucleation rate for first-order phase
  transitions},} \href{http://dx.doi.org/10.1103/PhysRevD.106.095026}{Phys.
  Rev. D {\bf 106} (2022) 095026}
  [\href{http://arxiv.org/abs/2205.05145}{{\ttfamily 2205.05145}}].

\bibitem{Gould:2022ran}
O.~Gould, S.~G\"uyer, and K.~Rummukainen, {\em{First-order electroweak phase
  transitions: A nonperturbative update},}
  \href{http://dx.doi.org/10.1103/PhysRevD.106.114507}{Phys. Rev. D {\bf 106}
  (2022) 114507} [\href{http://arxiv.org/abs/2205.07238}{{\ttfamily
  2205.07238}}].

\bibitem{Ekstedt:2022zro}
A.~Ekstedt, O.~Gould, and J.~L{\"o}fgren, {\em{Radiative first-order phase
  transitions to next-to-next-to-leading order},}
  \href{http://dx.doi.org/10.1103/PhysRevD.106.036012}{Phys. Rev. D {\bf 106}
  (2022) 036012} [\href{http://arxiv.org/abs/2205.07241}{{\ttfamily
  2205.07241}}].

\bibitem{Biondini:2022ggt}
S.~Biondini, P.~Schicho, and T.~V.~I. Tenkanen, {\em{Strong electroweak phase
  transition in t-channel simplified dark matter models},}
  \href{http://dx.doi.org/10.1088/1475-7516/2022/10/044}{JCAP {\bf 10} (2022)
  044} [\href{http://arxiv.org/abs/2207.12207}{{\ttfamily 2207.12207}}].

\bibitem{Schicho:2022wty}
P.~Schicho, T.~V.~I. Tenkanen, and G.~White, {\em{Combining thermal resummation
  and gauge invariance for electroweak phase transition},}
  \href{http://dx.doi.org/10.1007/JHEP11(2022)047}{JHEP {\bf 11} (2022) 047}
  [\href{http://arxiv.org/abs/2203.04284}{{\ttfamily 2203.04284}}].

\bibitem{Lofgren:2021ogg}
J.~L\"ofgren, M.~J. Ramsey-Musolf, P.~Schicho, and T.~V.~I. Tenkanen,
  {\em{Nucleation at Finite Temperature: A Gauge-Invariant Perturbative
  Framework},} \href{http://dx.doi.org/10.1103/PhysRevLett.130.251801}{Phys.
  Rev. Lett. {\bf 130} (2023) 251801}
  [\href{http://arxiv.org/abs/2112.05472}{{\ttfamily 2112.05472}}].

\bibitem{Gould:2023jbz}
O.~Gould and C.~Xie, {\em{Higher orders for cosmological phase transitions: a
  global study in a Yukawa model},}
  \href{http://dx.doi.org/10.1007/JHEP12(2023)049}{JHEP {\bf 12} (2023) 049}
  [\href{http://arxiv.org/abs/2310.02308}{{\ttfamily 2310.02308}}].

\bibitem{Kierkla:2023von}
M.~Kierkla, B.~Swiezewska, T.~V.~I. Tenkanen, and J.~van~de Vis,
  {\em{Gravitational waves from supercooled phase transitions: dimensional
  transmutation meets dimensional reduction},}
  \href{http://dx.doi.org/10.1007/JHEP02(2024)234}{JHEP {\bf 02} (2024) 234}
  [\href{http://arxiv.org/abs/2312.12413}{{\ttfamily 2312.12413}}].

\bibitem{Aarts:2023vsf}
G.~Aarts {\em et~al.}, {\em{Phase Transitions in Particle Physics}: {Results
  and Perspectives from Lattice Quantum Chromo-Dynamics},}
  \href{http://dx.doi.org/10.1016/j.ppnp.2023.104070}{Prog. Part. Nucl. Phys.
  {\bf 133} (2023) 104070} [\href{http://arxiv.org/abs/2301.04382}{{\ttfamily
  2301.04382}}].

\bibitem{Niemi:2024axp}
L.~Niemi, M.~J. Ramsey-Musolf, and G.~Xia, {\em{Nonperturbative study of the
  electroweak phase transition in the real scalar singlet extended standard
  model},} \href{http://dx.doi.org/10.1103/PhysRevD.110.115016}{Phys. Rev. D
  {\bf 110} (2024) 115016} [\href{http://arxiv.org/abs/2405.01191}{{\ttfamily
  2405.01191}}].

\bibitem{Chala:2024xll}
M.~Chala, J.~C. Criado, L.~Gil, and J.~L. Miras, {\em{Higher-order-operator
  corrections to phase-transition parameters in dimensional reduction},}
  \href{http://dx.doi.org/10.1007/JHEP10(2024)025}{JHEP {\bf 10} (2024) 025}
  [\href{http://arxiv.org/abs/2406.02667}{{\ttfamily 2406.02667}}].

\bibitem{Qin:2024idc}
R.~Qin and L.~Bian, {\em{First-order electroweak phase transition at finite
  density},} \href{http://dx.doi.org/10.1007/JHEP08(2024)157}{JHEP {\bf 08}
  (2024) 157} [\href{http://arxiv.org/abs/2407.01981}{{\ttfamily 2407.01981}}].

\bibitem{Gould:2024jjt}
O.~Gould and P.~M. Saffin, {\em{Perturbative gravitational wave predictions for
  the real-scalar extended Standard Model},}
  \href{http://dx.doi.org/10.1007/JHEP03(2025)105}{JHEP {\bf 03} (2025) 105}
  [\href{http://arxiv.org/abs/2411.08951}{{\ttfamily 2411.08951}}].

\bibitem{Chakrabortty:2024wto}
J.~Chakrabortty and S.~Mohanty, {\em{One Loop Thermal Effective Action},}
  \href{http://dx.doi.org/10.1016/j.nuclphysb.2025.117165}{Nucl. Phys. B {\bf
  1020} (2025) 117165} [\href{http://arxiv.org/abs/2411.14146}{{\ttfamily
  2411.14146}}].

\bibitem{Niemi:2024vzw}
L.~Niemi and T.~V.~I. Tenkanen, {\em{Investigating two-loop effects for
  first-order electroweak phase transitions},}
  \href{http://dx.doi.org/10.1103/PhysRevD.111.075034}{Phys. Rev. D {\bf 111}
  (2025) 075034} [\href{http://arxiv.org/abs/2408.15912}{{\ttfamily
  2408.15912}}].

\bibitem{Kierkla:2025qyz}
M.~Kierkla, P.~Schicho, B.~Swiezewska, T.~V.~I. Tenkanen, and J.~van~de Vis,
  {\em{Finite-temperature bubble nucleation with shifting scale hierarchies},}
  \href{http://dx.doi.org/10.1007/JHEP07(2025)153}{JHEP {\bf 07} (2025) 153}
  [\href{http://arxiv.org/abs/2503.13597}{{\ttfamily 2503.13597}}].

\bibitem{Bhatnagar:2025jhh}
A.~Bhatnagar, D.~Croon, and P.~Schicho, {\em{Interpreting the 95 GeV resonance
  in the Two Higgs Doublet Model: Implications for the Electroweak Phase
  Transition},} [\href{http://arxiv.org/abs/2506.20716}{{\ttfamily
  2506.20716}}].

\bibitem{Bernardo:2025vkz}
F.~Bernardo, P.~Klose, P.~Schicho, and T.~V.~I. Tenkanen,
  {\em{Higher-dimensional operators at finite temperature affect
  gravitational-wave predictions},}
  \href{http://dx.doi.org/10.1007/JHEP08(2025)109}{JHEP {\bf 08} (2025) 109}
  [\href{http://arxiv.org/abs/2503.18904}{{\ttfamily 2503.18904}}].

\bibitem{Chala:2025aiz}
M.~Chala and G.~Guedes, {\em{The high-temperature limit of the SM(EFT)},}
  \href{http://dx.doi.org/10.1007/JHEP07(2025)085}{JHEP {\bf 07} (2025) 085}
  [\href{http://arxiv.org/abs/2503.20016}{{\ttfamily 2503.20016}}].

\bibitem{Zhu:2025pht}
Y.~Zhu, J.~Liu, R.~Qin, and L.~Bian, {\em{Theoretical uncertainties in
  first-order electroweak phase transitions},}
  \href{http://dx.doi.org/10.1103/f4gr-hycg}{Phys. Rev. D {\bf 112} (2025)
  015018} [\href{http://arxiv.org/abs/2503.19566}{{\ttfamily 2503.19566}}].

\bibitem{Chala:2025oul}
M.~Chala, L.~Gil, and Z.~Ren, {\em{Phase transitions in dimensional reduction
  up to three loops},} \href{http://dx.doi.org/10.1088/1674-1137/adf322}{Chin.
  Phys. {\bf 49} (2025) 123105}
  [\href{http://arxiv.org/abs/2505.14335}{{\ttfamily 2505.14335}}].

\bibitem{Li:2025kyo}
X.-X. Li, M.~J. Ramsey-Musolf, T.~V.~I. Tenkanen, and Y.~Wu, {\em{An Effective
  Sphaleron Awakens},} [\href{http://arxiv.org/abs/2506.01585}{{\ttfamily
  2506.01585}}].

\bibitem{Annala:2025aci}
J.~Annala, K.~Rummukainen, and T.~V.~I. Tenkanen, {\em{Nonperturbative
  determination of the sphaleron rate for first-order phase transitions},}
  \href{http://dx.doi.org/10.1103/q1jq-gq9m}{Phys. Rev. D {\bf 113} (2026)
  016014} [\href{http://arxiv.org/abs/2506.04939}{{\ttfamily 2506.04939}}].

\bibitem{Navarrete:2025yxy}
P.~Navarrete, R.~Paatelainen, K.~Sepp{\"a}nen, and T.~V.~I. Tenkanen,
  {\em{Cosmological phase transitions without high-temperature expansions},}
  \href{http://dx.doi.org/10.1007/JHEP01(2026)113}{JHEP {\bf 01} (2026) 113}
  [\href{http://arxiv.org/abs/2507.07014}{{\ttfamily 2507.07014}}].

\bibitem{Chala:2025xlk}
M.~Chala, M.~C. Fiore, and L.~Gil, {\em{Phase diagram of the standard model
  effective field theory},} \href{http://dx.doi.org/10.1103/zxzs-d6zt}{Phys.
  Rev. D {\bf 113} (2026) 115061}
  [\href{http://arxiv.org/abs/2507.16905}{{\ttfamily 2507.16905}}].

\bibitem{Chala:2025cya}
M.~Chala, A.~Dashko, and G.~Guedes, {\em{Running couplings in high-temperature
  effective field theory},} \href{http://dx.doi.org/10.1103/ptqg-5g38}{Phys.
  Rev. D {\bf 113} (2026) 055026}
  [\href{http://arxiv.org/abs/2510.26878}{{\ttfamily 2510.26878}}].

\bibitem{Biekotter:2025npc}
T.~Biek{\"o}tter, A.~Dashko, M.~L{\"o}schner, and G.~Weiglein,
  {\em{Perturbative aspects of the electroweak phase transition with a complex
  singlet and implications for gravitational wave predictions},}
  [\href{http://arxiv.org/abs/2511.14831}{{\ttfamily 2511.14831}}].

\bibitem{Chakrabortty:2026swu}
J.~Chakrabortty, B.~S. Eduardo, S.~Karmakar, and P.~Schicho,
  {\em{Finite-temperature operator basis on $\mathbb{R}^3 \times S^1$ for
  SMEFT},} [\href{http://arxiv.org/abs/2605.02878}{{\ttfamily 2605.02878}}].

\bibitem{Arnold:1992rz}
P.~B. Arnold and O.~Espinosa, {\em{The Effective potential and first order
  phase transitions: Beyond leading-order},}
  \href{http://dx.doi.org/10.1103/PhysRevD.47.3546}{Phys. Rev. D {\bf 47}
  (1993) 3546} [\href{http://arxiv.org/abs/hep-ph/9212235}{{\ttfamily
  hep-ph/9212235}}].

\bibitem{Ekstedt:2024etx}
A.~Ekstedt, P.~Schicho, and T.~V.~I. Tenkanen, {\em{Cosmological phase
  transitions at three loops: The final verdict on perturbation theory},}
  \href{http://dx.doi.org/10.1103/PhysRevD.110.096006}{Phys. Rev. D {\bf 110}
  (2024) 096006} [\href{http://arxiv.org/abs/2405.18349}{{\ttfamily
  2405.18349}}].

\bibitem{Davies:2026cci}
J.~Davies, T.~Kaneko, C.~Marinissen, T.~Ueda, and J.~A.~M. Vermaseren,
  {\em{FORM Version 5.0}} [\href{http://arxiv.org/abs/2601.19982}{{\ttfamily
  2601.19982}}].

\bibitem{Laporta:2000dsw}
S.~Laporta, {\em{High-precision calculation of multiloop Feynman integrals by
  difference equations},}
  \href{http://dx.doi.org/10.1142/S0217751X00002159}{Int. J. Mod. Phys. A {\bf
  15} (2000) 5087} [\href{http://arxiv.org/abs/hep-ph/0102033}{{\ttfamily
  hep-ph/0102033}}].

\bibitem{Nishimura:2012ee}
M.~Nishimura and Y.~Schr{\"o}der, {\em{IBP methods at finite temperature},}
  \href{http://dx.doi.org/10.1007/JHEP09(2012)051}{JHEP {\bf 09} (2012) 051}
  [\href{http://arxiv.org/abs/1207.4042}{{\ttfamily 1207.4042}}].

\bibitem{Ghisoiu:2015uza}
I.~Ghisoiu, J.~Moller, and Y.~Schr{\"o}der, {\em{Debye screening mass of hot
  Yang-Mills theory to three-loop order},}
  \href{http://dx.doi.org/10.1007/JHEP11(2015)121}{JHEP {\bf 11} (2015) 121}
  [\href{http://arxiv.org/abs/1509.08727}{{\ttfamily 1509.08727}}].

\bibitem{Gil:2026cqz}
L.~Gil, J.~L{\'o}pez~Miras, and A.~Moreno-S{\'a}nchez, {\em{SIRENA --
  Sum-Integral REductioN Algorithm},}
  [\href{http://arxiv.org/abs/2605.06775}{{\ttfamily 2605.06775}}].

\bibitem{Jaeckel:2016jlh}
J.~Jaeckel, V.~V. Khoze, and M.~Spannowsky, {\em{Hearing the signal of dark
  sectors with gravitational wave detectors},}
  \href{http://dx.doi.org/10.1103/PhysRevD.94.103519}{Phys. Rev. D {\bf 94}
  (2016) 103519} [\href{http://arxiv.org/abs/1602.03901}{{\ttfamily
  1602.03901}}].

\bibitem{Addazi:2017gpt}
A.~Addazi and A.~Marciano, {\em{Gravitational waves from dark first order phase
  transitions and dark photons},}
  \href{http://dx.doi.org/10.1088/1674-1137/42/2/023107}{Chin. Phys. C {\bf 42}
  (2018) 023107} [\href{http://arxiv.org/abs/1703.03248}{{\ttfamily
  1703.03248}}].

\bibitem{Croon:2018erz}
D.~Croon, V.~Sanz, and G.~White, {\em{Model Discrimination in Gravitational
  Wave spectra from Dark Phase Transitions},}
  \href{http://dx.doi.org/10.1007/JHEP08(2018)203}{JHEP {\bf 08} (2018) 203}
  [\href{http://arxiv.org/abs/1806.02332}{{\ttfamily 1806.02332}}].

\bibitem{Breitbach:2018ddu}
M.~Breitbach, J.~Kopp, E.~Madge, T.~Opferkuch, and P.~Schwaller, {\em{Dark,
  Cold, and Noisy: Constraining Secluded Hidden Sectors with Gravitational
  Waves},} \href{http://dx.doi.org/10.1088/1475-7516/2019/07/007}{JCAP {\bf 07}
  (2019) 007} [\href{http://arxiv.org/abs/1811.11175}{{\ttfamily 1811.11175}}].

\bibitem{Christiansen:2025xhv}
M.~Christiansen, E.~Madge, C.~Puchades-Ib{\'a}{\~n}ez, M.~E. Ramirez-Quezada,
  and P.~Schwaller, {\em{Beyond the Daisy Chain: Running and the 3D EFT View of
  Supercooled Phase Transitions},}
  [\href{http://arxiv.org/abs/2511.02910}{{\ttfamily 2511.02910}}].

\bibitem{Halperin:1973jh}
B.~i. Halperin, T.~C. Lubensky, and S.-k. Ma, {\em{First order phase
  transitions in superconductors and smectic A liquid crystals},}
  \href{http://dx.doi.org/10.1103/PhysRevLett.32.292}{Phys. Rev. Lett. {\bf 32}
  (1974) 292}.

\bibitem{Dasgupta:1981zz}
C.~Dasgupta and B.~I. Halperin, {\em{Phase Transition in a Lattice Model of
  Superconductivity},}
  \href{http://dx.doi.org/10.1103/PhysRevLett.47.1556}{Phys. Rev. Lett. {\bf
  47} (1981) 1556}.

\bibitem{Kajantie:1998zn}
K.~Kajantie, M.~Laine, T.~Neuhaus, J.~Peisa, A.~Rajantie, and K.~Rummukainen,
  {\em{Vortex tension as an order parameter in three-dimensional U(1) + Higgs
  theory},} \href{http://dx.doi.org/10.1016/S0550-3213(99)00033-4}{Nucl. Phys.
  B {\bf 546} (1999) 351}
  [\href{http://arxiv.org/abs/hep-ph/9809334}{{\ttfamily hep-ph/9809334}}].

\bibitem{Lewicki:2024sfw}
M.~Lewicki, P.~Toczek, and V.~Vaskonen, {\em{Black holes and gravitational
  waves from phase transitions in realistic models},}
  \href{http://dx.doi.org/10.1016/j.dark.2025.102075}{Phys. Dark Univ. {\bf 50}
  (2025) 102075} [\href{http://arxiv.org/abs/2412.10366}{{\ttfamily
  2412.10366}}].

\bibitem{Franciolini:2025ztf}
G.~Franciolini, Y.~Gouttenoire, and R.~Jinno, {\em{Curvature Perturbations from
  First-Order Phase Transitions: Implications to Black Holes and Gravitational
  Waves},} \href{http://dx.doi.org/10.1103/tfcx-kzqx}{Phys. Rev. Lett. {\bf
  136} (2026) 171404} [\href{http://arxiv.org/abs/2503.01962}{{\ttfamily
  2503.01962}}].

\bibitem{Kierkla:2025vwp}
M.~Kierkla, N.~Ramberg, P.~Schicho, and D.~Schmitt, {\em{Thermodynamical
  uncertainties for primordial black holes from cosmological phase
  transitions},} \href{http://dx.doi.org/10.1103/nj27-ltyg}{Phys. Rev. D {\bf
  113} (2026) 095024} [\href{http://arxiv.org/abs/2506.15496}{{\ttfamily
  2506.15496}}].

\bibitem{Gould:2023ovu}
O.~Gould and T.~V.~I. Tenkanen, {\em{Perturbative effective field theory
  expansions for cosmological phase transitions},}
  \href{http://dx.doi.org/10.1007/JHEP01(2024)048}{JHEP {\bf 01} (2024) 048}
  [\href{http://arxiv.org/abs/2309.01672}{{\ttfamily 2309.01672}}].

\bibitem{Linde:1980ts}
A.~D. Linde, {\em{Infrared Problem in Thermodynamics of the Yang-Mills Gas},}
  \href{http://dx.doi.org/10.1016/0370-2693(80)90769-8}{Phys. Lett. B {\bf 96}
  (1980) 289}.

\bibitem{Kajantie:1995dw}
K.~Kajantie, M.~Laine, K.~Rummukainen, and M.~E. Shaposhnikov, {\em{Generic
  rules for high temperature dimensional reduction and their application to the
  standard model},} \href{http://dx.doi.org/10.1016/0550-3213(95)00549-8}{Nucl.
  Phys. B {\bf 458} (1996) 90}
  [\href{http://arxiv.org/abs/hep-ph/9508379}{{\ttfamily hep-ph/9508379}}].

\bibitem{Balui:2025yvd}
D.~Balui, T.~Biswas, J.~Chakrabortty, D.~Dey, C.~Englert, and S.~Mohanty,
  {\em{Gauge choices, infrared pitfalls, and thermal effects in effective
  potentials},} \href{http://dx.doi.org/10.1103/drsd-wfns}{Phys. Rev. D {\bf
  112} (2025) 056022} [\href{http://arxiv.org/abs/2507.22706}{{\ttfamily
  2507.22706}}].

\bibitem{Ghisoiu:2012yk}
I.~Ghisoiu and Y.~Schroder, {\em{A New Method for Taming Tensor
  Sum-Integrals},} \href{http://dx.doi.org/10.1007/JHEP11(2012)010}{JHEP {\bf
  11} (2012) 010} [\href{http://arxiv.org/abs/1208.0284}{{\ttfamily
  1208.0284}}].

\bibitem{Stevenson:1981vj}
P.~M. Stevenson, {\em{Optimized Perturbation Theory},}
  \href{http://dx.doi.org/10.1103/PhysRevD.23.2916}{Phys. Rev. D {\bf 23}
  (1981) 2916}.

\bibitem{Laine:2005ai}
M.~Laine and Y.~Schr{\"o}der, {\em{Two-loop QCD gauge coupling at high
  temperatures},} \href{http://dx.doi.org/10.1088/1126-6708/2005/03/067}{JHEP
  {\bf 03} (2005) 067} [\href{http://arxiv.org/abs/hep-ph/0503061}{{\ttfamily
  hep-ph/0503061}}].

\bibitem{Ghisoiu:2013zoj}
I.~Ghisoiu, {\em {Three-loop Debye mass and effective coupling in thermal
  QCD}}, PhD thesis, U. Bielefeld (main), 2013

\bibitem{Chala:2024llp}
M.~Chala, J.~L{\'o}pez~Miras, J.~Santiago, and F.~Vilches, {\em{Efficient
  on-shell matching},}
  \href{http://dx.doi.org/10.21468/SciPostPhys.18.6.185}{SciPost Phys. {\bf 18}
  (2025) 185} [\href{http://arxiv.org/abs/2411.12798}{{\ttfamily 2411.12798}}].

\bibitem{Camargo-Molina:2024sde}
E.~Camargo-Molina, R.~Enberg, and J.~L{\"o}fgren, {\em{A catalog of first-order
  electroweak phase transitions in the Standard Model Effective Field Theory},}
  \href{http://dx.doi.org/10.1007/JHEP08(2025)113}{JHEP {\bf 08} (2025) 113}
  [\href{http://arxiv.org/abs/2410.23210}{{\ttfamily 2410.23210}}].

\bibitem{Hirvonen:2021zej}
J.~Hirvonen, J.~L{\"o}fgren, M.~J. Ramsey-Musolf, P.~Schicho, and T.~V.~I.
  Tenkanen, {\em{Computing the gauge-invariant bubble nucleation rate in finite
  temperature effective field theory},}
  \href{http://dx.doi.org/10.1007/JHEP07(2022)135}{JHEP {\bf 07} (2022) 135}
  [\href{http://arxiv.org/abs/2112.08912}{{\ttfamily 2112.08912}}].

\bibitem{Kajantie:1995kf}
K.~Kajantie, M.~Laine, K.~Rummukainen, and M.~E. Shaposhnikov, {\em{The
  Electroweak phase transition: A Nonperturbative analysis},}
  \href{http://dx.doi.org/10.1016/0550-3213(96)00052-1}{Nucl. Phys. B {\bf 466}
  (1996) 189} [\href{http://arxiv.org/abs/hep-lat/9510020}{{\ttfamily
  hep-lat/9510020}}].

\bibitem{Farakos:1994xh}
K.~Farakos, K.~Kajantie, K.~Rummukainen, and M.~E. Shaposhnikov, {\em{3-d
  physics and the electroweak phase transition: A Framework for lattice Monte
  Carlo analysis},} \href{http://dx.doi.org/10.1016/0550-3213(95)80129-4}{Nucl.
  Phys. B {\bf 442} (1995) 317}
  [\href{http://arxiv.org/abs/hep-lat/9412091}{{\ttfamily hep-lat/9412091}}].

\bibitem{Kleinert:1986jp}
H.~Kleinert and W.~Miller, {\em{Renormalization of Charge in Villain Lattice
  Gauge Theory},} \href{http://dx.doi.org/10.1103/PhysRevLett.56.11}{Phys. Rev.
  Lett. {\bf 56} (1986) 11}.

\bibitem{Mo:2001fi}
S.~Mo, J.~Hove, and A.~Sudbo, {\em{The Order of the metal to superconductor
  transition},} \href{http://dx.doi.org/10.1103/PhysRevB.65.104501}{Phys. Rev.
  B {\bf 65} (2002) 104501}
  [\href{http://arxiv.org/abs/cond-mat/0109260}{{\ttfamily cond-mat/0109260}}].

\bibitem{Jansen:1985cq}
K.~Jansen, J.~Jersak, C.~B. Lang, T.~Neuhaus, and G.~Vones, {\em{Phase
  Structure of U(1) Gauge - Higgs Theory on $D=4$ Lattices},}
  \href{http://dx.doi.org/10.1016/0370-2693(85)90652-5}{Phys. Lett. B {\bf 155}
  (1985) 268}.

\bibitem{Kleinert:1986te}
H.~Kleinert, {\em{Tricritical Ratio of Length Scales in the $D=4$ Abelian Higgs
  Model},} \href{http://dx.doi.org/10.1103/PhysRevLett.56.1441}{Phys. Rev.
  Lett. {\bf 56} (1986) 1441}.

\bibitem{Coleman:1973jx}
S.~R. Coleman and E.~J. Weinberg, {\em{Radiative Corrections as the Origin of
  Spontaneous Symmetry Breaking},}
  \href{http://dx.doi.org/10.1103/PhysRevD.7.1888}{Phys. Rev. D {\bf 7} (1973)
  1888}.

\bibitem{Lewicki:2023ioy}
M.~Lewicki, P.~Toczek, and V.~Vaskonen, {\em{Primordial black holes from strong
  first-order phase transitions},}
  \href{http://dx.doi.org/10.1007/JHEP09(2023)092}{JHEP {\bf 09} (2023) 092}
  [\href{http://arxiv.org/abs/2305.04924}{{\ttfamily 2305.04924}}].

\bibitem{Carr:2026hot}
B.~Carr, A.~J. Iovino, G.~Perna, V.~Vaskonen, and H.~Veerm{\"a}e,
  {\em{Primordial black holes: constraints, potential evidence and prospects},}
  \href{http://dx.doi.org/10.1007/s40766-026-00080-z}{Riv. Nuovo Cim. {\bf 49}
  (2026) 225} [\href{http://arxiv.org/abs/2601.06024}{{\ttfamily 2601.06024}}].

\bibitem{Ghisoiu:2012kn}
I.~Ghisoiu and Y.~Schr{\"o}der, {\em{A new three-loop sum-integral of mass
  dimension two},} \href{http://dx.doi.org/10.1007/JHEP09(2012)016}{JHEP {\bf
  09} (2012) 016} [\href{http://arxiv.org/abs/1207.6214}{{\ttfamily
  1207.6214}}].

\bibitem{Curtin:2022ovx}
D.~Curtin, J.~Roy, and G.~White, {\em{Gravitational waves and tadpole
  resummation: Efficient and easy convergence of finite temperature QFT},}
  \href{http://dx.doi.org/10.1103/PhysRevD.109.116001}{Phys. Rev. D {\bf 109}
  (2024) 116001} [\href{http://arxiv.org/abs/2211.08218}{{\ttfamily
  2211.08218}}].

\bibitem{Collins:2016aya}
J.~C. Collins and J.~A.~M. Vermaseren, {\em{Axodraw Version 2},}
  [\href{http://arxiv.org/abs/1606.01177}{{\ttfamily 1606.01177}}].

\bibitem{Laine:2016hma}
M.~Laine and A.~Vuorinen,
  \href{http://dx.doi.org/10.1007/978-3-319-31933-9}{{\em {Basics of Thermal
  Field Theory}}}, vol.~925.
\newblock Springer, 2016, [\href{http://arxiv.org/abs/1701.01554}{{\ttfamily
  1701.01554}}].

\bibitem{Davydychev:2023jto}
A.~I. Davydychev, P.~Navarrete, and Y.~Schr{\"o}der, {\em{Factorizing two-loop
  vacuum sum-integrals},} \href{http://dx.doi.org/10.1007/JHEP02(2024)104}{JHEP
  {\bf 02} (2024) 104} [\href{http://arxiv.org/abs/2312.17367}{{\ttfamily
  2312.17367}}].

\bibitem{Davydychev:2022dcw}
A.~I. Davydychev and Y.~Schr{\"o}der, {\em{Recursion-free solution for two-loop
  vacuum integrals with {\textquotedblleft}collinear{\textquotedblright}
  masses},} \href{http://dx.doi.org/10.1007/JHEP12(2022)047}{JHEP {\bf 12}
  (2022) 047} [\href{http://arxiv.org/abs/2210.10593}{{\ttfamily 2210.10593}}].

\bibitem{Schicho:2020xaf}
P.~Schicho, \href{http://dx.doi.org/10.24442/BORISTHESES.1988}{{\em {Multi-loop
  investigations of strong interactions at high temperatures}}}, PhD thesis, U.
  Bern, 2020

\bibitem{Smirnov:2010hn}
A.~V. Smirnov and A.~V. Petukhov, {\em{The Number of Master Integrals is
  Finite},} \href{http://dx.doi.org/10.1007/s11005-010-0450-0}{Lett. Math.
  Phys. {\bf 97} (2011) 37} [\href{http://arxiv.org/abs/1004.4199}{{\ttfamily
  1004.4199}}].

\bibitem{Arnold:1994eb}
P.~B. Arnold and C.-x. Zhai, {\em{The Three loop free energy for high
  temperature QED and QCD with fermions},}
  \href{http://dx.doi.org/10.1103/PhysRevD.51.1906}{Phys. Rev. D {\bf 51}
  (1995) 1906} [\href{http://arxiv.org/abs/hep-ph/9410360}{{\ttfamily
  hep-ph/9410360}}].

\bibitem{Arnold:1994ps}
P.~B. Arnold and C.-X. Zhai, {\em{The Three loop free energy for pure gauge
  QCD},} \href{http://dx.doi.org/10.1103/PhysRevD.50.7603}{Phys. Rev. D {\bf
  50} (1994) 7603} [\href{http://arxiv.org/abs/hep-ph/9408276}{{\ttfamily
  hep-ph/9408276}}].

\bibitem{Catani:2008xa}
S.~Catani, T.~Gleisberg, F.~Krauss, G.~Rodrigo, and J.-C. Winter, {\em{From
  loops to trees by-passing Feynman's theorem},}
  \href{http://dx.doi.org/10.1088/1126-6708/2008/09/065}{JHEP {\bf 09} (2008)
  065} [\href{http://arxiv.org/abs/0804.3170}{{\ttfamily 0804.3170}}].

\bibitem{Seppanen:2025owq}
K.~Sepp{\"a}nen, {\em {Quark Matter Thermodynamics from High-Order Perturbative
  QCD}}, PhD thesis, Helsinki U., 2025

\bibitem{Moller:2012chx}
J.~Moller and Y.~Schr{\"o}der, {\em{Three-loop matching coefficients for hot
  QCD: Reduction and gauge independence},}
  \href{http://dx.doi.org/10.1007/JHEP08(2012)025}{JHEP {\bf 08} (2012) 025}
  [\href{http://arxiv.org/abs/1207.1309}{{\ttfamily 1207.1309}}].

\bibitem{Gynther:2007bw}
A.~Gynther, M.~Laine, Y.~Schr{\"o}der, C.~Torrero, and A.~Vuorinen,
  {\em{Four-loop pressure of massless O(N) scalar field theory},}
  \href{http://dx.doi.org/10.1088/1126-6708/2007/04/094}{JHEP {\bf 04} (2007)
  094} [\href{http://arxiv.org/abs/hep-ph/0703307}{{\ttfamily
  hep-ph/0703307}}].

\bibitem{Moller:2010xw}
J.~Moller and Y.~Schr{\"o}der, {\em{Open problems in hot QCD},}
  \href{http://dx.doi.org/10.1016/j.nuclphysbps.2010.08.046}{Nucl. Phys. B
  Proc. Suppl. {\bf 205-206} (2010) 218}
  [\href{http://arxiv.org/abs/1007.1223}{{\ttfamily 1007.1223}}].

\bibitem{Andersen:2008bz}
J.~O. Andersen and L.~Kyllingstad, {\em{Four-loop Screened Perturbation
  Theory},} \href{http://dx.doi.org/10.1103/PhysRevD.78.076008}{Phys. Rev. D
  {\bf 78} (2008) 076008} [\href{http://arxiv.org/abs/0805.4478}{{\ttfamily
  0805.4478}}].

\bibitem{Schroder:2012hm}
Y.~Schr{\"o}der, {\em{A fresh look on three-loop sum-integrals},}
  \href{http://dx.doi.org/10.1007/JHEP08(2012)095}{JHEP {\bf 08} (2012) 095}
  [\href{http://arxiv.org/abs/1207.5666}{{\ttfamily 1207.5666}}].

\bibitem{Farakos:1994kx}
K.~Farakos, K.~Kajantie, K.~Rummukainen, and M.~E. Shaposhnikov, {\em{3-D
  physics and the electroweak phase transition: Perturbation theory},}
  \href{http://dx.doi.org/10.1016/0550-3213(94)90173-2}{Nucl. Phys. B {\bf 425}
  (1994) 67} [\href{http://arxiv.org/abs/hep-ph/9404201}{{\ttfamily
  hep-ph/9404201}}].

\bibitem{Laine:1995np}
M.~Laine, {\em{Exact relation of lattice and continuum parameters in
  three-dimensional SU(2) + Higgs theories},}
  \href{http://dx.doi.org/10.1016/0550-3213(95)00356-W}{Nucl. Phys. B {\bf 451}
  (1995) 484} [\href{http://arxiv.org/abs/hep-lat/9504001}{{\ttfamily
  hep-lat/9504001}}].

\bibitem{Abbott:1980hw}
L.~F. Abbott, {\em{The Background Field Method Beyond One Loop},}
  \href{http://dx.doi.org/10.1016/0550-3213(81)90371-0}{Nucl. Phys. B {\bf 185}
  (1981) 189}.

\bibitem{Gynther:2005dj}
A.~Gynther and M.~Vepsalainen, {\em{Pressure of the standard model at high
  temperatures},} \href{http://dx.doi.org/10.1088/1126-6708/2006/01/060}{JHEP
  {\bf 01} (2006) 060} [\href{http://arxiv.org/abs/hep-ph/0510375}{{\ttfamily
  hep-ph/0510375}}].

\bibitem{Gynther:2005av}
A.~Gynther and M.~Vepsalainen, {\em{Pressure of the standard model near the
  electroweak phase transition},}
  \href{http://dx.doi.org/10.1088/1126-6708/2006/03/011}{JHEP {\bf 03} (2006)
  011} [\href{http://arxiv.org/abs/hep-ph/0512177}{{\ttfamily
  hep-ph/0512177}}].

\bibitem{Ekstedt:2022bff}
A.~Ekstedt, P.~Schicho, and T.~V.~I. Tenkanen, {\em{DRalgo: A package for
  effective field theory approach for thermal phase transitions},}
  \href{http://dx.doi.org/10.1016/j.cpc.2023.108725}{Comput. Phys. Commun. {\bf
  288} (2023) 108725} [\href{http://arxiv.org/abs/2205.08815}{{\ttfamily
  2205.08815}}].

\bibitem{Born:2024mgz}
L.~Born, J.~Fuentes-Mart{\'\i}n, S.~Kvedarait{\.{e}}, and A.~E. Thomsen,
  {\em{Two-loop running in the bosonic SMEFT using functional methods},}
  \href{http://dx.doi.org/10.1007/JHEP05(2025)121}{JHEP {\bf 05} (2025) 121}
  [\href{http://arxiv.org/abs/2410.07320}{{\ttfamily 2410.07320}}].

\bibitem{Abbott:1981ke}
L.~F. Abbott, {\em{Introduction to the Background Field Method},} Acta Phys.
  Polon. B {\bf 13} (1982) 33.

\end{thebibliography}

}
\end{document}